\documentclass[twocolumn,aps,prb,superscriptaddress,floatfix,showpacs]{revtex4}

\usepackage{graphicx,subfigure}
\usepackage{epstopdf}
\usepackage{amssymb,amsmath,amsfonts,hyperref,wasysym}
\usepackage{placeins} 
\usepackage{verbatim} 
\usepackage{color}
\usepackage{longtable}

\begin{document}

\newcommand{\myint}{{\!\int\!}}
\newcommand{\myintr}{{\!\int\!\! d^2 \mathbf{r}\:}}
\newcommand{\fld}[1]{{#1(\mathbf{r})}}
\newcommand{\SAVE}[1] {{}}  
\newcommand{\UNSURE}[1] {{\textbf{????}#1\textbf{????}}}  
\newcommand{\BULLET}[1] {{$\bullet$#1}}  
\newcommand{\rvec}{{\mathbf{r}}}
\newcommand{\kvec}{{\mathbf{k}}}
\newcommand{\pol}{{\mathbf{P}}}
\newcommand{\kpol}{{\tilde{\mathbf{P}}}}
\newcommand{\krho}{{\tilde{\rho}}}
\newcommand{\polkd}{{\mathbf{B}}} 
\newcommand{\kpolkd}{{\tilde{\mathbf{B}}}}

\newcommand{\SP}[1] {{\textcolor{red}{SP: #1}}}  
\newcommand{\ND}[1] {{\textcolor{blue}{ND: #1}}}  
\newcommand{\SPHIDE}[1] {{}}  
\newcommand{\NDHIDE}[1] {{}}  

\title{Bilayer Coulomb phase of two dimensional dimer models: Absence of power-law columnar order}

\author{Nisheeta Desai}
\affiliation{Dept. of Theoretical Physics,
Tata Institute of Fundamental Research, Mumbai 400 005, India.}
\author{Sumiran Pujari}
\affiliation{Dept. of Physics,
IIT Bombay, Powai, Mumbai, MH 400076, India.}
\author{Kedar Damle}
\affiliation{Dept. of Theoretical Physics,
Tata Institute of Fundamental Research, Mumbai 400 005, India.}

\begin{abstract}
Using renormalization group (RG) analyses and Monte Carlo (MC) simulations, we study the fully-packed dimer model on the bilayer square lattice with fugacity equal to $z$ ($1$) for inter-layer (intra-layer) dimers, and intra-layer interaction $V$ between neighbouring parallel dimers on any elementary plaquette in either layer. For a range of not-too-large $z> 0$ and repulsive interactions $0< V < V_s$ (with $V_s \approx 2.1$), we demonstrate the existence of a {\em bilayer Coulomb phase} with purely dipolar two-point functions, {\em i.e.}, without the power-law columnar order that characterizes the usual Coulomb phase of square and honeycomb lattice dimer models. The transition line $z_{c}(V)$ separating this bilayer Coulomb phase from a large-$z$ disordered phase is argued to be in the inverted Kosterlitz-Thouless universality class. Additionally, we argue for the possibility of a tricritical point at which the bilayer Coulomb phase, the large-$z$ disordered phase and the large-$V$ staggered phase meet in the large-$z$, large-$V$ part of the phase diagram. In contrast, for the attractive case with $ V_{cb} < V \leq 0$ ($V_{cb} \approx -1.2$), we argue that any $z > 0$ destroys the power-law correlations of the $z=0$ decoupled layers, and leads immediately to a short-range correlated state, albeit with a slow crossover for small $|V|$. For $V_{c} < V < V_{cb}$ ($V_{c} \approx -1.55$), we predict that any small nonzero $z$ immediately gives rise to long-range {\em bilayer columnar order} although the $z=0$ decoupled layers remain power-law correlated in this regime; this implies a non-monotonic $z$ dependence of the columnar order parameter for fixed $V$ in this regime.  Further, our RG arguments predict that this bilayer columnar ordered state is separated from the large-$z$ disordered state by a line of Ashkin-Teller transitions $z_{\rm AT}(V)$. Finally, for $V< V_{c}$, the $z=0$ decoupled layers are already characterized by long-range  columnar order, and a small nonzero $z$ leads immediately to a locking of the order parameters of the two layer, giving rise to the same bilayer columnar ordered state for small nonzero $z$.
\end{abstract}

\maketitle

\tableofcontents

\section{Introduction}
\label{sec:Introduction}
Dimer models on two and three dimensional bipartite lattices such as the square, the honeycomb,  and the cubic lattice represent paradigmatic classical examples of long-wavelength physics controlled by the fluctuations of an emergent 
gauge field~\cite{youngblood_axe_prb1981,youngblood_axe_mccoy_prb1980,Henley_review,huse_etal_prl2003,fradkin_etal_prb2004,papanikolaou_luijten_fradkin_prb2007,alet_etal_pre2006}
Specifically, the long-distance correlations between dimers are well-described on the square/honeycomb (cubic) lattice in terms of the Gaussian fluctuations of a divergence-free two-component (three-component) ``magnetic field'' parameterized by the corresponding ``vector potential''. This {\em Coulomb phase} phenomenology provides a simple classical example of the role of emergent degrees of freedom and entropic interactions in determining the long-wavelength properties of systems with a macroscopic degeneracy of low-energy configurations.\begin{figure}[!] 
 \includegraphics[width=0.8\columnwidth]{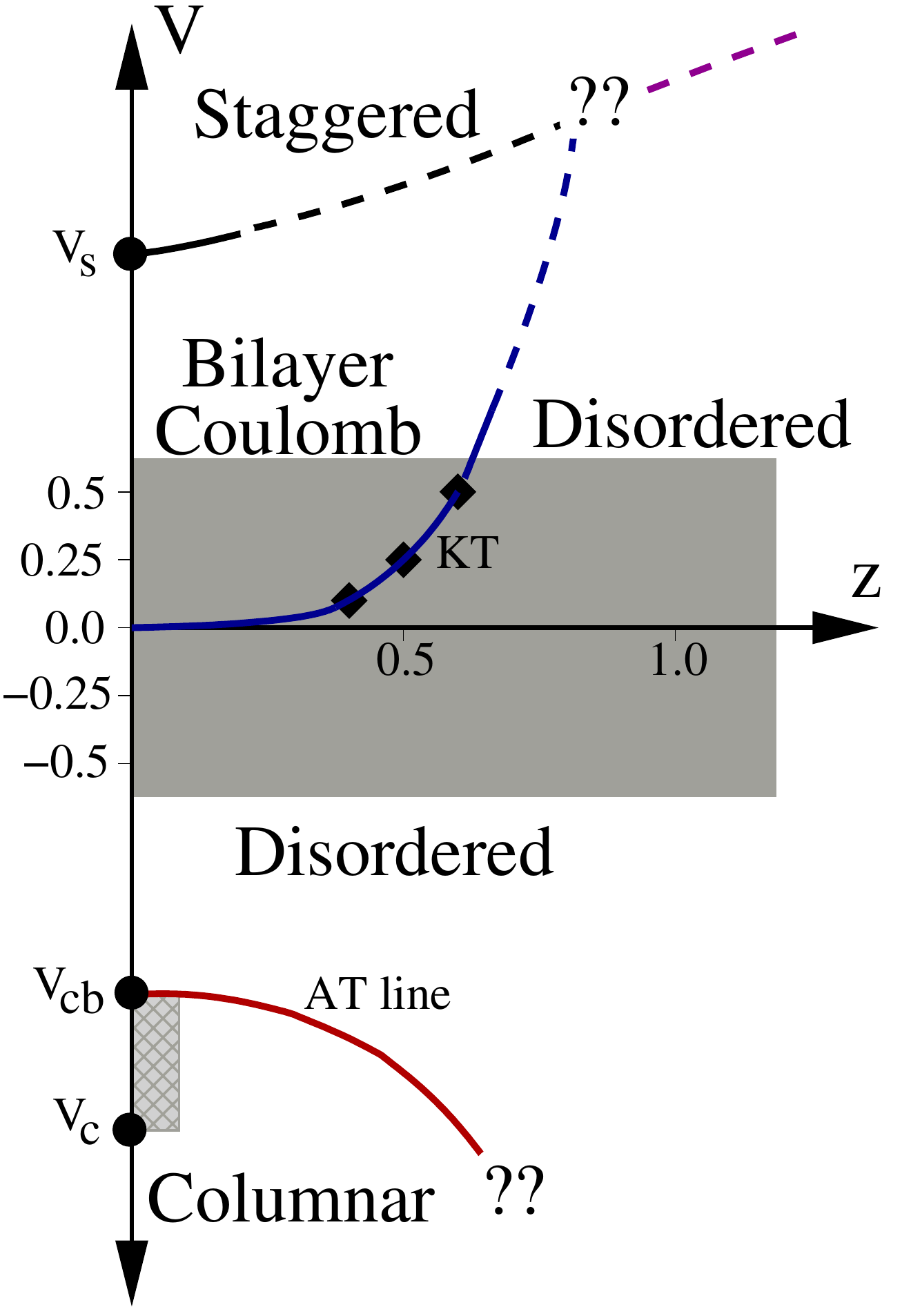}
\caption{\label{fig:SchematicPhaseDiagram}
Schematic phase diagram summarizing the results of our renormalization group
analysis and Monte Carlo studies. As described in Sec.~\ref{sec:model}, $V$ is the intralayer interaction between neighbouring parallel dimers within each layer of a fully-packed dimer model on the bilayer square lattice, and $z$ is the fugacity of interlayer dimers measured in units of the intralayer dimer fugacity. The ``AT line'' is a line of phase transitions in the Ashkin-Teller universality class, separating a phase with bilayer columnar order from a large-$z$ disordered phase; this critical line could potentially connect to a first order transition at larger attractive $|V|$. Likewise, ``KT'' labels a phase boundary in the inverted Kosterlitz-Thouless transition universality class, separating the bilayer Coulomb phase from the large-$z$ disordered phase. The shaded box is a schematic representation of the region of parameter space scanned by our Monte Carlo studies (thus, other parts of the phase diagram are displayed based exclusively on conclusions drawn from our detailed renormalization group analysis). The hatched strip near $z=0$ inside the bilayer columnar ordered phase represents our prediction for a nonmonotonic $z$ dependence of the columnar order parameter for fixed $V$ in this regime; this is due to the vanishing of the columnar order parameter at both $z=0$ and $z=z_{\rm AT}(V)$. The phase at large $V>0$ and small $z$ has staggered dimer order. This points to the possible existence of a multicritical point at which this staggered phase, the bilayer Coulomb phase, and the large-$z$ disordered phase meet. For a detailed discussion of the scaling picture for each of these phases, see Secs.~\ref{sec:ScalingpictureV>0}, \ref{sec:ScalingpictureVleq0}, and \ref{sec:Ashkin-Tellercriticality}}
\end{figure}

On the square and honeycomb lattice, the ``vector potential'' is nothing but a scalar height field $h$, and the two components $B_{\mu}$ of the magnetic field are given by transverse derivatives of this height field: $B_\mu = \epsilon_{\mu \nu} \partial_\nu h$ (here $\epsilon_{\mu \nu}$ is the totally antisymmetric tensor in two dimensions, with $\epsilon_{xy} = +1$). On the square lattice, the resulting momentum-space structure factor $S_{\mu \mu}$ of dimers oriented in direction $\mu$ has a characteristic {\em pinch-point} singularity in the vicinity of  wavevector ${\mathbf Q} \equiv (\pi, \pi)$. The corresponding fluctuations of the dimer density $n_{\mu}$ (at ${\mathbf Q}$ and nearby wavevectors) are represented in this effective theory by the long-wavelength fluctuations of $B_{\mu}$: $\hat{n}_{\mu}({\mathbf Q} + {\mathbf q}) \sim \hat{B}_{\mu}({\mathbf q})$ (where the hat represents the Fourier transform)
~\cite{fradkin_etal_prb2004,papanikolaou_luijten_fradkin_prb2007,alet_etal_pre2006,ramola_damle_dhar_prl2015,patil_etal_prb2014}. The honeycomb lattice has similar pinch-point phenomenology, albeit with a different pinch-point wavevector ${\mathbf Q} \equiv 0$~\cite{fradkin_etal_prb2004,papanikolaou_luijten_fradkin_prb2007,alet_etal_pre2006,ramola_damle_dhar_prl2015,patil_etal_prb2014}.

On the cubic lattice, the dipolar fluctuations represented by the three-dimensional analog of this pinch-point structure provide the {\em sole} power-law contribution to the long-distance correlations~\cite{huse_etal_prl2003}. In contrast, the two-dimensional Coulomb phase of square and honeycomb lattice dimer models exhibits a {\em second} power-law contribution to the correlations of $n_\mu$, which can, in certain regimes (for instance with attractive interacions) dominate over the dipolar contribution of the pinch-point which always falls of as $1/r^2$ in two dimensions. This is understood in the height phenomenology to be a consequence of the compact nature of the height field, whereby $h(r) \rightarrow h(r) +1$ represents a redundancy in the height description, which allows {\em vertex operators} like $\exp(2 \pi i h)$ in the height description. 

On the square lattice, this additional contribution has weight only in the vicinity of ${\mathbf K}_x \equiv (\pi, 0)$ (${\mathbf K}_y \equiv (0,\pi)$) for $\mu = x$ ($\mu = y$). The corresponding exponent $\eta$ depends on the value of the stiffness to height fluctuations and can be tuned by the strength and nature of interactions between dimers. This power-law contribution to the two-point function of dimers signals the presence of power-law columnar order\cite{fradkin_etal_prb2004,papanikolaou_luijten_fradkin_prb2007,alet_etal_pre2006,ramola_damle_dhar_prl2015,patil_etal_prb2014}. On the honeycomb lattice, the analogous vertex operator contribution leads to power-law correlations at the three-sublattice wavevector of the underlying triangular Bravais lattice, and again signals the presence of power-law columnar order~\cite{fradkin_etal_prb2004,papanikolaou_luijten_fradkin_prb2007,alet_etal_pre2006,ramola_damle_dhar_prl2015,patil_etal_prb2014}.

On the square lattice, we thus write 
\begin{eqnarray}
n_x({\mathbf r}) -1/4 &\sim & (-1)^x {\mathcal A} \cos(2 \pi h({\mathbf r})) + (-1)^{x+y} \partial_y h 
\label{eq:operatorcorrespondencex}
\\
n_y({\mathbf r})-1/4 & \sim & (-1)^y {\mathcal A} \sin(2 \pi h({\mathbf r})) -(-1)^{x+y} \partial_x h \; ,
\label{eq:operatorcorrespondencey}
\end{eqnarray}
 where the first term at the columnar wavevector arises from contributions of the vertex operator, and the second term represents the dipolar contribution of modes in the neighbourhood of wavevector ${\mathbf Q}$.
A crucial aspect of this two-dimensional Coulomb phenomenology is thus the presence of two different power-law contributions to the long-distance correlations of $n_{x}$ ($n_y$): A dipolar contribution that falls off as $(-1)^{x+y}/r^2$ and another power-law contribution that falls off as $(-1)^x/r^{\eta}$ ($(-1)^y/r^{\eta}$), with tunable exponent $\eta$.
\begin{figure}
    \includegraphics[width=0.7\columnwidth]{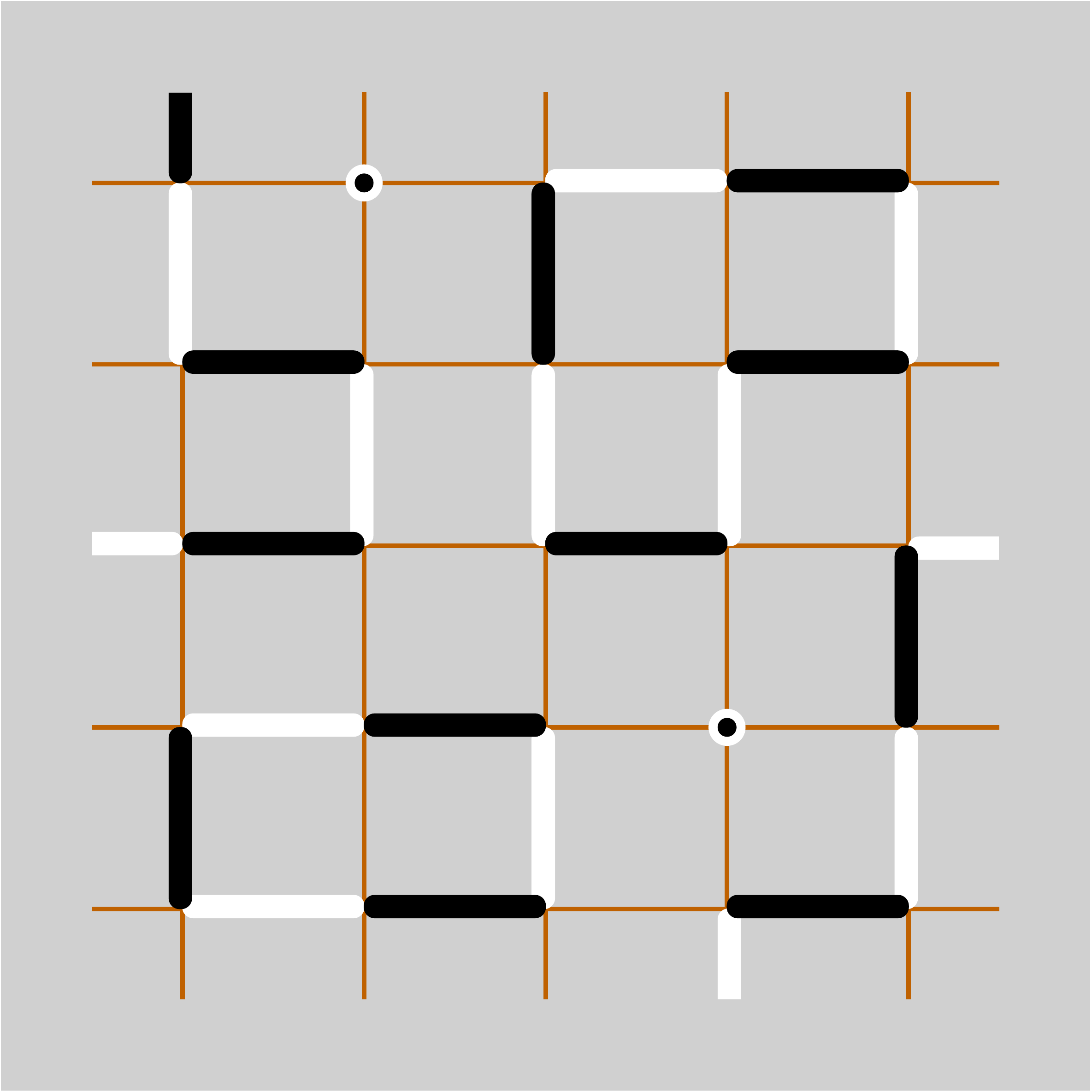}
    \caption{When the dimer configurations (black dimers on one layer, white on the other) of the two layers are laid on top of each other, they define an ensemble of loops on a square lattice with annealed vacancy disorder corresponding to locations of interlayer dimers (black circles). See Sec.~\ref{sec:ScalingpictureV>0} and Sec.~\ref{sec:numerics} for a detailed discussion.}
    \label{fig:overlaploopdefinition}
\end{figure}

This understanding leads to a natural and interesting question: Can two-dimensional dimer models support a different kind of stable Coulomb phase with {\em purely dipolar} long-distance correlations like in three dimensions, {\em i.e.} {\em without} the second contribution and associated nonuniversal exponent $\eta$?

Here, we answer this question in the affirmative using a combination of classical Monte Carlo (MC) simulations and renormalization group (RG) analysis. Our work provides a simple realization of such a Coulomb phase of a two-dimensional dimer model.
An appealing aspect of our construction is that this kind of Coulomb phase is realized on a simple variant of the square lattice, namely the bilayer square lattice, and preserves much of the simplicity of the square lattice dimer model (with the exception of exact solvability). 

More specifically, we study the fully-packed dimer model on the bilayer square lattice with fugacity equal to $z$ ($1$) for inter-layer (intra-layer) dimers, and intra-layer interaction $V$ between neighbouring parallel dimers on any elementary plaquette in either layer. For weak repulsive interactions ($V>0$) we present RG arguments and Monte Carlo results that establish the presence of a qualitatively different kind of (bilayer) Coulomb phase. The two-point dimer correlation functions in this phase are purely dipolar in character. Within the coarse-grained effective field-theory framework we develop here, this arises in the following way (for details, see Sec.~\ref{subsec:BilayerCoulombphase}):  The coarse-grained theory decomposes into two independent sectors, one gapped, and the other critical. The two-point correlation function at the columnar ordering wavevector ${\mathbf K}$ is a {\em product} of a power-law factor arising from the critical sector of this effective field theory, and an exponentially-decaying factor arising from the gapped sector. Whereas the two-point correlation function at the dipolar pinch point wavevector ${\mathbf Q}$ is a {\em sum} of a a dipolar power-law term arising from the critical sector, and a short-ranged correlated piece arising from the gapped sector.
 
For stronger repulsive interactions, our RG analysis also points to the possible existence of an interesting multicritical point, which represents the confluence of three phases: a disordered large-$z$ phase, the bilayer Coulomb phase, and a phase with staggered dimer order in each layer.
\begin{figure}
    \includegraphics[width=\columnwidth]{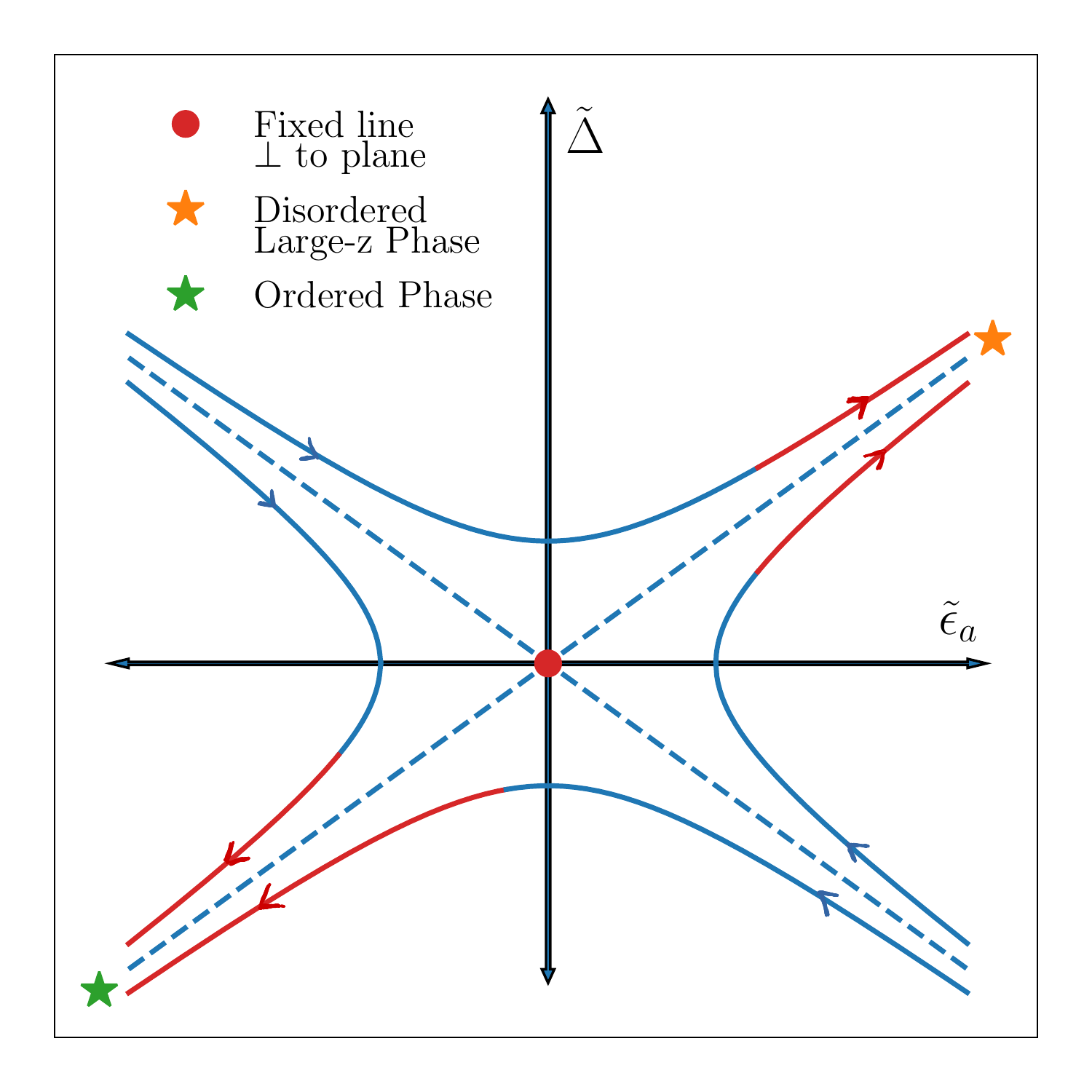}
    \caption{Schematic of flows in the vicinity of the Ashkin-Teller fixed line. The red dot at the origin schematically represents a fixed-line perpendicular to the plane of the figure, with the coupling $\tilde{\epsilon}_s$ serving as the coordinate along this fixed line. See Sec.~\ref{sec:Ashkin-Tellercriticality}   for the definitions of these variables and a more detailed discussion.}
    \label{fig:RGflows}
\end{figure}
Our analysis also predicts that a nonzero $z$ immediately destroys the critical state of of the decoupled system when the system is  noninteracting or has weak intralayer attractive interactions.

In contrast, for a range of moderately strong attractive intralayer interactions $V <0$, our RG analysis predicts the presence of a critical line of Ashkin-Teller transitions separating a bilayer columnar-ordered phase from a disordered phase. In part of this bilayer columnar-ordered phase, we predict that the columnar order parameter at fixed $V$ has an unusual non-monotonic dependence on the interlayer fugacity $z$, vanishing both at $z=0$ and at the phase boundary $z_{\rm AT}(V)$, and peaking for intermediate values of $z$. For stronger attractive interactions larger than a threshold, we predict that this nonmonotonic behaviour is eliminated by the presence of long range columnar order at $z=0$ for the two decoupled layers. In this latter regime, a small nonzero $z$ merely causes these pre-existing columnar ordering patterns of each layer to line up with each other. 
A detailed summary of these results appears for ready reference in the schematic phase diagram displayed in Fig.~\ref{fig:SchematicPhaseDiagram}, as well as in Sec.~\ref{sec:model}. 

The rest of this article, beyond Sec.~\ref{sec:model} is devoted to a detailed discussion of this interesting physics: In Sec.~\ref{sec:Coarsegraineddescription}, we develop the coarse-grained description that provides us the theoretical starting point for studying this system using renormalization group (RG) techniques. In Sec.~\ref{sec:RGflows-general}, we derive the general renormalization group flow equations for the coupling constants of the coarse-grained description, working in Coulomb gas language, and then specialize to linearized flows in the vicinty of a fixed-plane that controls much of the interesting physics. 
\begin{figure*}[!]
    \includegraphics[width=\linewidth]{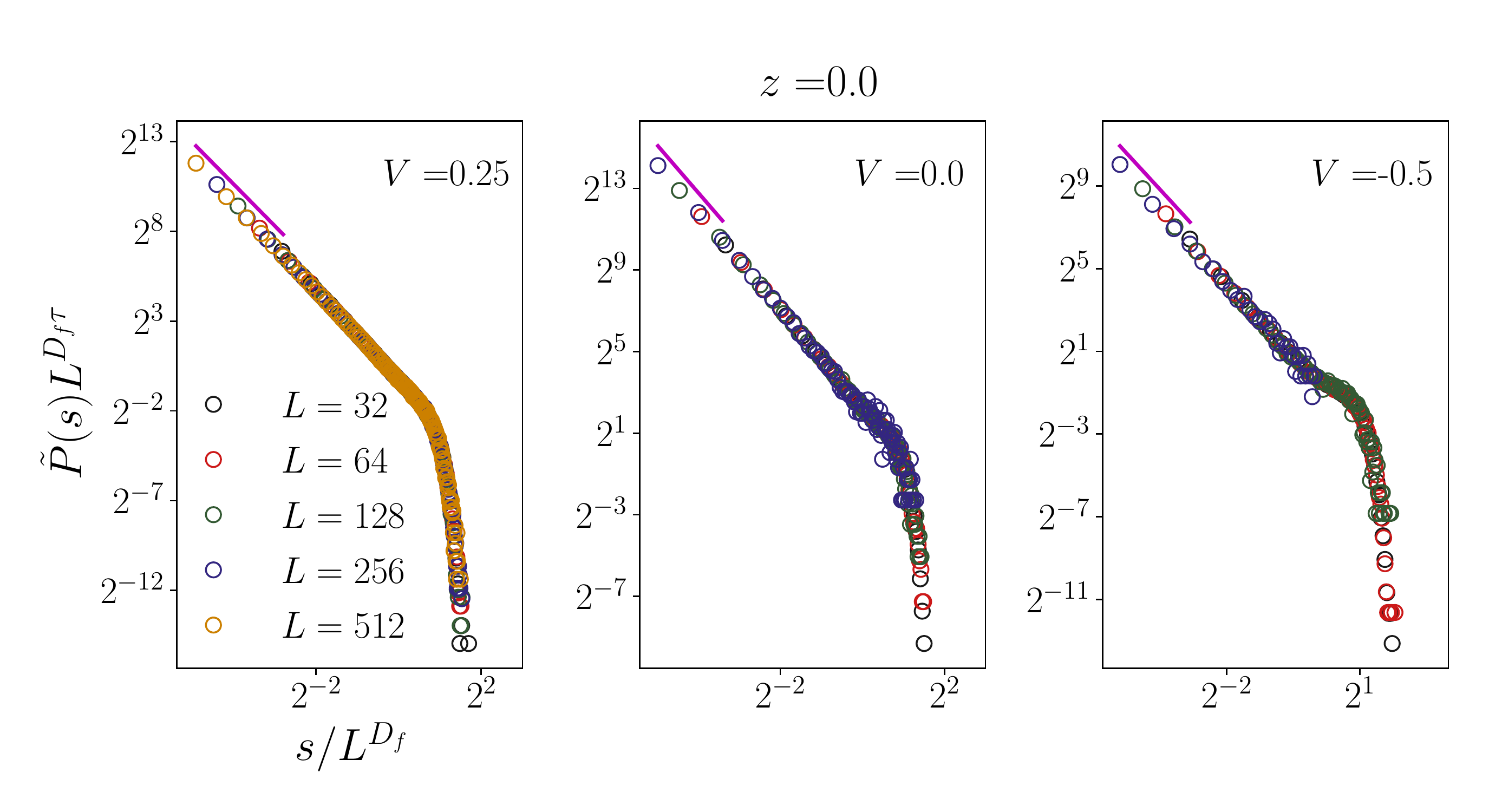}
    \caption{The probability distribution $\tilde{P}(s,L)$ for non-winding overlap loops of length $s$ in a $L\times L$ sample with periodic boundary conditions collapses well onto the postulated scaling form Eq.~{\protect{\ref{eq:scalingformforoverlaps}}} for $V=0.25, 0, -0.5$ at fugacity $z=0$. Further, the scaling function $\Phi(x) $is seen to have the expected power-law behaviour $x^{-\tau}$ with $\tau = 7/3$ for $x \ll 1$. The magenta lines with slope $7/3$ provide visual confirmation of this behaviour. This is behaviour characteristic of contour lines of the Gaussian free field that describes the long-wavelength physics of two dimensional fully-packed dimer models. See Sec.~\ref{sec:ScalingpictureV>0} and Sec.~\ref{sec:numerics} for a detailed discussion.}
    \label{fig:scalingformforz_zero_overlaps}
\end{figure*}
In Sec.~\ref{sec:ScalingpictureV>0}, we use these leading order flow equations to establish the presence of a novel bilayer Coulomb phase in the presence of small repulsive interactions $V$. We also explore the possibility of realizing an interesting multicritical point in the large $V$, large $z$ part of the phase diagram. In Sec.~\ref{sec:ScalingpictureVleq0}, we study the effect of a nonzero $z$ for weak attractive interactions, establishing the fact that any $z$ however small immediately drives the system to a large-$z$ disordered phase for nonzero but weak attractive interactions $V$. We also demonstrate very similar behaviour for the non-interacting problem. In Sec.~\ref{sec:Columnarorder}, we establish for moderately strong attractive interactions, the presence of an unusual regime in which the bilayer is columnar ordered at small nonzero $z$, although the decoupled layers at $z=0$ are critical. We also analyze how this regime is continuoulsy connected, at stronger attractive interactions, to a columnar ordered phase in which the columnar order goes to a nonzero limit at $z=0$. In  Sec.~\ref{sec:Ashkin-Tellercriticality}, we argue that the transition from this bilayer columnar ordered phase to the large-$z$ disordered phase is in the Ashkin-Teller universality class, and provides an unusual example of an Ashkin-Teller critical line. In Sec.~\ref{sec:numerics}, we provide detailed numerical evidence that supports our prediction of a bilayer Coulomb phase for weak repulsive interactions and small $z$, and also establishes the presence of a disordered phase even at small $z$ for weak attractive interactions. Finally, we close with a brief discussion split into two parts, an aside in Sec.~\ref{Aside} comparing our results with the recent results of Wilkins and Powell~\cite{wilkins_powell2020} for a closely related system, and a discussion in Sec.~\ref{Outlook} of the outlook in terms of directions for follow-up work.

\section{Lattice model and summary of results}
\label{sec:model}
 
We consider fully-packed dimer configurations of a bilayer square lattice, with partition function
\begin{equation}
  Z = z^{N_v} e^{-VN_f}
\label{eq:bilayer_cdm_hamil}
\end{equation}
where $N_{v}$ is the number of interlayer ``vertical'' dimers, $N_{f}$ is the total number of ``flippable'' intra-layer plaquettes in either layer with two parallel dimers on links of the plaquette, and $V$ is the interaction between such parallel dimers.

At $z =0$, this reduces to two statistically independent fully-packed square lattice dimer models which are in the usual two-dimensional Coulomb phase for a range of $V \in (V_{c}, V_{s})$ straddling $V=0$. The values of $V_c$ and $V_s$, which determine the extent of the Coulomb phase, have been estimated in previous computational studies~\cite{papanikolaou_luijten_fradkin_prb2007,alet_etal_pre2006,Castelnovo_Chamon_Mudry_Pujol_Annals,Otsuka_PRE2009}. From these studies, the value of $V_c$ is known reasonably accurately to be $V_c \approx -1.55$. The various estimates of $V_s$ have a larger spread, with Ref.~\onlinecite{Castelnovo_Chamon_Mudry_Pujol_Annals} quoting $V_s \approx +1.4$, and other studies~\cite{Otsuka_PRE2009,wilkins_powell2020} finding in favour of a larger value $V_s \approx 2.1$. In our work described here, we will focus mainly on values of $V$ significantly below the lower end of this range for $V_s$, rendering this discrepancy unimportant as far as our conclusions are concerned.

In the limit $z \to \infty$ (with $V$ fixed to a finite value), the partition sum is dominated by a single configuration in which inter-layer vertical dimers cover all sites of the bilayer. Expanding about this limit in a systematic ``strong-coupling'' expansion in $1/z$, it is easy to see that this yields a stable large $z$ phase with short ranged correlations between dimers.
For $V \in (V_{c}, V_{s})$, the question then is whether the $z=0$ Coulomb phase is separated from the large-$z$ short-range correlated phase by an intermediate bilayer phase, or whether the system is in this short-ranged correlated phase for any $z >0$ however small.

In our work, we address this by formulating an RG analysis starting with the $z=0$ Coulomb phase for $V \in (V_{c}, V_{s})$. For repulsive interactions $0<V<V_{s}$, we conclude (as advertised earlier) that a small $z>0$ leads to the purely dipolar bilayer Coulomb phase. As noted in the introduction, we find that this is expected to undergo an inverted Kosterlitz-Thouless transition at $z_{\rm inv. KT}(V) > 0$ to the large-$z$ short-range correlated phase. For $V>V_s$, the decoupled layers at $z=0$ undergo a first-order transition to a phase with staggered long-range order~\cite{Castelnovo_Chamon_Mudry_Pujol_Annals}. This, in conjunction with our RG analysis strongly suggests the possibility of an interesting multicritical point in the large $z$, large positive $V$ part of the phase diagram, at which the large-$z$ short-range correlated phase, the bilayer Coulomb phase, and the staggered phase all meet.\begin{figure}[!]
\includegraphics[width=0.7\columnwidth]{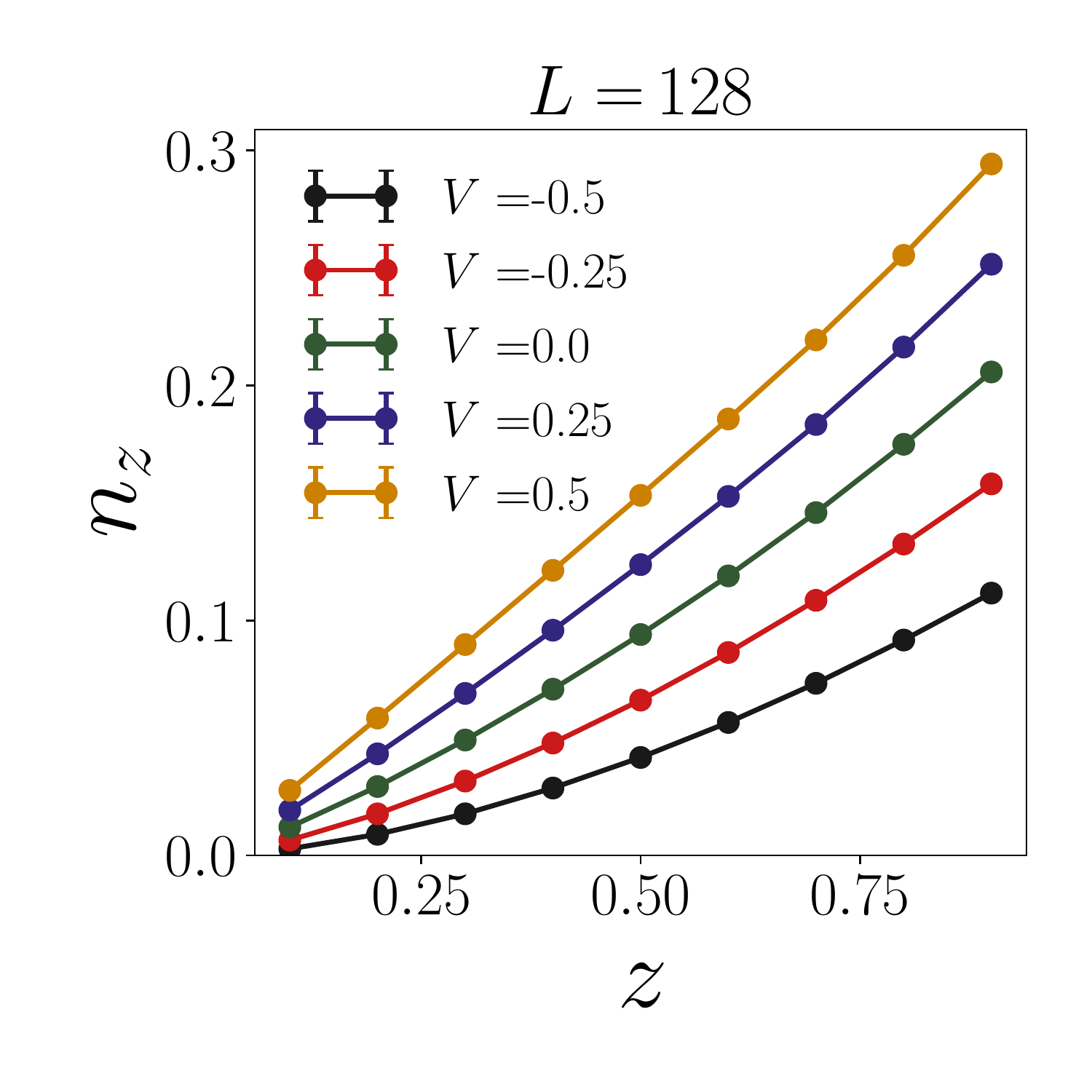}
\caption{\label{fig:interlayer_dimer_corr}
Density of interlayer dimers as
a function of their fugacity $z$ (for the definition of the model parameters, see Sec.~\ref{sec:model}). Note that this is monotonically increasing as expected, and the data shown for $L=128$ is already saturated to the thermodynamic limit. This $z$ dependence shows no indication of the different phases that exist in the phase diagram of the bilayer (see Sec.~\ref{sec:numerics} for a discussion).
}
\end{figure}

For small $V$ on the attractive side, {\em i.e.} for $V_{cb} < V \leq 0$, our RG analysis predicts that a small $z >0$ leads immediately to the short-range correlated phase, albeit with a slow crossover. Combining the results of an earlier numerical study~\cite{alet_etal_pre2006} of the single layer system with our own RG analysis, we estimate $V_{cb} \approx -1.2$. 
For stronger attractive interactions $V \in (V_c, V_{cb})$, our RG analysis predicts the existence of a bilayer columnar ordered phase for nonzero $z$ so long as $z <z_{\rm AT}(V)$, where $z_{\rm AT}(V)$ represents a critical line of Ashkin-Teller transitions from this bilayer columnar ordered phase to the short-range correlated large-$z$ phase. Another outcome of our RG analysis is that the columnar order parameter is predicted to vanish both at $z=0$ and at $z=z_{\rm AT}(V)$ in this regime, implying an interesting nonmonotonic $z$ dependence of the columnar order parameter for fixed $V$ in this regime.

For even stronger attractive interactions $V < V_c$, each decoupled layer at $z=0$ develops long-range columnar order. A small nonzero $z$ is then predicted to immediately lock together the order parameters of the two layers, leading again to the same bilayer columnar phase as above. Our analysis does not directly shed light on the nature of the phase transition from the bilayer columnar phase to the large-$z$ disordered phase in this regime of stronger attractive interactions. One possibility is that the line of Ashkin-Teller transitions ends in a tricritical point, beyond which the phase boundary has first-order character. Another possibility is that the Ashkin-Teller character of the phase boundary remains unchanged for all finite $V$. 

Note that our analysis implies that the usual Coulomb phase (in which the dimer correlator is a sum of a dipolar piece and a term corresponding to power-law columnar order) occurs only at $z=0$ in our bilayer dimer model, both in the noninteracting case, and for not-too-strong intralayer interactions of either sign. In the non-interacting case and with not-too-strong attractive intralayer interactions, our RG analysis shows that an infinitesimally small nonzero $z$ immediately renders this usual Coulomb phase unstable, and the system immediately goes into a disordered phase continuously connected with the large-$z$ disordered regime. For not-too-strong repulsive intralayer interactions, an infinitesimally small nonzero $z$ again renders the usual Coulomb phase unstable, but now the system goes into the qualitatively different bilayer Coulomb phase. 

The focus of the computational part of our work here is the predicted existence of the bilayer Coulomb phase with its unusual purely dipolar correlations. Therefore, we do not pursue here a detailed numerical study of either the bilayer columnar ordered phase and transitions out of it for strong attractive interactions, or the intriguing possibility of a multicritical point 
in the large $V>0$, large $z$ part of the phase diagram.

Instead, our computational study focuses mainly on relatively small $|V|$ of either sign. 
Using Monte Carlo simulations, we study the dimer structure factor and correlation functions, the correlation function of test monomers, as well as the random geometry of fully-packed loops defined by the overlap of upper and lower layer dimer configurations in equilibrium. Our results for these observables are seen to be consistent with expectations from our RG analysis, confirming the predicted presence of a stable bilayer Coulomb phase with purely dipolar correlations $z < z_{\rm inv. KT}(V)$ for small $V>0$. \begin{figure*}[!]
    \includegraphics[width=\linewidth]{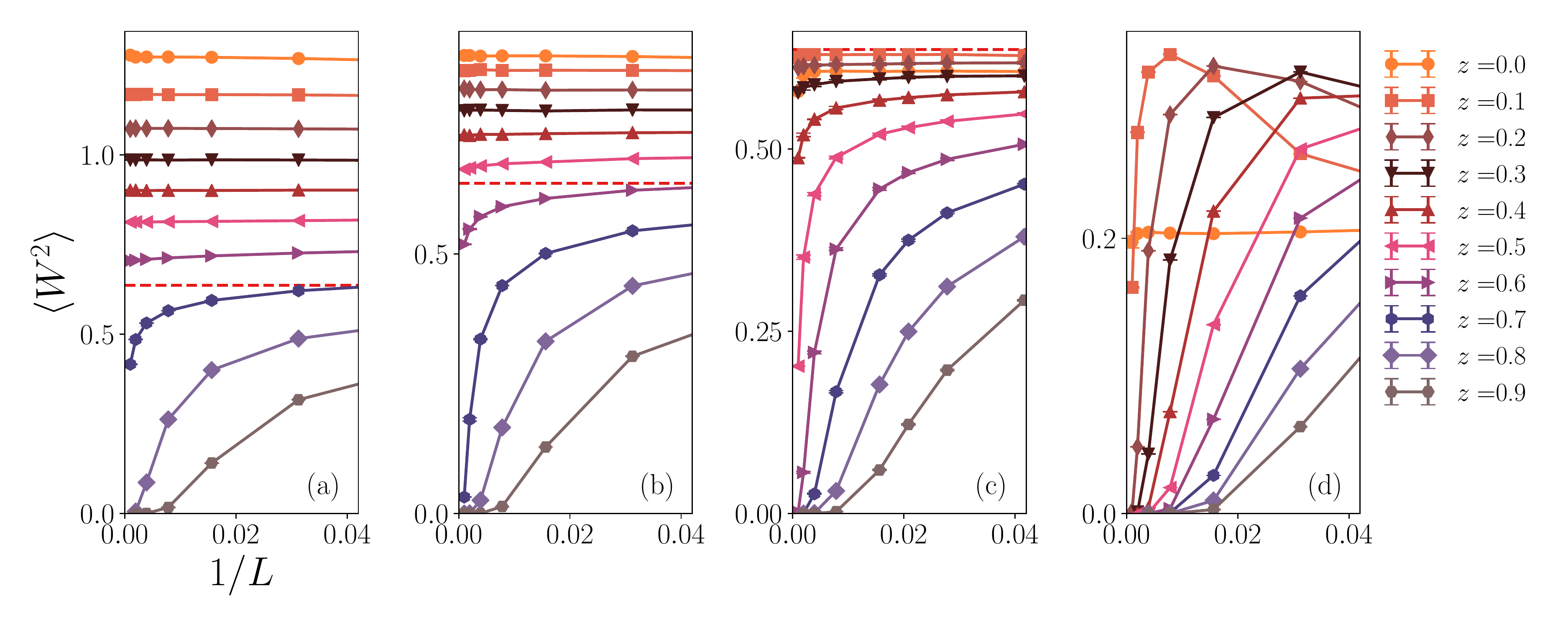}
    \caption{Mean square winding, $\langle W^2 \rangle \equiv \langle W_x^2 + W_y^2\rangle/2$, plotted against $1/L$ for interaction strength (a) $V=0.5$ (b) $V=0.25$ (c) $V=0.0$ and (d) $V=-0.5$. $\langle W^2 \rangle$ appears to extrapolate to a finite number for small $z$ in the non-interacting and repulsive cases, but goes to 0 for large $z$. In contrast, this quantity goes to $0$ for all $z>0$ in the attractive case. The red dashed line in (a), (b) and (c) 
	indicates the value $\langle W^2 \rangle_{\rm inv. KT} = {\mathcal J}(0.25) = 0.6365\ldots$, where ${\mathcal J}(g_-)$ is the theoretically expected value (see Eq.~\ref{eq:Jdefn} and Eq.~\ref{eq:W2fromJ})at the transition from the bilayer Coulomb phase to the large-$z$ disordered phase (see Sec.~\ref{sec:ScalingpictureV>0} and Sec.~\ref{sec:numerics} for details). 
	}
    \label{fig:w2vsL}
\end{figure*}

We close this overview by noting that our work is closely related to very recent work by Wilkins and Powell~\cite{wilkins_powell2020} who study a bilayer with intralayer interactions identical to those considered here, but no interlayer dimers at all. Instead, Wilkins and Powell consider the effects of interlayer interactions of strength $K$ that couple the dimers on corresponding links of the two layers. A comparison of our results and theirs is very instructive, in that it provides us a natural way to highlight and emphasize the key physical effect that is responsible for the emergence of the novel bilayer Coulomb phase in the system studied here. This is described in Sec.~\ref{Aside}.

\section{Coarse-grained description}
\label{sec:Coarsegraineddescription}

We begin by generalizing the coarse-grained height description of the square lattice dimer model to the bilayer case. This is written in terms of two height fields $h_{1/2}(\vec{r})$, and generalizes the well-known single-layer effective theory\cite{fradkin_etal_prb2004,papanikolaou_luijten_fradkin_prb2007,alet_etal_pre2006,ramola_damle_dhar_prl2015,patil_etal_prb2014}. to the bilayer case by adding the leading-order coupling terms allowed by symmetries and consistent with the microscopic features of our system. Thus, we write
\begin{eqnarray}
Z & \propto & \int {\mathcal D}h_{1} {\mathcal D}h_2 \exp(-S)
\end{eqnarray}
where ${\mathcal D}h_{1/2}$ denotes the functional integral over configurations of $\fld{h_1}$ and $\fld{h_2}$ defined on a square lattice (which hosts a convenient re-discretization of the coarse-grained contiuum action) and this re-discretized version of the coarse-grained action $S$ reads:
\begin{align}
& S =  \\
& \: \pi g \sum_{{\mathbf r}} \left[\left(\Delta_\mu \fld{h_1}) + \fld{C_\mu} \right)^2                     
+ \left(\Delta_\mu \fld{h_2} + \fld{C_\mu} \right)^2 \right] \nonumber \\
& - 2\pi g_{12} \sum_{{\mathbf r}} \left[\Delta_\mu \fld{h_1} + \fld{C_\mu} \right]                
\cdot \left[\Delta_\mu \fld{h_2} + \fld{C} \right] \nonumber \\
& - \log y_v \sum_{{\mathbf r}} \fld{m^2} - \lambda \sum_{{\mathbf r}} \left[\cos(8 \pi \fld{h_1}) + \cos(8 \pi \fld{h_2})) \right] \nonumber \\
& -\lambda_{-} \sum_{{\mathbf r}} \cos(2 \pi (\fld{h_1} - \fld{h_2})) \nonumber \\
& -\lambda_+ \sum_{{\mathbf r}} \cos(4 \pi (\fld{h_1} + \fld{h_2})) + \ldots 
\label{eq:coupledsinegordon}
\end{align}
where $\Delta_{\mu}$ denotes the $\mu$ component of the lattice gradient, $C_{\mu}$ is an integer-valued vector field on links of the the square lattice, which satisfies
$\epsilon_{z\mu \nu} \Delta_{\mu} C_{\nu} = \fld{m}$ and can be chosen for instance to be nonzero only on the $y$ links of the lattice. Note that both layers are described by the same site coordinate ${\mathbf r}$, and thus $C_\mu$ and $m$ are common to both layers. Here $\fld{m}$ is an integer-valued field on the faces of a square lattice, which, by a slight abuse of notation, we represent as $\fld{m}$. \begin{figure}[!]
 \subfigure[]{\includegraphics[width=\columnwidth]{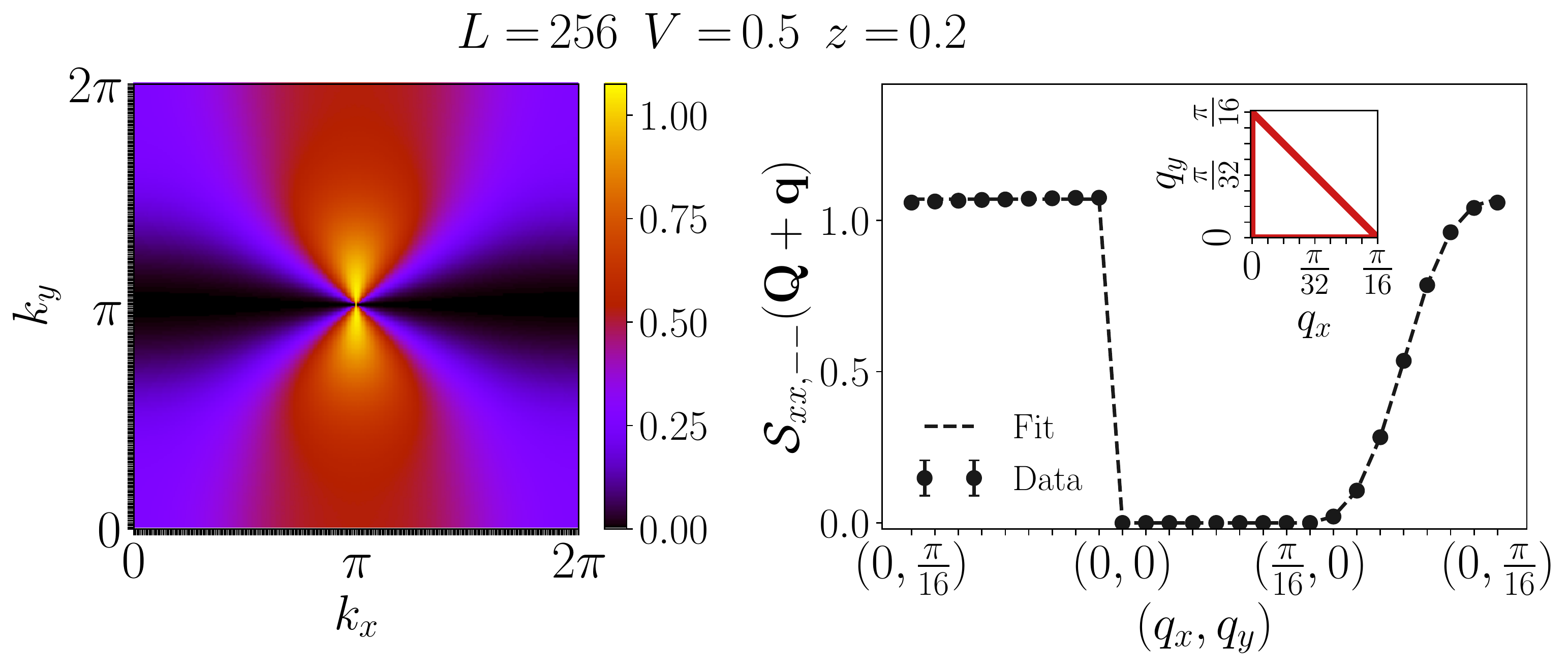}}
 \subfigure[]{\includegraphics[width=\columnwidth]{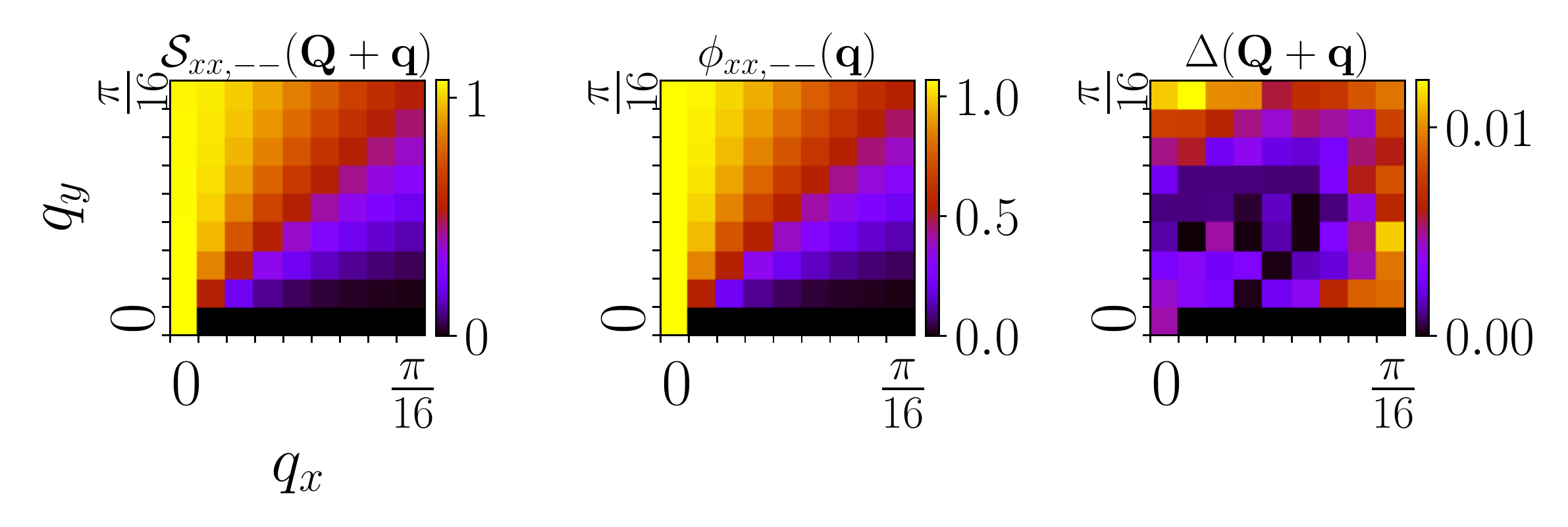}}
 \subfigure[]{\includegraphics[width=\columnwidth]{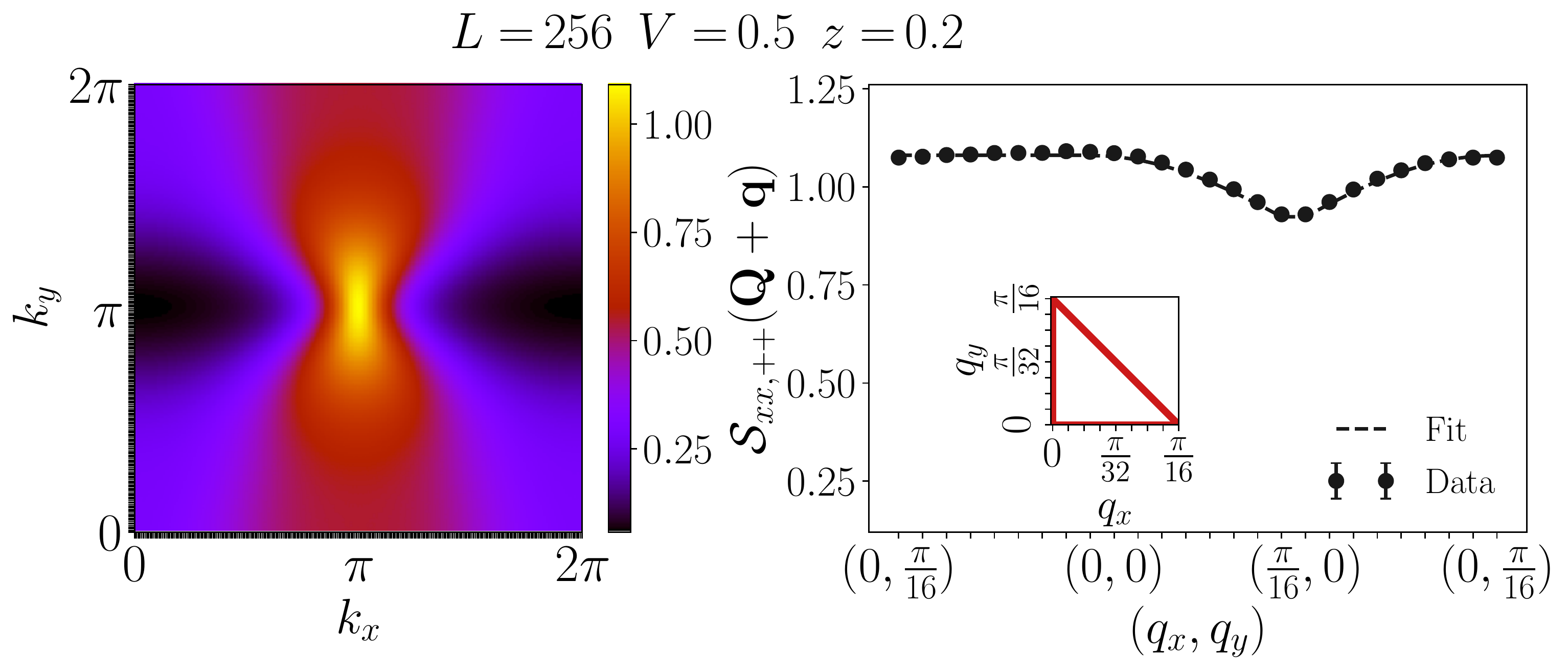}}
 	\caption{ \label{fig:struc_fac_repulsive}
	 (a) Structure factor of $n_{x,-}(\mathbf{r})$,
	 $S_{xx,--}(\mathbf{k}) \equiv \langle \tilde{n}_{x,-}(-\mathbf{k}) \tilde{n}_{x,-}(\mathbf{k}) \rangle$ 
	 for repulsive interaction $V=0.5$ and $z=0.2$, depicted in a heat map on the left. On the right is the same quantity plotted close to the pinch point, $\mathbf{Q}$, along the path shown in the inset: $(\pi,\pi) \rightarrow (\pi+\frac{\pi}{16},\pi) \rightarrow (\pi,\pi+\frac{\pi}{16}) \rightarrow (\pi,\pi)$. The data can be seen to fit to the form derived in Eq.~\ref{eq:bminuscorrfit}, where $g_{-}^{*}=0.148$ is extracted from Gaussian fits to histograms of winding fluctuations (see Fig~\ref{fig:w2gcomparison_rep}). (b) Heat map of the structure factor zoomed into a $\frac{\pi}{16} \times \frac{\pi}{16}$ grid in the Brillouin zone around the pinch point. In the middle is the heat map of the fitting form, $S^{\text{fit}}_{xx,--}(\mathbf{Q}+\mathbf{q})=\phi_{xx,--}(\mathbf{q})$. In the right is the heat map of $\Delta(\mathbf{Q}+\mathbf{q})=\frac{|S^{\text{fit}}_{xx,--}(\mathbf{Q}+\mathbf{q})-S_{xx,--}(\mathbf{Q}+\mathbf{q})|}{S^{\text{fit}}_{xx,--}(\mathbf{Q}+\mathbf{q})}$. The structure factor values obtained from Monte Carlo data lie within an average of 0.3\% of the values predicted by the fitting form. (c) Structure factor of $n_{x,+}(\mathbf{r})$, $S_{xx,++}(\mathbf{k}) \equiv \langle \tilde{n}_{x,+}(-\mathbf{k}) \tilde{n}_{x,+}(\mathbf{k}) \rangle$ for the same parameters as above depicted in a heatmap on the left. On the right is the same quantity plotted close to $\mathbf{Q}$ along the path shown in the inset: $(\pi,\pi) \rightarrow (\pi+\frac{\pi}{16},\pi) \rightarrow (\pi,\pi+\frac{\pi}{16}) \rightarrow (\pi,\pi)$. It is shown to fit to a form derived in Eq.~\ref{eq:bpluscorrfit} with $g_{+}=0.172$ and $\log(1/y_0)=32.4$. See Sec.~\ref{sec:numerics} for a detailed discussion.}
\end{figure}

Before we proceed, it is useful to understand the microscopic origin of various terms included here.
Generalizing from the discussion of a single layer in Ref~\onlinecite{alet_etal_prl2006,alet_etal_prl2005,alet_etal_pre2006}, we first note that the quadratic terms in $(\Delta_\mu h_{\alpha} + C_\mu)$ proportional to $g$ simply represent the
quadratic part of the coarse-grained height action for two independent dimer model on two uncoupled layers, both of which allow monomers to exist, but only if the monomers are at exactly the same locations on both layers; this constraint is reflected in the fact that exactly the same field $C_\mu$ enters both terms proportional to $g$. It models the fact that vertical interlayer dimers of our bilayer are ``seen'' as monomers by the intra-layer dimers of each layer. The monomer fugacity parameter $y_v$ is thus related to the microscopic fugacity of interlayer dimers, and expected to be linear in $z$ for small $z$.

Additionally, the quadratic term proportional to $g_{12}$ has a simple interpretation that follows from the operator correspondence Eqs.~\ref{eq:operatorcorrespondencex}, \ref{eq:operatorcorrespondencey} discussed in the Introduction. Generalizing this correspondence to our bilayer case, we see that $\Delta_\mu h_{\alpha} + C_\mu$ represents the ``dipolar'' (composed of Fourier modes near ${\mathbf Q}$) part of the dimer density $n_{\mu}^{(\alpha)}$ in layer $\alpha$. Thus, this term tries to align the dipolar part of the dimer density in both layers. What about a similar tendency of the ``columnar'' part of the dimer density fields 
$n_{\mu}^{(\alpha)}$ to align with each other? From the operator correspondence, we see that this is clearly represented by the cosine term proportional to $\lambda_{-}$. Thus $g_{12}$ and $\lambda_{-}$ taken together represent a tendency for dimers in the two layers to line up. The underlying reason for this
tendency to align is entropic: Clearly, there are many more choices for the positions of interlayer dimers if the intralayer dimer configurations of the two layers match up. Therefore, we expect both $g_{12}> 0$ and $\lambda_{-}> 0$ to turn on as soon as $z$ becomes nonzero, since these terms correctly encode this entropic advantage. Since the underlying entropic attraction has its roots in the fact that two interlayer dimers on adjacent links can be traded in for a pair of intralayer dimers on identical links of both layers, we expect both $g_{12}$ and $\lambda_{-}$ to scale as $z^2$ for small $z$.

Next we note that the symmetry analysis of Ref.~\onlinecite{alet_etal_pre2006,alet_etal_prl2005,alet_etal_prl2006,Henley_review} goes through unchanged, so long as the symmetry operations considered are applied to both layers simultaneously. 
This means that the heights of both layers must transform simultaneously for the transformation to be a symmetry of the coupled bilayer. In other words, the relevant symmetry operations are: i) $h_1 \to -h_1$ and $h_2 \to -h_2$ simultaneously,  ii) $h_1 \to h_1+1/4$ and $h_2 \to h_2 +1/4$ simultaneously.
Additionally, we note that the operator correspondence described in Sec.~\ref{sec:Introduction} that relates microscopic dimer density operators to operators in the coarse-grained height theory naturally remains essentially unchanged when using the height description to analyze the properties of the bilayer.

These symmetry transformations allow the cosine terms proportional to $\lambda$ which exist in the single-layer system as well. As in Ref.~\onlinecite{alet_etal_pre2006,alet_etal_prl2005,alet_etal_prl2006,Henley_review}, these represent the entropic advantage of configurations that are proximate to perfectly columnar ordered states in each layer. One expects $\lambda$ to be positive, although our analysis does not rely on this in any crucial way. The parameter $\lambda_{+}$ controls the strength of an additional cosine interaction, which has been included because it is the leading allowed term of this type. 

We close by explicitly reiterating an important symmmetry distinction between the couplings $\lambda$ and $g$ on the one hand, and the couplings $g_{12}$, $\lambda_{+}$, $y_v$ and $\lambda_{-}$ on the other hand. These latter four couplings can only exist at nonzero $z$, {\em i.e.}
only when the two layers are coupled by interlayer dimers. This is because the decoupled system has a larger symmetry (of {\em independent} translations of $h_1$ and $h_1$ by a $1/4$ and independent changes of sign of $h_1$ and $h_2$) which forbids these additional term. Indeed, as we have already argued, we have:
\begin{eqnarray}
y_v & \propto & z \nonumber \\
\lambda_{-} & \propto & z^2 \nonumber \\
g_{12} & \propto & z^2 \nonumber \\
\lambda_{+} &\to &0 
\label{eq:smallzasymptotics}
\end{eqnarray}
as $z \to 0$. \begin{figure}[!]
 \includegraphics[width=\columnwidth]{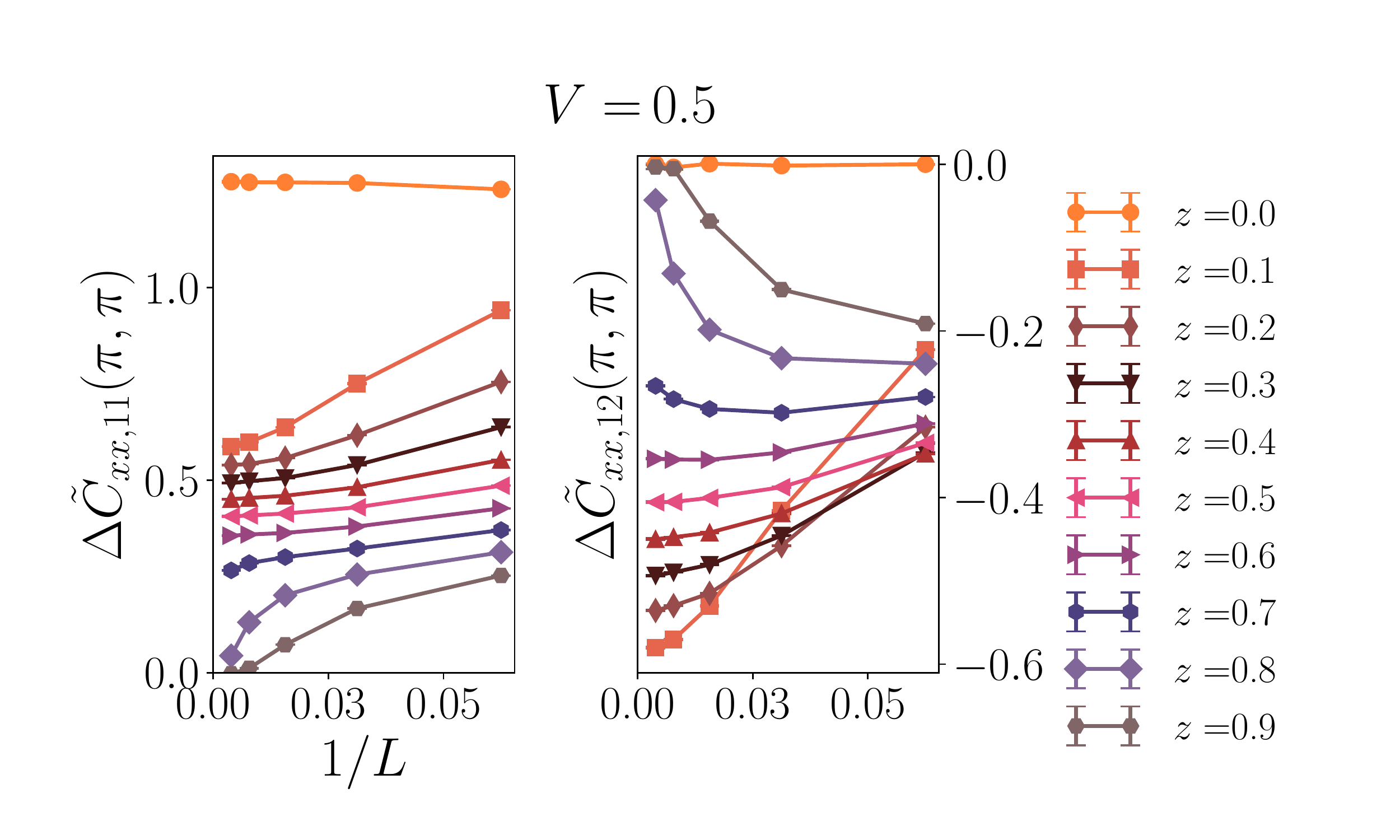}
 \caption{\label{fig:pnchpntintra_rep} This figure displays the strength of the pinchpoint singularity as a function of size $L$  in the intralayer (left) and interlayer (right) structure factors at $V=0.5$. This strength is defined as the difference between structure factor values at $\mathbf{Q}$ and the next allowed momentum-space grid point in the $\mathbf{e}_x$ direction: $S_{xx,11}(\mathbf{Q})-S_{xx,11}(\mathbf{Q}+\frac{2\pi}{L}\mathbf{e}_x)$ and $S_{xx,12}(\mathbf{Q})-S_{xx,12}(\mathbf{Q}+\frac{2\pi}{L}\mathbf{e}_x)$ respectively. Note the singular nature of the limit $z \rightarrow 0$, as discussed in Sec.~\ref{sec:numerics}.}
\end{figure}
\begin{figure}[!]
    \includegraphics[width=\columnwidth]{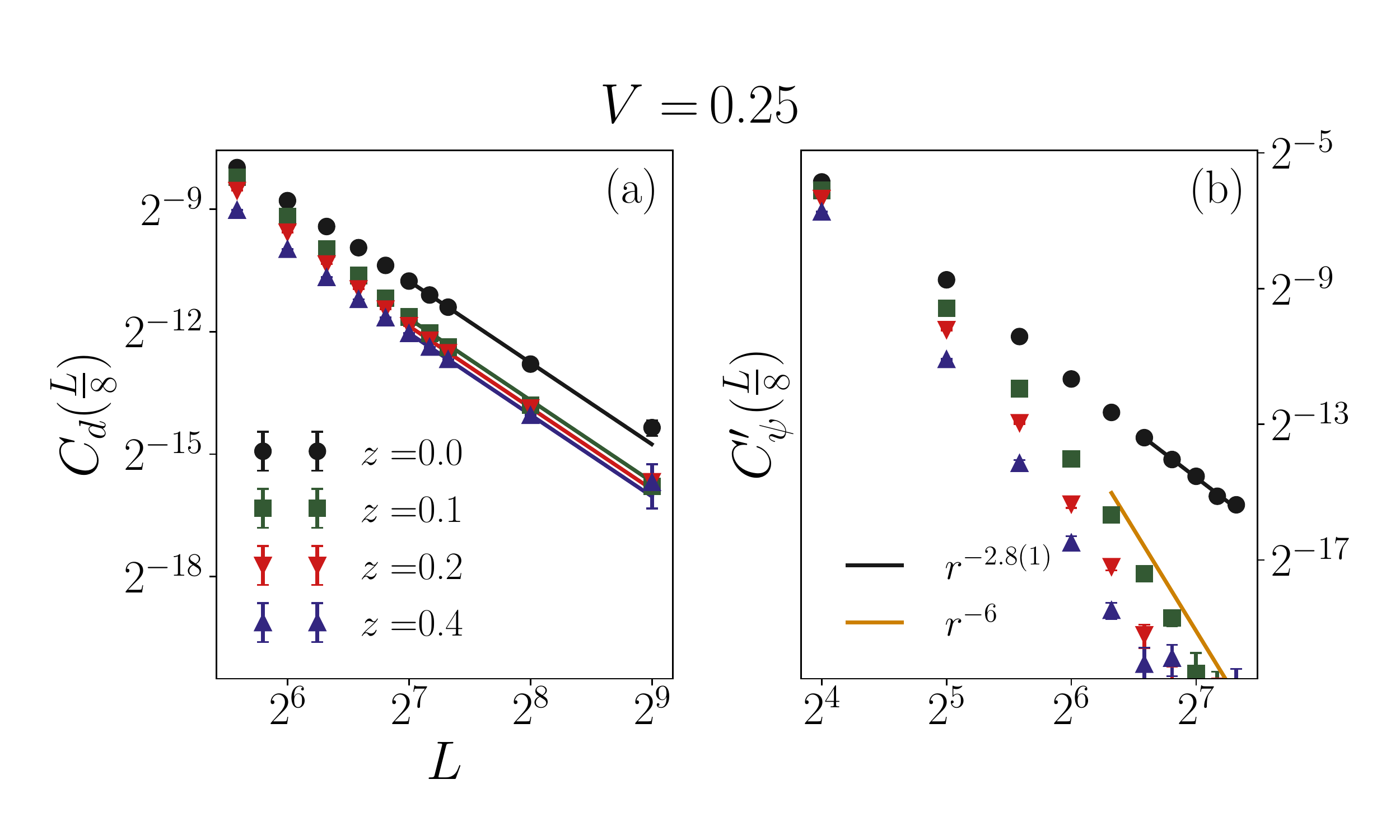}
    \includegraphics[width=\columnwidth]{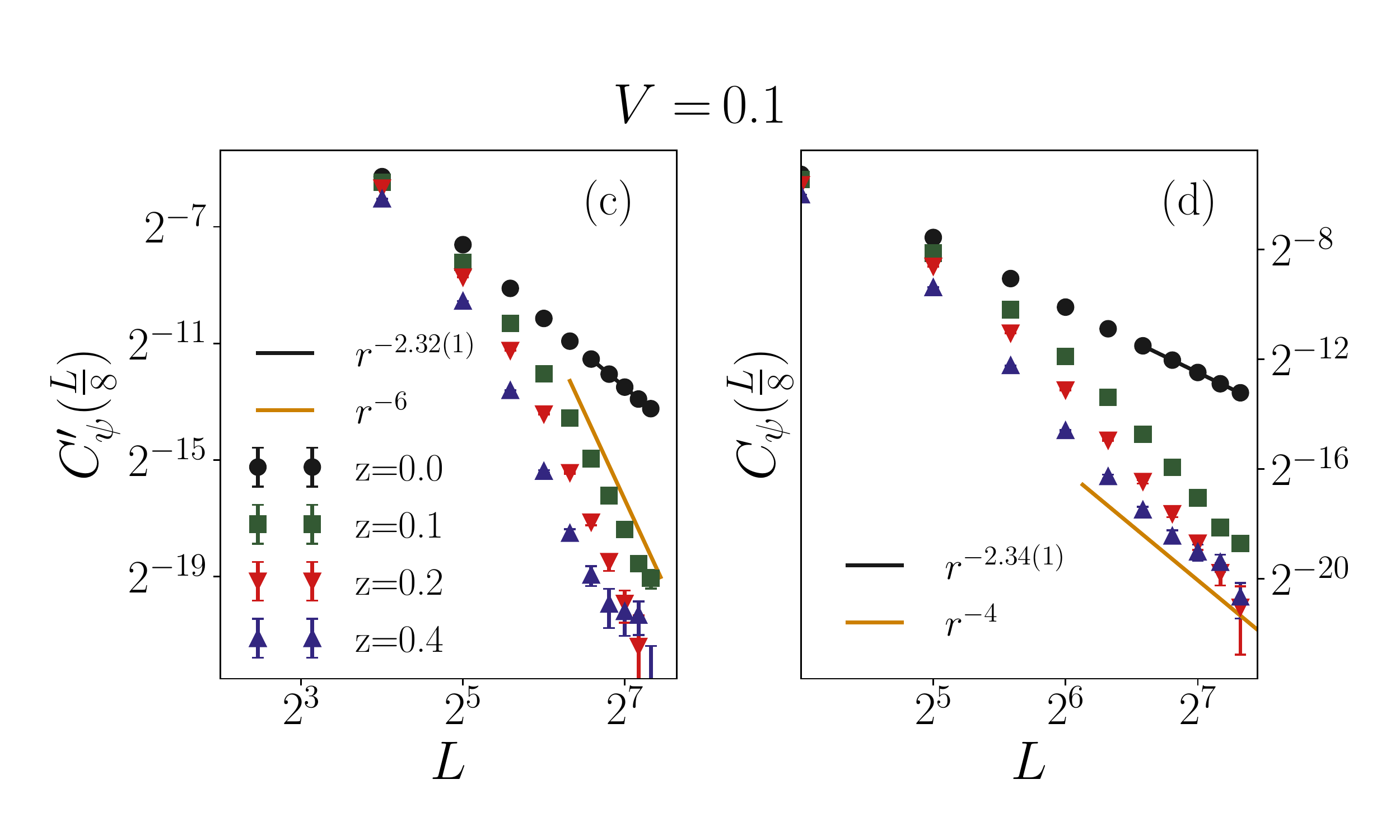}
    \caption{(a) The dipolar component of intralayer dimer correlations (as defined in Eq.~\ref{eq:dipolar_linear_combo}) for $V=0.25$ fits to the expected form $a L^{-2}$. (b) The corresponding
    columnar component at $V=0.25$ (as defined in Eq.~\ref{eq:columnar_linear_combo_2}) is seen to decay with a power law of $r^{-2.8(1)}$ at $z=0$. However, for $z>0$, it decays to zero much more rapidly, falling off as $\sim L^{-6}$. As explained in Sec.~\ref{sec:numerics}, this is the expected behaviour of $C'_{\psi}(\mathbf{r}_L)$ in the bilayer Coulomb phase when dimer correlations are purely dipolar in nature. (c) This columnar component $C'_{\psi}(\mathbf{r}_L)$ given by Eq.~\ref{eq:columnar_linear_combo_2} for V=0.1 again scales as $\sim L^{-6}$ for nonzero $z$, while (d) an alternate definition of columnar component, $C_{\psi}(\mathbf{r}_L)$ given by Eq.~\ref{eq:columnar_linear_combo_1}, scales as $\sim L^{-4}$ for the same nonzero $z$. This is consistent with the absence of power-law columnar order in the bilayer Coulomb phase (see Sec.~\ref{sec:numerics} for details).}
    \label{fig:psicorr_rep}
\end{figure}

\section{RG flows of coarse-grained theory}
\label{sec:RGflows-general}

With this out of the way, we now proceed to analyse the perturbative stability of the Gaussian fixed plane described by the quadratic terms in the action. This is motivated by the following considerations: First we note that each decoupled layer at $z=0$ remains in a two-dimensional Coulomb phase for $V \in (V_c,V_s)$ (with $V_c \approx -1.55$,\cite{papanikolaou_luijten_fradkin_prb2007,alet_etal_pre2006} and $V_s$ in the range $1.4$~\cite{Castelnovo_Chamon_Mudry_Pujol_Annals} to $2.1$~\cite{Otsuka_PRE2009,wilkins_powell2020}). This Coulomb phase is described by a line of Gaussian fixed points parametrized by the value of $g$, with the $\lambda$ term being an irrelevant perturbation of this Gaussian fixed line. In this Gaussian action, $g_{12} = 0$ and $C_{\mu} = 0$. Turning on a nonzero but small $z$ corresponds to turning on the couplings $\lambda_{-}$, $y$, $g_{12}$ and $\lambda_{+}$. Therefore, to study the effect of a small interlayer dimer fugacity $z$, one must analyze the perturbative stablility of the Gaussian fixed line. 
In fact, one may incorporate $g_{12}$ in the Gaussian theory exactly, and then study the renormalization group flows of the other couplings in the vicinity of the Gaussian fixed points parametrized by $g$ and $g_{12}$.
However, it must be remembered that $g_{12}(z) \to 0$ as $z \to 0$, while $g$ is tuned by the intralayer interaction $V$. 

\subsection{Coulomb-gas formulation}
\label{subsec:Coulombgasformulation}

This RG analysis is greatly facilitated by going over to the equivalent electromagnetic Coulomb gas formulation (in which our `electric' charges correspond to the cosine terms, and our `magnetic' charge corresponds to the monomer numbers $m$). As a prelude to this, we first rewrite the effective theory in Villain form
\begin{eqnarray}
Z & \propto  &  \sum_{\left\{m,u,l,q,p\right\} }  \int {\mathcal D} h_1 {\mathcal D} h_2  e^{-S_{\rm Villain}}
\end{eqnarray}
where ${\mathcal D}h_{1/2}$ again denotes the functional integral over configurations of $h_1$ and $h_2$ defined on a square lattice, the sum is over configurations $\left\{m,u,l,q,p\right\}$ of integer-valued fields $\fld{m}$, $\fld{u}$, $\fld{l}$, $\fld{q}$, and $\fld{p}$ and the coarse-grained action $S_{\rm Villain}$ reads:
\begin{align}
& S_{\rm Villain} = \\
& \: \pi g \sum_{{\mathbf r}} \left[\left(\Delta_\mu \fld{h_1}) + \fld{C_\mu} \right)^2 
+ \left(\Delta_\mu \fld{h_2} + \fld{C_\mu} \right)^2 \right] \nonumber \\
& - 2\pi g_{12} \sum_{{\mathbf r}} \left[\Delta_\mu \fld{h_1} + \fld{C_\mu} \right] 
\cdot \left[\Delta_\mu \fld{h_2} + \fld{C} \right] \nonumber \\
& - \sum_{{\mathbf r}} \log (Y_v[\fld{m}]) -2 \pi i \sum_{{\mathbf r}} \fld{q} \left( \fld{h_1} - \fld{h_2} \right)\nonumber \\
& -4 \pi i \sum_{{\mathbf r}} \fld{p} \left( \fld{h_1} + \fld{h_2} \right) \nonumber \\
& -8\pi i \sum_{{\mathbf r}} \left(\fld{u} \fld{h_1} + \fld{l} \fld{h_2} \right) \nonumber \\
& -\sum_{{\mathbf r}} \log (Y_{-}[\fld{q}]) - \sum_{{\mathbf r}} \log (Y_{+}[\fld{p}])\nonumber \\
& -\sum_{{\mathbf r}}\log (Y_{\lambda}[\fld{u}])-\sum_{{\mathbf r}}\log (Y_{\lambda}[\fld{l}])
\label{eq:villainformulation}
\end{align}
where $\epsilon_{z\mu \nu} \Delta_{\mu} C_{\nu} = \fld{m}$ and
\begin{eqnarray}
&&\log(Y_v[m]) = m^2\log y_v \; \; \; \ \log(Y_{-}[q]) = q^2\log(\lambda_{-}/2) \nonumber \\
&&\log(Y_\lambda[x]) = x^2\log(\lambda/2) \; \; \; (x=u,l)\nonumber \\
&&\log(Y_+[p]) = p^2\log(\lambda_{+}/2) 
\label{eq:microscopiccouplings}
\end{eqnarray}

Defining
\begin{align}
g_{-} =  \pi(J + K) = \frac{g + g_{12}}{2} \; \; ; \; \; \theta_{-} = 2\pi(h_1-h_2) \\
g_{+} = \pi(J - K) = \frac{g - g_{12}}{2} \; \; ; \; \; \theta_{+} = 2 \pi(h_1+h_2)
\label{eq:variablechangeofcouplings}
\end{align}
we may rewrite this as
\begin{align}
& S_{\rm Villain} = \\
& \frac{g_{+}}{4\pi} \sum_{{\mathbf r}} \left(\Delta_\mu \fld{\theta_+} + 4\pi \fld{C_\mu} \right)^2  + \frac{g_{-}}{4\pi} \sum_{{\mathbf r}} \left(\Delta_\mu \fld{\theta_-} \right)^2 \nonumber \\
& - i \sum_{{\mathbf r}} \fld{q} \fld{\theta_-} - 2i \sum_{{\mathbf r}} \fld{p} \fld{\theta_+} - \sum_{{\mathbf r}} \log (Y_v[\fld{m}]) \nonumber \\
& - 2i \sum_{{\mathbf r}} \left(\fld{u} + \fld{l} \right) \fld{\theta_+}
- 2i \sum_{{\mathbf r}} \left(\fld{u} - \fld{l} \right) \fld{\theta_-} \nonumber \\
& -\sum_{{\mathbf r}} \log (Y_{-}[\fld{q}]) - \sum_{{\mathbf r}} \log (Y_{+}[\fld{p}])\nonumber \\
& -\sum_{{\mathbf r}}\log (Y_{\lambda}[\fld{u}])-\sum_{{\mathbf r}}\log (Y_{\lambda}[\fld{l}])
\label{eq:villainformulationvariablechange}
\end{align}
Thus the angular variable $\fld{\theta_+}$ has 2-fold anisotropy and doubled vortices, whereas the angular variable
$\fld{\theta_-}$ has external
field and 2-fold anisotropy. \begin{figure}[!]
        \includegraphics[width=\columnwidth]{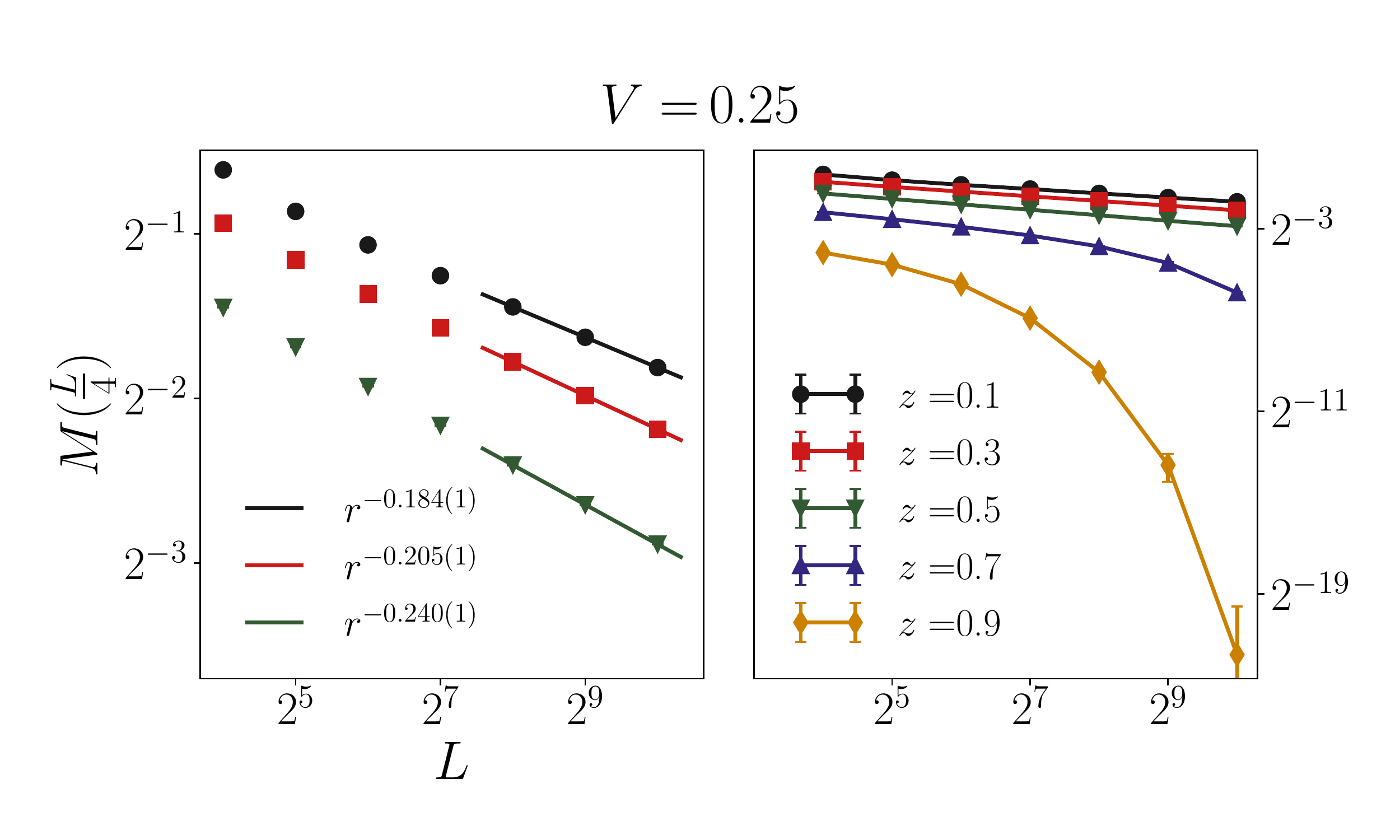}
    \caption{The monomer-antimonomer correlation function (normalized to be unity at separation $\mathbf{r} = 0$) at separation $\mathbf{r}=(\frac{L}{4},0)$ for repulsive interaction $V=0.25$  fits well to a power law form $L^{-\eta_m}$, with best-fit values of $\eta_m(z)$ extracted in this way shown in the legend for various values of small but nonzero $z$. Our theoretical expectation is that $\eta_m = g^{*}_-$, thus allowing us obtain an estimate of $g^{*}_-(z)$ from this measurement. In contrast, the data at larger $z$ shows faster-than-power-law decay (right panel). See Sec.~\ref{sec:ScalingpictureV>0} and  Sec.~\ref{sec:numerics} for a detailed discussion.}
    \label{fig:monomercorr_rep}
\end{figure}

Following the standard procedure (described for instance in Ref.~\onlinecite{Nienhuis_review}) for switching to a Coulomb gas representation in the continuum, we now arrive at:
\begin{eqnarray}
Z & \propto  &  \sum_{\left\{m,u,l,q,p\right\} }  C_{\left\{m,u,l,q,p\right\}}e^{-S'_{\rm Coulomb}}
\end{eqnarray}
where the sum is over configurations of the integer-valued charges, $C_{\left\{m,u,l,q,p\right\}}$
is the usual combinatorial factor that accounts for the fact that all charges of a particular type are indistinguishable particles,
and the action reads:
\begin{align}
& S'_{\rm Coulomb} = \\
&  - \frac{1}{2g_{-}} \sum_{i \neq j} \left(q_i + 2 \tilde{Q}_i \right) 
\log \left( \frac{|\mathbf{r}_i - \mathbf{r}_j|}{a} \right)
\left(q_j + 2 \tilde{Q}_j \right) \nonumber \\
&  - \frac{1}{2g_{+}} \sum_{i \neq j} \left(2p_i + 2 \tilde{P}_i \right) 
\log \left( \frac{|\mathbf{r}_i - \mathbf{r}_j|}{a} \right)
\left(2p_j + 2 \tilde{P}_j \right) \nonumber\\
&  - \frac{g_{+}}{2} \sum_{i \neq j} \left(2 m_i \right) 
\log \left( \frac{|\mathbf{r}_i - \mathbf{r}_j|}{a} \right)
\left(2 m_j \right) \nonumber \\
& -i \sum_{i \neq j} \left( 2 m_i \right) \Phi(\mathbf{r}_i - \mathbf{r}_j) 
\left(2 p_j + 2 \tilde{P}_j \right) -\sum_i \log(Y_{v}[m_i]) \nonumber \\
& -\sum_{i} \log (Y_{-}[q_i]) - \sum_{i} \log (Y_{+}[p_i])\nonumber \\
& -\sum_{i}\log (Y_{\lambda}[u_i])-\sum_{i}\log (Y_{\lambda}[l_i]) 
\label{eq:continuumcoulombgas}
\end{align}
with $\Phi(\mathbf{r}) = \text{Im} (\log(x + iy))$, $\tilde{P} = u + l$,
$\tilde{Q} = u -l$. This is a multi-component electromagnetic Coulomb gas, with one magnetic charge $m$, and four kinds of electric charges $V$, $l$, $q$, and $p$, all of which can take on any integer value. The interactions of these charges however have a very specific structure, which dictates the outcome of much of the subsequent analysis.

To proceed further, we first note that the form of the interactions implies three global charge-neutrality conditions in the thermodynamic limit: 
\begin{eqnarray}
&& E_{\rm tot} \equiv q_{\rm tot}+2\tilde{Q}_{\rm tot} = 0 \; , \nonumber \\
&& F_{\rm tot} \equiv p_{\rm tot} + \tilde{P}_{\rm tot} = 0 \; , \nonumber \\
&& m_{\rm tot} = 0
\label{eq:chargeneutralityconstraints}
\end{eqnarray}
Guided by the form of the interactions in this action, we now switch to an equivalent formulation in terms of the charge-vector $(E,F,m)$, where $E_j = q_j+2\tilde{Q}_j$ and $F_j = p_j + \tilde{P}_j$:
\begin{eqnarray}
Z & \propto  &  \sum_{\left\{E,F,m\right\} }  C_{\left\{E,F,m\right\}}e^{-S_{\rm Coulomb}}
\end{eqnarray}
where the action has the form:
\begin{align}
& S_{\rm Coulomb} = \\
&  - \frac{1}{2g_{-}} \sum_{i \neq j} E_i
\log \left( \frac{|\mathbf{r}_i - \mathbf{r}_j|}{a} \right)
E_j \nonumber \\
&  - \frac{1}{2g_{+}} \sum_{i \neq j} \left(2 F_i \right)
\log \left( \frac{|\mathbf{r}_i - \mathbf{r}_j|}{a} \right)
\left(2 F_j \right) \nonumber\\
&  - \frac{g_{+}}{2} \sum_{i \neq j} \left(2 m_i \right)
\log \left( \frac{|\mathbf{r}_i - \mathbf{r}_j|}{a} \right)
\left(2 m_j \right) \nonumber \\
& -i \sum_{i \neq j} \left( 2 m_i \right) \Phi(\mathbf{r}_i - \mathbf{r}_j)
\left(2 F_j \right) 
-\sum_i \log Y(E_i,F_i,m_i)
\label{eq:continuumcoulombgasfinalform}
\end{align}
Here, the partition sum is now over configurations of integer-vector charges $(E,F,m)$, and $C_{\left\{E,F,m\right\}}$ denotes the usual combinatorial factor that accounts for the indistinguishability of charges with identical charge vectors $(E,F,m)$. The charge neutrality condition that is operative in the thermodynamic limit is of course $E_{\rm tot} = F_{\rm tot} = m_{\rm tot} = 0$. 

In formulating it in this manner, we have attempted to be somewhat more general than strictly necessary for the bare theory we started with, in which the charge vectors $(E,F,m)$ are of just
five types, which represent five ``rays'' in the three-dimensional charge lattice labeled by the coordinates
$(E,F,m)$: $(2u,u,0)$, $(-2l,l,0)$, $(q,0,0)$, $(0,p,0)$, and $(0,0,m)$ with corresponding fugacities given by $Y(2u,u,0)=Y_\lambda(u)$, $Y(-2l,l,0)=Y_\lambda(l)$,
$Y(q,0,0)=Y_{-}(q)$, $Y(0,p,0)=Y_{+}(p)$, and $Y(0,0,m)=Y_v(m)$. This is the initial condition for the flow equations we derive below. \begin{figure}[!]
    \includegraphics[width=\columnwidth]{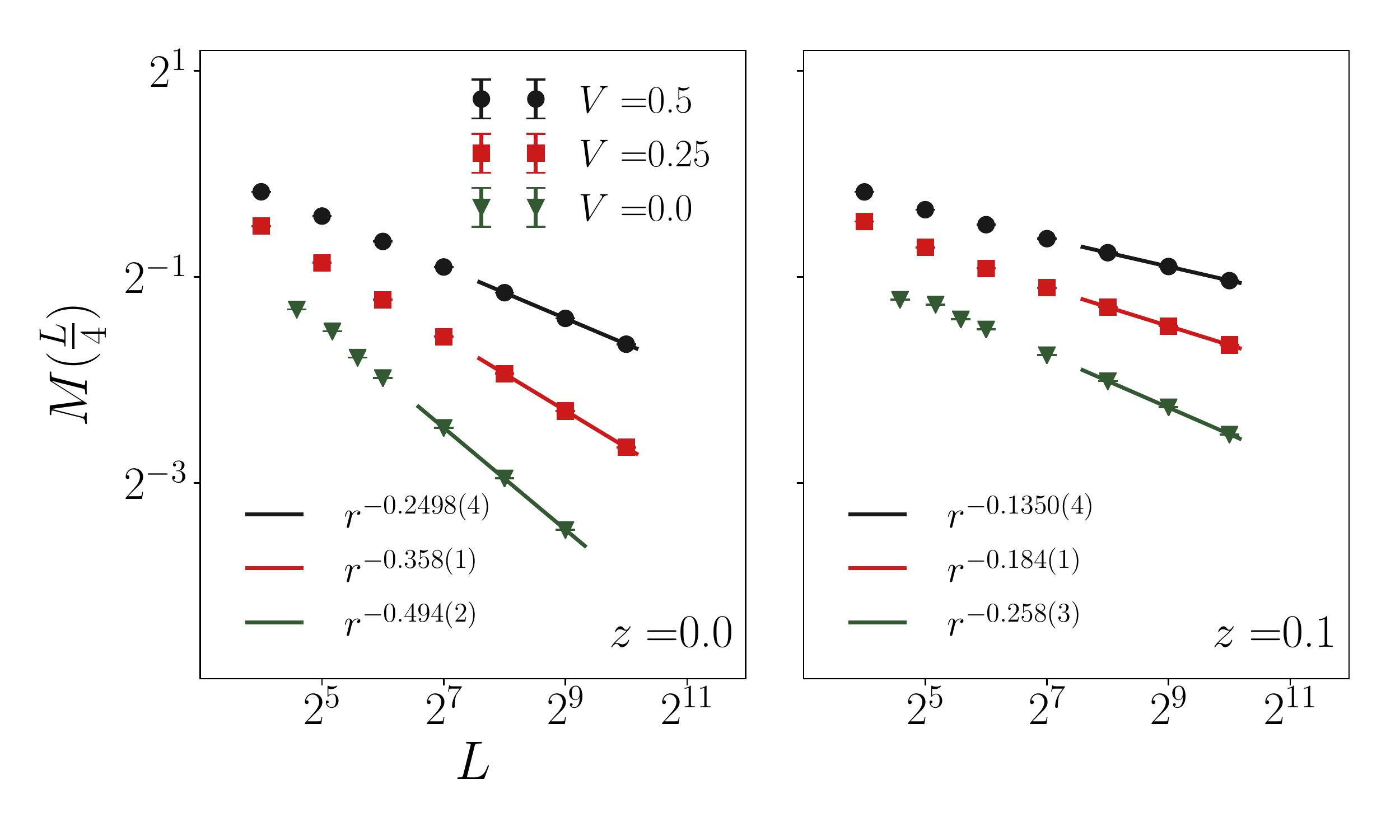}
    \caption{The monomer-antimonomer correlation function (normalized to be unity at separation $\mathbf{r} = 0$) for separation $\mathbf{r}=(\frac{L}{4},0)$ over a range of interaction strengths $V \geq 0$  at fugacity $z=0$ (left panel) and $z=0.1$ (right panel). Data in both panels fits well to a power law form $L^{-\eta_m}$, with best-fit values of $\eta_m(z=0,V)$ (left panel) and $\eta_m(z=0.1,V)$ (right panel) extracted in this way shown in the legend for various values of $V$. Our theoretical expectation is that $\eta_m(z=0,V)=g^{*}(V)$, where $g^{*}(V)$ is the long-wavelength value of the stiffness of the decoupled layers. Similarly, our theory predicts that $\eta_m(z,V) = g^{*}_{-}(z,V)$ for nonzero $z$ in the bilayer Coulomb phase; here $g^{*}_{-}(z,V)$ is the long-wavelength stiffness that characterizes a point in the bilayer Coulomb phase. Note that $\eta_m(z=0.1,V)$ obtained here obeys the approximate relation $\eta_m(z=0.1,V) \approx \eta_m(z=0,V)/2$ consistent with the theoretical expectation that $g^{*}_{-} \to g^{*}/2$ in the  $z \to 0$ limit. This seems to also be the case at $V=0$ when the data is fit over our range of accessible sizes, although we expect an eventual crossover from the bilayer Coulomb phase to the disordered large-$z$ phase in this case at inaccessibly large length scales. See Sec.~\ref{sec:ScalingpictureV>0}, Sec.~\ref{sec:ScalingpictureVleq0} and Sec.~\ref{sec:numerics} for a detailed discussion.}
    
    \label{fig:monomerLby4fug0}
\end{figure}

\subsection{General flow equations}
\label{subsec:Generalfloweqns}
The RG flow equations for $S_{\rm Coulomb}$ can be derived in a fairly straightforward manner using Kosterlitz's renormalization group procedure (see for instance the review by Nienhuis~\cite{Nienhuis_review}), adapted suitably to account for the unusual features of our Coulomb gas, which consists of two flavours of electric charges, and a single flavour of magnetic charge that is conjugate to one of these two electric charges. Equivalent results can presumably be obtained by working instead with the corresponding coupled sine-Gordon field theory, but we have not checked this directly for this particular problem. Since the structure of the interactions in $S_{\rm Coulomb}$ is somewhat unusual and does not appear to have been studied earlier, we first display the full RG flow equations obtained within this approach, before focusing on the behaviour of the couplings $Y(2u,u,0) \equiv Y_\lambda(u)$, $Y(-2l,l,0)=Y_\lambda(l)$,
$Y(q,0,0)=Y_{-}(q)$, $Y(0,p,0)=Y_{+}(p)$, and $Y(0,0,m)=Y_v(m)$ of particular interest to us.

As is well-known, the basic idea in Kosterlitz's RG procedure is to increase slightly the microscopic cutoff length scale from $a$ to $ae^{\delta l}$, so that $ae^{\delta l}$ represents the minimum separation between two charges of the renormalized Coulomb gas after one step of the RG procedure, and work out how the form of $S_{\rm Coulomb}$ changes if we reexpress the partition function only in terms of effective charges that now have a minimum separation $ae^{\delta l}$. There are two processes that change the configuration of charges under this operation: Either a charge can combine with another charge to give rise to an effective charge with a different charge vector, or two charges with equal and opposite charge vectors can annihilate each other. Apart from keeping track of these two possibilities, one must also account for the change of
length scale in the dimensionless arguments of the logarithms that govern the interaction between charges. 
In effect, this is a low-density expansion, valid in the vicinity of the charge vacuum, {\em i.e.} the Gaussian fixed plane parameterized by $g_{\pm}$.

The leading contribution to the flow of $g_{-}$ comes from the annihilation of a pair of charges with charge vectors of the form $(E,F,0)$ or $(E,0,m)$ with nonzero $E$. Similarly, the leading contribution to the flow of $g_{+}$ comes from the annihilation of charge vectors which are either of the form $(E,F,0)$ with nonzero $F$,
or of the form $(E,0,m)$ with nonzero $m$. Finally, the fugacity $Y(E,F,m)$ flows at leading order due to just the rescaling of $a$. Higher order contributions to the flow of $Y$ come from the merger of two charges, as well as the annihilation of two charges.
Keeping track of all these processes, one arrives at flow equations that control the scale dependence of
$g_{-}$, $g_{+}$ and the fugacities $Y(E,F,m)$.

The equation for $g_{-}$ reads
\begin{align} 
&\frac{dg_{-}}{dl} = \\
& 2 \pi^2 \Bigg[ \sum^{'}_{\left( E_p,0,m_p \right)} E^2_p \: Y(E_p,0,m_p) \: Y(-E_p,0,-m_p)  \nonumber \\
& + \sum^{'}_{\left( E_p,F_p,0 \right)} E^2_p \: Y(E_p,F_p,0) \: Y(-E_p,-F_p,0) \; , \nonumber
\Bigg]
\label{eq:generalflowgminus}
\end{align}
where charge vectors of the type $(E_p,0,0)$ are by convention included only in the first sum, and the prime on both summations indicate a restriction to terms with $E_p > 0$. As is evident from the structure of the right hand side of this equation, the flow of $g_{-}$ is controlled
by the annihilation of charges of the type $(E,0,m)$ or $(E,F,0)$ (with nonzero $E$) as the cutoff length scale is progressively increased.
Similarly, the flow equation for $g_{+}$ reads
\begin{align}
&\frac{dg_{+}}{dl} = \\
& 2 \pi^2 \Bigg[ \sum^{'}_{\left( E_p,F_p,0 \right)} 4 F^2_p \: Y(E_p,F_p,0) \: Y(-E_p,-F_p,0)  \nonumber \\
& - 4 g^2_{+} \sum^{'}_{\left( E_p,0,m_p \right)} m^2_p \: Y(E_p,0,m_p) \: Y(-E_p,0,-m_p) \; .\nonumber 
\Bigg]
\label{eq:generalflowgplus}
\end{align}
This flow is controlled by the annihilation of charges with vectors that have one of $F$ or $m$ nonzero, and the prime on the summations again denotes a restriction to the relevant half-plane ($F_p >0$ in the first sum and $m_p > 0$ in the second).

Finally, the flow of the fugacities is governed by 
\begin{align}
&\frac{dY(E,F,m)}{dl} = \\
& \left(2 - \frac{E^2}{2g_{-}} -\frac{4F^2}{2g_{+}} - \frac{g_{+}}{2} (4m^2) \right) Y(E,F,m) \nonumber \\
& + \pi \sum''_{(E',F',m')} Y(E',F',m') Y(E - E', F - F', m - m') \times \nonumber \\
& \:\:\:\:\:\:\:\:\:\:\:\:\:\:\:\:\:\:\:\:\:\:\:\:\:\:\:\:\:\:\:\:\:\delta(m F' + m' F - 2m' F') \nonumber \\
& - {\mathcal C} Y(E,F,m) \Bigg[ \sum^{'}_{(E',F')} Y(E',F',0) Y(-E',-F',0) \nonumber \\
& \:\:\:\:\:\:\:\:\:\:\:\:\:\:\:\:\:\:\:\:\:\:\:\:\:\:\:\:\:\:\:\:\:
+ \sum^{'}_{(E',m')} Y(E',0,m') Y(-E',0,-m') \Bigg]
\label{eq:generalflowY}
\end{align}
where the double primes on the sum indicate that $(0,0,0)$ and $(E,F,m)$ are to be left out of its ambit, the single primes indicate that the corresponding sum is over the appropriate half-plane, and 
${\mathcal C}=\sqrt{3}\pi+ 8\pi^2/3$. Here, the first term is simply the leading change in the fugacities due to a rescaling of the cutoff, the second term term captures the effect of the merger of two charges, while the 
third term accounts for the renormalization due to the annihilation of two charges. \begin{figure}[!]
 \includegraphics[width=\columnwidth]{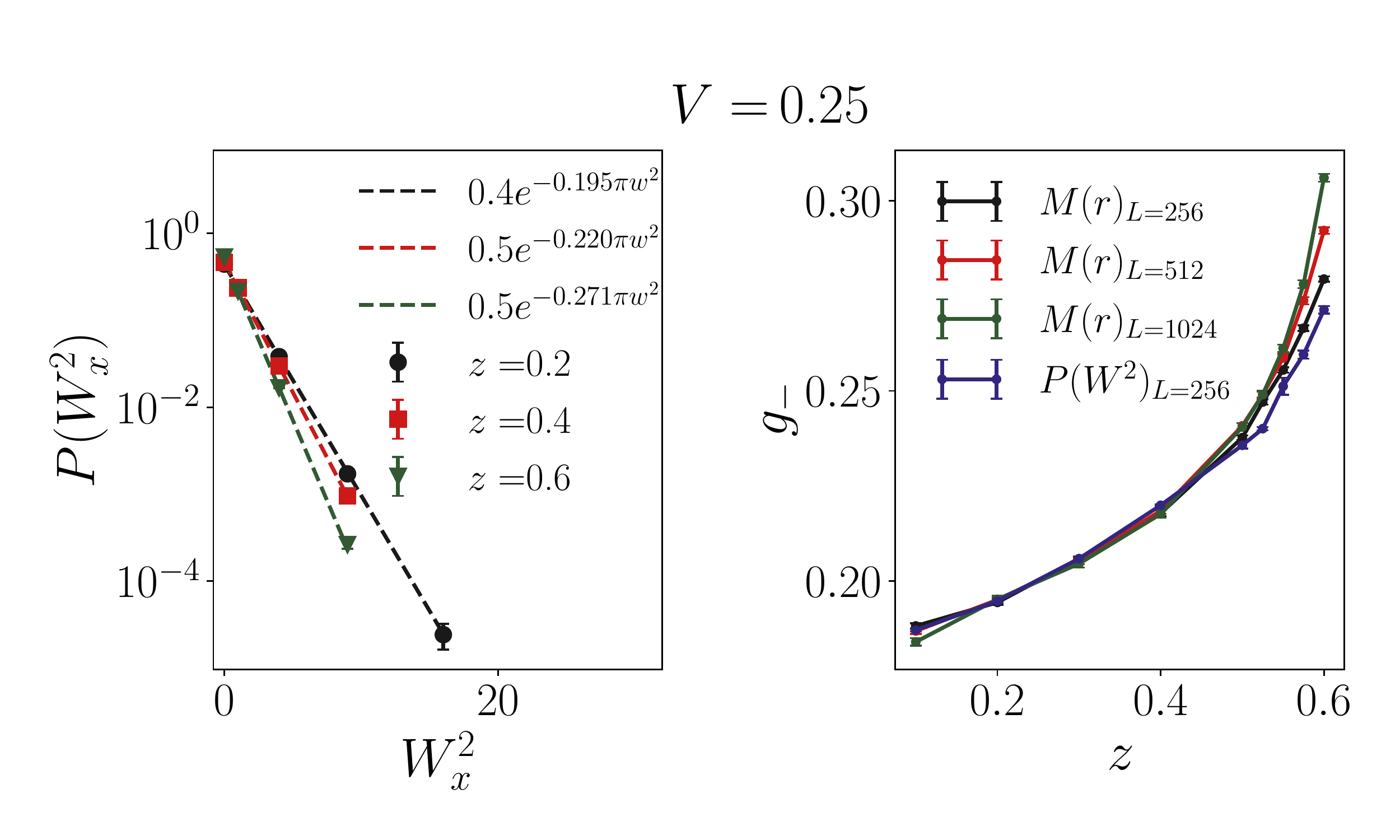}
 \includegraphics[width=\columnwidth]{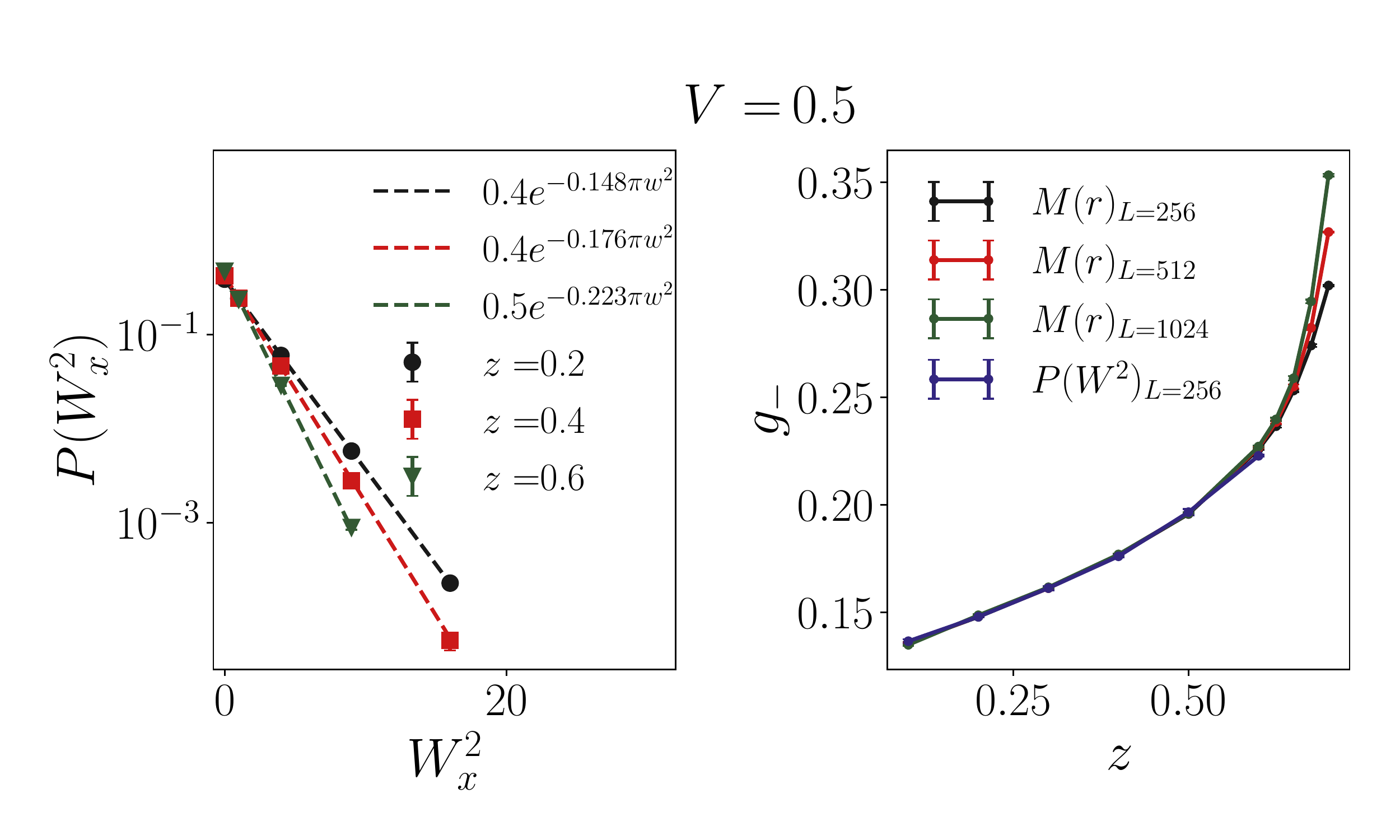}
	\caption{ \label{fig:w2gcomparison_rep}
	 Comparison of $g_{-}^{*}$ extracted from two different analyses: (1) fitting monomer correlations to a power law, $M(r) = A r^{-g_{-}^{*}}$ and (2) fitting histograms of $W_x^2$ and $W_y^{2}$ to a common functional form $P(W^2) = C e^{- \pi g^{*}_{-} W^2}$ for $V=0.25$ (top panels) and $V=0.5$ (bottom panels). The actual Gaussian fits are shown in the left panels. The monomer power law exponent has been extracted from fits to monomer-antimonomer correlations at separation $\mathbf{r}= (L/4,0)$ over a range of sizes up to $L= L_{max}$. The right panels display a comparison between the values of $g^{*}_{-}$ obtained in these two ways, for a range of choices of $L_{\max}$. The legends $M(r)_{L=L_{max}}$ in the right panels give the value of $L_{max}$ in each case. As is clear from the right panels, these estimates of $g^{*}_{-}$ are all consistent with each other for a range of nonzero $z$ for both $V=0.25$ and $V=0.5$. This provides compelling evidence in favour of a bilayer Coulomb phase that extends over a sizeable range of $z$ for not-too-large repulsive interactions $V$. Beyond a threshold value of fugacity $z$, we see deviations between the different estimates of $g^{*}_{-}$, which reflect the fact that this Coulomb phenomenology no longer provides a consistent account of the long-distance behaviour as the system transitions into the large-$z$ disordered phase. For even larger values of $z$, the monomer correlations decay faster than a power law, and there are no appreciable winding fluctuations. See Sec.~\ref{sec:ScalingpictureV>0} and Sec.~\ref{sec:numerics} for further details.}
\end{figure}

\subsection{Leading order flow equations near fixed plane}
\label{subsec:Leadingorderfloweq}
In order to develop a scaling theory for the behaviour of the system and use it to understand the physical picture at large length scales, we begin with the observation that all points on the $(g_{-}^{*},g_{+}^{*})$ plane, i.e with all renormalized fugacities $Y^{*}$ set to zero, are fixed points of the RG flows. The microscopic tuning parameters  $V$ and $z$ and the geometry of the square lattice control the bare values $g_{+}$, $g_{-}$ as well as all the bare fugacities $Y$ that determine the initial
conditions for this flow. 

At $V=z=0$, which corresponds to two decoupled square lattice dimer models, we have a bare theory with $g_{+} = g_{-} = g/2$. The only nonzero fugacity in the bare theory is $Y_\lambda > 0$. This is because symmetry dictates that $g_{12}$, $Y_{+}$, $Y_v$ and $Y_{-}$ can only be nonzero for nonzero $z$, {\em i.e.}
only when the two layers are coupled by interlayer dimers: The decoupled system at $z=0$ has a larger symmetry (of {\em independent} translations of $h_1$ and $h_2$ by a $1/4$, and independent changes of sign of $h_1$ and $h_2$) which forbids these terms. 
In this noninteracting $z=0$ system, one expects $Y_{\lambda}(x)$ to flow to zero, and $g$ to flow to a fixed point value $g^{*} = 1/2$, corresponding to the known behaviour of the square lattice dimer model without interactions.\citep{Fisher_Stephenson_dimercorrelations}

Further, one expects the bare values of $g$ and hence $g_{\pm} = (g \mp g_{12})/2$ to increase if an attractive interaction $V<0$ is turned on, and decrease if a repulsive $V>0$ is ramped up. Indeed, as $V$ varies in the range $(V_c,V_s)$, the decoupled layers at $z=0$ are expected to be described by a fixed point with
$g_{-}^{*} = g_{+}^{*} =  g^{*}/2$, where $g^{*}(V)$ is a decreasing function of $V$ that takes values in the range 
$(0,4)$, with $g^{*} \to 0$ as $V \to V_s$ and $g^{*} \to 4$ as $V \to V_c$. 

Turning on the interlayer fugacity $z$ is expected to lead to an increase in the bare value of $g_{-}$ and a concomitant decrease in $g_{+}$ (since a nonzero $z$ is expected to give rise to a bare $g_{12} >0$ of order ${\mathcal O}(z^2)$). A nonzero $z$ also gives rise in general to nonzero fugacities $Y_v(m)$, $Y_+(p)$, and $Y_{-}(q)$ in the bare theory (in addition to the $Y_\lambda(x)$ that is already present at $z=0$). More precisely, we expect the bare values of the $Y_{\pm}$ to be of order ${\mathcal O}(z^2)$, while the bare value of the  $Y_v$ is expected to be of order ${\mathcal O}(z)$, as already noted in Eq.~\ref{eq:smallzasymptotics}.

Next, we note from the structure of the quadratic terms in the general flow equations Eq.~\ref{eq:generalflowY} that no such quadratic terms can arise in the flow equations for the leading-order couplings of each symmetry class, {\em i.e.} for $y_v \equiv Y_v(m=\pm 1)$, $y_{-} \equiv Y_{-}(q = \pm 1)$, $y_{+} \equiv Y_{+}(p = \pm 1)$, and $y_{\lambda} \equiv Y_{\lambda}(x=\pm 1)$, so long as we do not include the effects of additional couplings corresponding to new types of charge vectors that are generated by the renormalization flows. 
Since these effects, and the effects of the cubic terms are both systematically small in the vicinity of the fixed-plane, we can develop a scaling picture for the behaviour of the system by working with the linearized flow equations for  $y_v$, $y_q$, $y_{+}$, and $y_{\lambda}$, in conjunction with the leading-order flow equations for the two stiffnesses $g_{+}$ and $g_{-}$, in which we only include the contributions of $y_v$, $y_q$, $y_{+}$, and $y_{\lambda}$ to the flow of $g_{\pm}$.\begin{figure}[!]
\includegraphics[width=\linewidth]{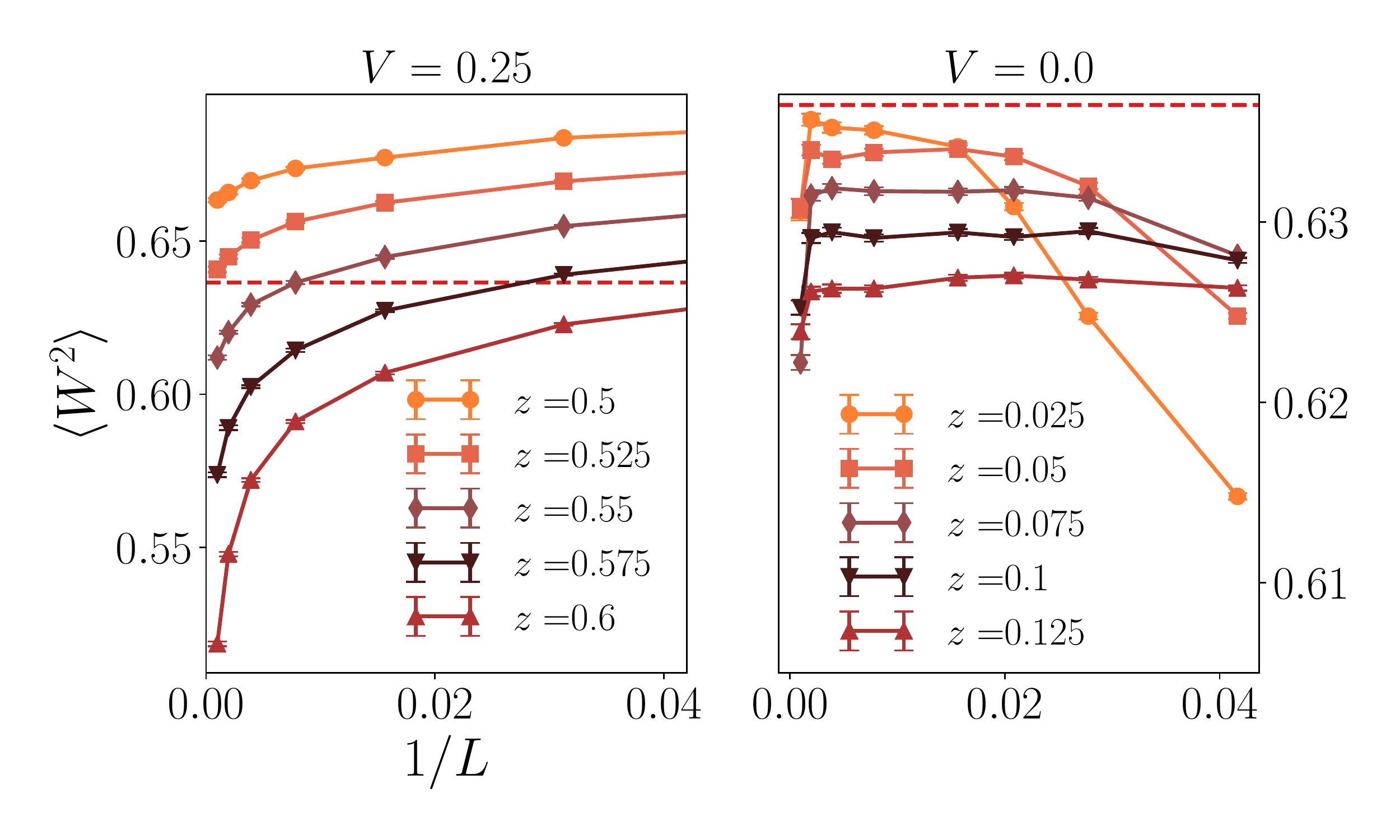}
\caption{\label{fig:repulsivezoom} Left panel: A closer look at the mean square winding $\langle W^2 \rangle = \langle W_x^2+W_y^2 \rangle/2$ for repulsive interaction $V=0.25$ and $z$ in the vicinity of the transition from the bilayer Coulomb phase to the large-$z$ disordered phase (this is a close-up of the relevant regime in the data set already displayed in Fig~\ref{fig:w2vsL} (b)). Right panel: The corresponding data for a range of $z$ close to $z=0$ for the noninteracting $V=0$ system. The dashed line denotes the value $\langle W^2 \rangle = {\mathcal J}(g_{-}^{*} = 1/4) =0.6365\dots$ which is the expected value of $\langle W^2 \rangle$, corresponding to the inverted Kosterlitz-Thouless transition that separates the bilayer Coulomb phase from the large-$z$ disordered phase.
The $V=0.25$ dataset in the left panel is seen to clearly have a separatrix that coincides with this critical value of $\langle W^2 \rangle$, while the $V=0$ dataset lies entirely below this critical value, consistent with our expectation that there is no stable bilayer Coulomb phase at $V=0$.
See Sec.~\ref{sec:ScalingpictureV>0} and Sec.~\ref{sec:numerics} for a detailed discussion.}
\end{figure}
\begin{figure}[!]
    \includegraphics[width=0.5\textwidth]{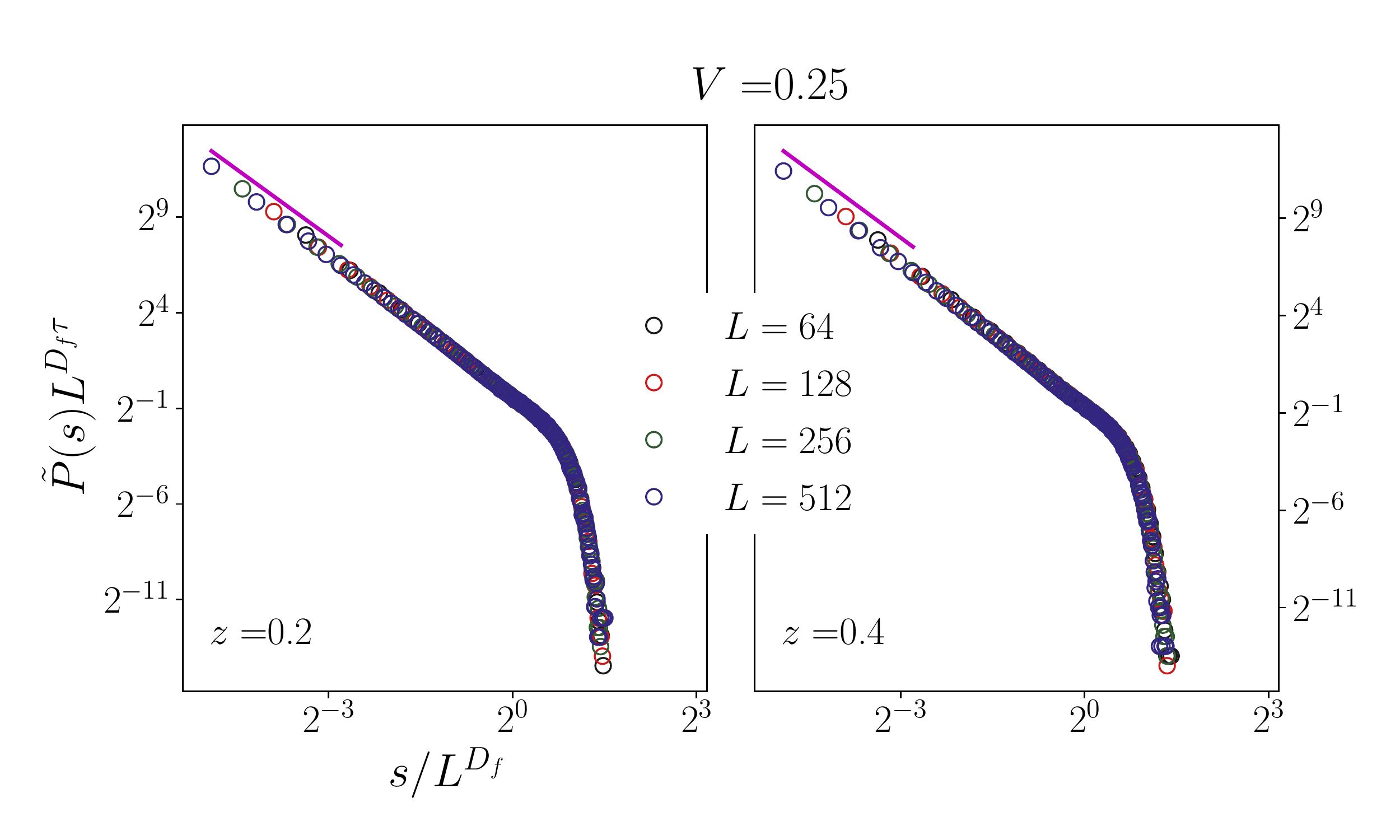}
    \caption{The probability distribution $\tilde{P}(s,L)$ for non-winding overlap loops of length $s$ in a $L\times L$ sample with periodic boundary conditions collapses well onto the postulated scaling form Eq.~{\protect{\ref{eq:scalingformforoverlaps}}} for $V=0.25$ at fugacity $z=0.2$ and $z=0.4$. Further, the scaling function $\Phi(x) $is seen to have the expected power-law behaviour $x^{-\tau}$ with $\tau = 7/3$ for $x \ll 1$. The magenta lines with slope $7/3$ provide visual confirmation of this behaviour. This provides further evidence for the existence of a stable bilayer Coulomb phase in this parameter regime. See Sec.~\ref{sec:ScalingpictureV>0} and Sec.~\ref{sec:numerics} for a detailed discussion.}
    \label{fig:scalingformforoverlaps}
\end{figure}

These observations motivate the following leading-order flow equations:
\begin{align}
&\frac{dy_v}{dl} = \left(2 - 2g_{+} \right) y_v\nonumber \\
&\frac{dy_{-}}{dl} = \left(2 - \frac{1}{2g_{-}} \right) y_{-}\nonumber \\
&\frac{dy_{+}}{dl} = \left(2 - \frac{2}{g_{+}} \right) y_{+}\nonumber \\
&\frac{dy_{\lambda}}{dl} = \left(2 - \frac{2}{g_{-}} -\frac{2}{g_{+}} \right) y_{\lambda} 
\label{eq:leadingflowY}
\end{align}

\begin{align} 
&\frac{dg_{-}}{dl} = 2 \pi^2 \Bigg[ y_{-}^2   + 8 y_{\lambda}^2 \Bigg] \nonumber \\
&\frac{dg_{+}}{dl} = 2 \pi^2 \Bigg[ 8y_{\lambda}^2 + 4y_+^2 -  4g^2_{+} y_v^2\Bigg]
\label{eq:leadingflowg}
\end{align}

These leading order equations are valid as long as all the fugacities remain small enough. This is indeed the case for the bare values of the fugacities at small $z$, as we have noted in Eq.~\ref{eq:smallzasymptotics}.The linear (`tree-level') terms in these equations could have of course been obtained simply by working out the power-law exponents that govern the long-distance behaviour of the correlation functions of the corresponding operators in the Gaussian theory.
The more elaborate analysis sketched in the foregoing is however needed to obtain the form of the higher order terms in the full flow equations written down earlier, as well as the leading second order contributions to the flow of the stiffnesses $g_{\pm}$.

In our subsequent analysis, we use this system of flow equations and the information about the initial conditions for these
flows summarized above to determine the asymptotic behaviour of the system. This analysis separates quite naturally into three parts, corresponding to systems with attractive interations $V < 0$, noninteracting systems with $V=0$, and systems with repulsive interactions $V>0$. Below, we consider each in turn.

\section{Scaling picture: $V>0$}
\label{sec:ScalingpictureV>0}
In this case, the decoupled layers at $z=0$ are described by a fixed point with $g^{*} <  1/2$, which translates to fixed-point values $g_{-}^{*} = g_{+}^{*} < 1/4$ at $z=0$. Turning on a small $z$ leads to a correspondingly small value for $g_{12}$, resulting in a small increase in the bare value of $g_{-}$, and a corresponding reduction in the bare value of $g_{+}$. It also leads to nonzero bare values for $y_v$, $y_{+}$ and $y_{-}$ since the symmetries of the $z > 0$ system permit these terms. In addition, we also have a nonzero $y_{\lambda}$ in the bare theory even at $z=0$.
\begin{figure}[!]
    \includegraphics[width=\columnwidth]{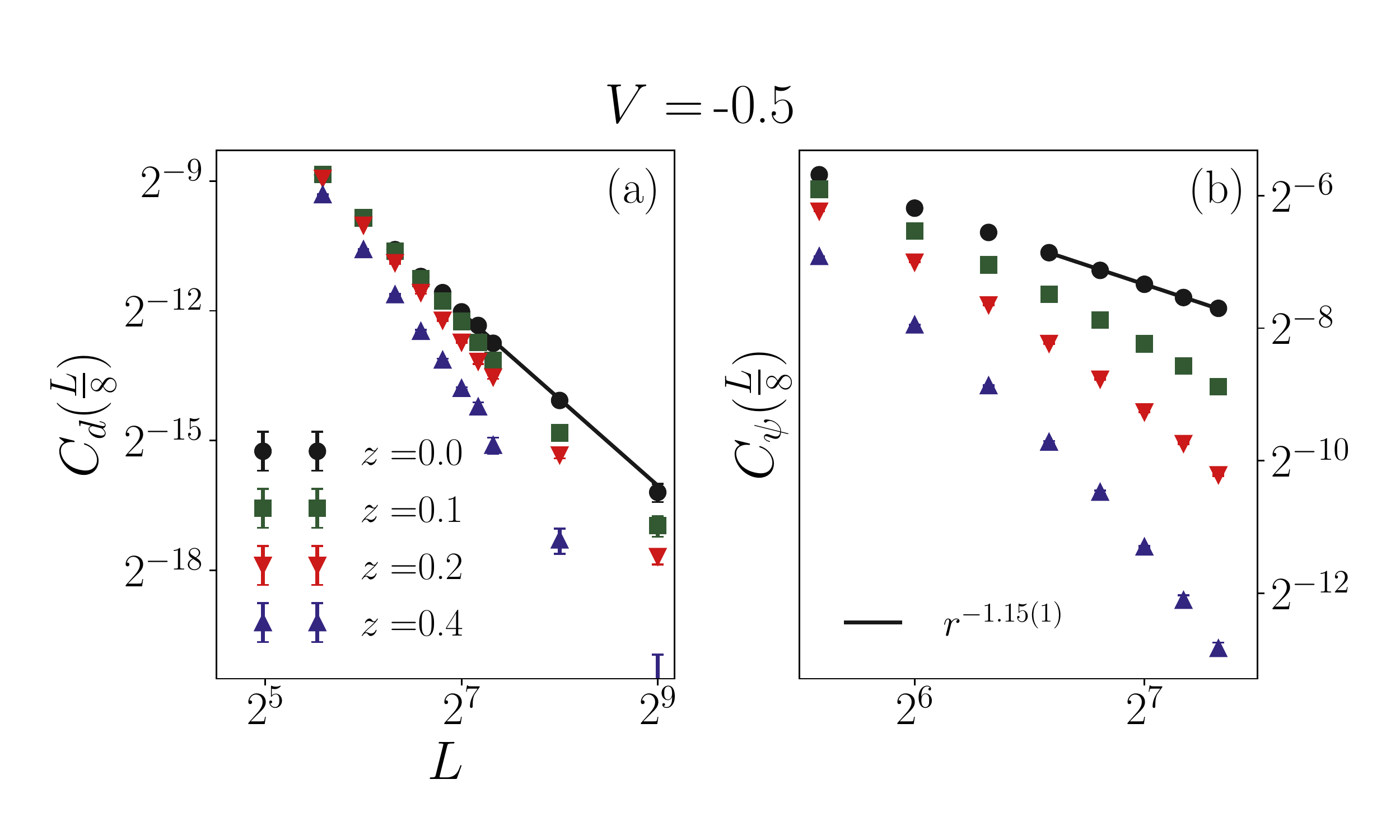}
    \caption{
Left panel: The dipolar component of intralayer dimer correlations (as defined in Eq.~\ref{eq:dipolar_linear_combo}) for $V=-0.5$ fits to the expected form $a L^{-2}$ only for $z=0$, but decays faster than a power law for nonzero $z$. Right panel: The corresponding
    columnar component at $V=-0.5$ (as defined in Eq.~\ref{eq:columnar_linear_combo_1}) is seen to decay with a power law of $r^{-1.15(1)}$ at $z=0$. However, for $z>0$, it decays to zero much more rapidly. Taken together, these behaviours show that there is no bilayer Coulomb phase for nonzero $z$ in this regime. See Sec.~\ref{sec:numerics} for details.}    
    \label{fig:psicorr_att}
\end{figure}
\begin{figure}
    \includegraphics[width=\columnwidth]{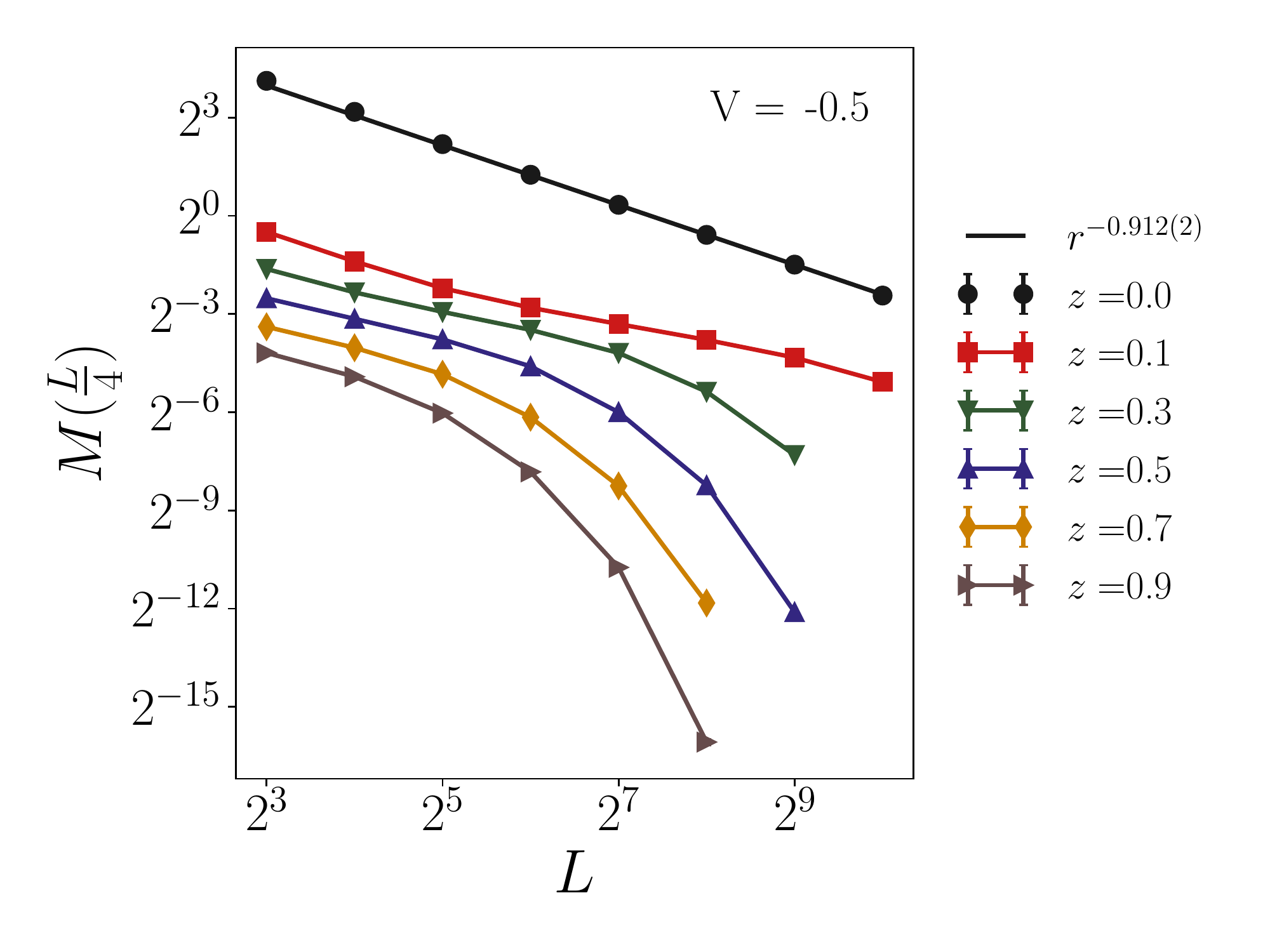}
    \caption{The monomer-antimonomer correlation function (normalized to be unity at separation $\mathbf{r} = 0$) for separation $\mathbf{r}=(\frac{L}{4},0)$ for $V = -0.5$  at fugacity $z=0$ fits well to a power law form $L^{-\eta_m}$, with best-fit value $\eta_m(V=-0.5,z=0) = 0.912(2)$. Our theoretical expectation is that $\eta_m(z=0,V)=g^{*}(V)$, where $g^{*}(V)$ is the long-wavelength value of the stiffness of the decoupled layers. For nonzero $z$, we see that the data deviates from power-law behaviour, falling off faster. This is consistent with our theoretical expectation that there is no stable bilayer Coulomb phase for any nonzero $z$ when $V < 0$. See Sec.~\ref{sec:ScalingpictureVleq0} and Sec.~\ref{sec:numerics} for a detailed discussion.}
    \label{fig:monomercorr_att}
\end{figure}

\subsection{$V \in (0,V_s)$; $z< z_c(V)$: Bilayer Coulomb phase}
\label{subsec:BilayerCoulombphase}
From the leading order scaling equations, it is clear that $y_{+}$ is strongly irrelevant and flows rapidly to zero in this regime, which corresponds to $g^{*}$ in the range $(0,1/2)$. Thus, it does not directly influence long-distance behaviour. This is also true of $y_\lambda$, which is strongly irrelevant and flows rapidly to zero, thereby decoupling the $\theta_{+}$ sector of the theory from the $\theta_{-}$ sector as far as the long-wavelength physics is concerned. The coupling $y_{-}$ induced by nonzero $z$ is also irrelevant, but flows to zero more slowly than $y_{+}$, scaling as $y_{-} \sim z^2 \exp\left((2-1/g^{*})l\right)$ as a function of the RG scale $l$. In contrast, $y_v$ induced by small nonzero $z$ is relevant, and scales as $y_v \sim z \exp( (2-g^{*})l)$. This flow of $y_v$ to strong coupling implies that $\theta_{+}$ is disordered beyond the length scale $\xi_v \sim (1/z)^{1/(2-g^{*})}$. Likewise, beyond a length scale $\xi_{-} \sim z^{-1/(1-1/2g^{*})}$, $y_{-}$ is negligible, implying that long-wavelength fluctuations of $\theta_{-}$ continue to be described by the Gaussian fixed point form of the action for $\theta_{-}$, albeit with a slightly altered value of $g_{-}^{*}$, which is induced by these flows. This signals the existence of an entirely new kind of Coulomb phase, which we dub the {\em bilayer Coulomb phase}. This bilayer Coulomb phase is characterized by the striking absence of power-law columnar order for the dimers in either layer, and an altered pattern of coefficients for the pinch-point singularities at the dipolar wavevector.

To see this, we recall, from Eq.~\ref{eq:operatorcorrespondencex} and Eq.~\ref{eq:operatorcorrespondencey},  that the dimer density operators $n_{\mu a}(r)$ (where $\mu = x,y$ denotes orientation of dimer and $a=1,2$ the layer index) have a representation consisting of two terms, one oscillating at the columnar wavevector and proportional to the real or imaginary parts of $\exp(i \theta_a(r))$, and the other oscillating at the dipolar wavevector and proportional to $\epsilon_{\mu \nu} \partial_\nu \theta_a$. 
In this description, the power-law columnar order which characterizes the decoupled $z=0$ limit of our bilayer is a consequence of power-law correlations of $\exp(i\theta_1)$ and $\exp(i\theta_2)$ in the effective field theory. To analyze the correlators of these vertex operators when $z$ becomes nonzero in this regime, it is useful to write $\theta_{1/2}$ as linear combinations of $\theta_+$ and $\theta_{-}$ since the $\theta_{+}$ sector is decoupled from the $\theta_{-}$ sector at long-wavelengths when $z > 0$. 

In this manner,  we immediately see that the correlation functions
$\langle e^{i\theta_{a}(r)} e^{-i\theta_{b}(0)} \rangle$ for $a,b = 1,2$ all factorize into a {\em product} of two factors: a short-ranged factor contributed by correlations of $\exp(i\theta_{+}/2)$, and power-law decay with a floating exponent $\eta_{-} = 1/4g^{*}_{-}$ contributed by correlations of $\exp(i\theta_{-}/2)$. In spite of this power-law contribution, the short-ranged nature of the other factor causes the product to remain short-ranged.
In sharp contrast, the dimer density correlations in the vicinity of the dipolar wavevector remain of the dipolar form since these correlations are a {\em sum} of two terms, a short-ranged piece arising from the $\theta_{+}$ sector, and a dipolar power-law arising from the $\theta_{-}$ sector. As a result of this additive structure, the dipolar contribution arising from the $\theta_{-}$ sector controls the long-distance behaviour of these correlations in the vicinity of the dipolar wavevector. This also leads to an altered pattern of coefficients for the pinch-point singularities in the structure factors of $n_{1}$ and $n_{2}$, which we characterise presently. 
Thus, this regime represents a qualitatively distinct {\em bilayer Coulomb phase} characterized by purely dipolar power-law correlation functions of the dimer density operators of each layer. 

A long-wavelength description of this physics is readily obtained from the fixed-point description in the $\theta_{-}$ sector, augmented by a simple phenomenological description of the short-ranged correlations in the $\theta_{+}$ sector. For the $\theta_{-}$ sector, this fixed-point description of a bilayer dimer system with periodic boundary conditions
on a $L_x \times L_y$ torus may be summarized as follows:
\begin{eqnarray}
Z &=& \int {\mathcal D}h_{-}(\mathbf{r}) \exp \left(-\pi g^{*}_{-}\int d^2x (\nabla h_{-})^2 \right)\; ,
\label{eq:fixedpointdefinition1}
\end{eqnarray}
where the functional integral is over field configurations $h_{-}(\mathbf{r})$ with ``winding boundary conditions'' on the torus. To see this, we note that periodic boundary conditions on the bilayer dimer system do {\em not} translate simply to periodic boundary conditions on $h_{-}(\mathbf{r})$. Instead, they translate to a sum over winding sectors labeled by winding numbers $W_x$ and $W_y$. In any particular winding sector, the path integral is over field configurations  $h_{-}(\mathbf{r})$ that obey winding boundary conditions:
\begin{eqnarray}
h_{-}(L_x,0) &=& W_x + h_{-}(0,0) \nonumber \\
h_{-}(\mathbf(0,L_y) &=& W_y + h_{-}(0,0) \; .
\label{eq:definition_windingBCS}
\end{eqnarray}

We now change variables 
\begin{equation}
h(\mathbf{r}) = \tilde{h}(\mathbf{r}) + W_x x/L_x + W_y y/L_y \;,
\end{equation}
where $\tilde{h}(\mathbf{r})$ now has periodic boundary conditions regardless of winding sector.
With this change of variables, we can conveniently rewrite the partition sum as
\begin{eqnarray}
Z &=& {\mathcal Z} (g^{*}_{-})\int {\mathcal D}\tilde{h}_{-}(\mathbf{r}) \exp \left (-\pi g^{*}_{-}\int d^2x (\nabla \tilde{h}_{-})^2 \right) \; ,
\label{eq:fixedpointdefinition2}
\end{eqnarray}
where the functional integral is now over field configurations $\tilde{h}_{-}(\mathbf{r})$ with periodic boundary conditions on the torus, and we have used an integration by parts to arrive at this factorized description. Here, the first factor ${\mathcal Z}$ accounts for the sum over winding sectors:
\begin{eqnarray}
{\mathcal Z}(g^{*}_-)&=&\sum_{W_x,W_y} e^{-\pi g^{*}_- (W_x^2+W_y^2)}
\label{eq:windingpartitionsum}
\end{eqnarray}

Thus, we see that the mean square winding $\langle W^2 \rangle = \langle W_x^2+W_y^2\rangle/2$, which can be measured in Monte Carlo simulations, can be readily obtained from this fixed-point description as:
\begin{eqnarray}
\langle W^2 \rangle &=& {\mathcal J}(g^{*}_-) \nonumber \\
{\mathcal J}(g^{*}_-)&=& -\frac{1}{2\pi} \frac{\partial \log{\mathcal Z}(g^{*}_-)}{\partial g^{*}_{-}} \; .
\label{eq:Jdefn}
\end{eqnarray}
This fixed point description also implies (see Appendix for details) the usual dipolar form of Coulomb-phase correlators for $n_{-}$, valid for nonzero but small $q$:
\begin{align}
	\langle \hat{n}_{\mu,-}(-\mathbf{Q}-\mathbf{q}) \hat{n}_{\nu,-}(\mathbf{Q}+\mathbf{q}) \rangle & =
\frac{1}{2 \pi g^{*}_-} 
\left( \delta_{\mu\nu} - \frac{q_\mu q_\nu}{q^2}\right) \; . 
\label{eq:transverse_structure_factor_k12_insec}
\end{align}
Additionally, we see that this structure factor {\em at} $\mathbf{Q}$ is given exactly by the mean square winding ${\mathcal J}(g^{*}_{-})$.

As described in detail in the Appendix, this may be supplemented by a simple phenomenological description of the correlations of $n_{+}$ in the vicinity of pinch-point wavevector $\mathbf{Q}$, to arrive at the following prediction:
\begin{eqnarray}
\langle \hat{n}_{\mu,+}(-\mathbf{Q}-\mathbf{q}) \hat{n}_{\nu,+}(\mathbf{Q}+\mathbf{q}) \rangle & = &
\frac{1}{2 \pi g_+}  \: \delta_{\mu\nu} \; ,
\end{eqnarray}
that reflects the short-ranged non-singular nature of $n_{+}$ correlations. Putting this together, we see that the bilayer Coulomb phase is expected to have an altered singularity structure in the layer-resolved structure factor for nonzero but small $q$:
\begin{align}
\text{intralayer:  } \langle \hat{n}_{\mu,a} & (-\mathbf{Q}-\mathbf{q}) \hat{n}_{\mu,a}(\mathbf{Q}+\mathbf{q}) \rangle \nonumber \\
& = \frac{1}{4} 
\left[ \frac{1}{2 \pi g_+}    
\delta_{\mu\nu} + \frac{1}{2 \pi g^{*}_-} \left( \delta_{\mu\nu} - \frac{q_\mu q_\nu}{q^2} \right)
 \right]  
\label{eq:structure_factor1_k12a_insec} \\
\text{interlayer:  } \langle \hat{n}_{\mu,a} & (-\mathbf{Q}-\mathbf{q}) \hat{n}_{\nu,b}(\mathbf{Q}+\mathbf{q}) \rangle \nonumber \\
& = \frac{1}{4} 
\left[ \frac{1}{2 \pi g_+}    
\delta_{\mu\nu} - \frac{1}{2 \pi g^{*}_-} \left( \delta_{\mu\nu} - \frac{q_\mu q_\nu}{q^2} \right)
 \right]  
\label{eq:structure_factor1_k12b_insec}
\end{align}
Note that these expressions do not carry over smoothly to $z=0$, since the $z \to 0$ limit does not commute with the $q \to 0$ limit.

Our fixed-point description of the bilayer Coulomb phase also has interesting implications for the random geometry of {\em overlap loops}, which we now discuss. If one superimposes the dimer configuration of one layer on to the corresponding configuration of the second layer (leaving out interlayer dimers), this defines a configuration of non-intersecting fully-packed loops (including loops of length $2$, corresponding to two intralayer dimers on corresponding links of the two layers) on a square lattice with {\em annealed vacancy disorder} (which encodes the fluctuating locations of the interlayer dimers). This is depicted in the example shown in Fig.~\ref{fig:overlaploopdefinition}.             Thought of in this way, this is an apparently complicated lattice model of non-intersecting fully-packed loops on a lattice with annealed vacancy disorder, with each loop configuration having weight $2^{N_{\rm loop}}$, where $N_{\rm loop}$ is the number of distinct loops of length larger than $2$.

However, in the bilayer Coulomb phase, we can think in terms of the Gaussian fixed point action for the coarse-grained height field $h_{-}$. In this language, these loops correspond to contour lines of a Gaussian free field with stiffness $g_{-}^{*}$. This insight allows us to connect the geometry of these overlap loops to that of contour lines of a Gaussian free field in two dimensions, which represents the height fluctuations of a Gaussian random surface. This has been studied in earlier work by Henley and Kondev.\cite{Henley_Kondev} To understand what to expect for the statistics of loop lengths $s$ in a finite $L \times L$ sample, we may use the results of Ref.~\onlinecite{Henley_Kondev} for the power-law distribution of loop lengths and the fractal dimension $D_f$ of these loops. As argued by Henley and Kondev, one expects
that the distribution of lengths $l$ of such contour lines scales as $\tilde{P}(s) \sim 1/s^{\tau}$ where $\tau=7/3$.\cite{Henley_Kondev} Also, the fractal dimension of these contour lines is given by $D_f = 3/2$.\cite{Henley_Kondev}

Here, we use finite-size scaling ideas to build on these results to arrive at a prediction for the corresponding finite-size behaviour:
\begin{eqnarray}
\tilde{P}(s,L) &=& \frac{{\mathcal C}}{L^{D_f\tau}} \Phi \left(\frac{s}{L^{D_f}} \right )
\label{eq:scalingformforoverlaps_insec}
\end{eqnarray}
where $\tau = 7/3$, $D_f = 3/2$, $\Phi(x) \sim x^{-\tau}$ for $x \ll 1$, and $\Phi(x)$ decays rapidly for $x \gg 1$. This finite-size scaling ansatz assembles information about both exponents $\tau$ and $D_f$ into a finite-size scaling form that provides a potentially useful framework for analysing the statistics of these overlap loops.

\subsection{Phases and transitions at large $V$ and $z$}
\label{subsec:LargeVandz}
As $V$ increases, $g^{*}$ for each decoupled layer at $z=0$ is expected to decrease monotonically, until it finally goes to zero at $V_{s}$---this signals a freezing transition into a staggered state, as noted in previous work.\cite{Castelnovo_Chamon_Mudry_Pujol_Annals} 

As we turn on $z$ for our bilayer in this vicinity, we expect this $z=0$ transition to continue into the $V-z$ plane as a line of transitions from the bilayer Coulomb phase to the staggered phase. Since we expect that the fixed-point value $g^{*}_{-}$ increases with increasing $z$ for small $z$, we expect at a qualitative level that this phase boundary will tilt upwards in the $V-z$ plane. This is depicted in Fig.~\ref{fig:SchematicPhaseDiagram}. Further, we note that a large enough $z$ will of course lead to a large-$z$ disordered phase even in this regime. Thus, there are three distinct phases in this region of parameter space, with large $V$ and large $z$: a bilayer Coulomb phase below a threshold value $V_s(z)$,
a large-$z$ disordered phase above a threshold value $z_c(V)$ and a frozen staggered phase above $V_s(z)$ for not-too-large $z$. 

This points to the interesting possibility of a multicritical point at which these three phases meet. We leave further study of this possibility to future work, and turn next to the case of attractive interactions.\begin{figure}[!]
    \includegraphics[width=\columnwidth]{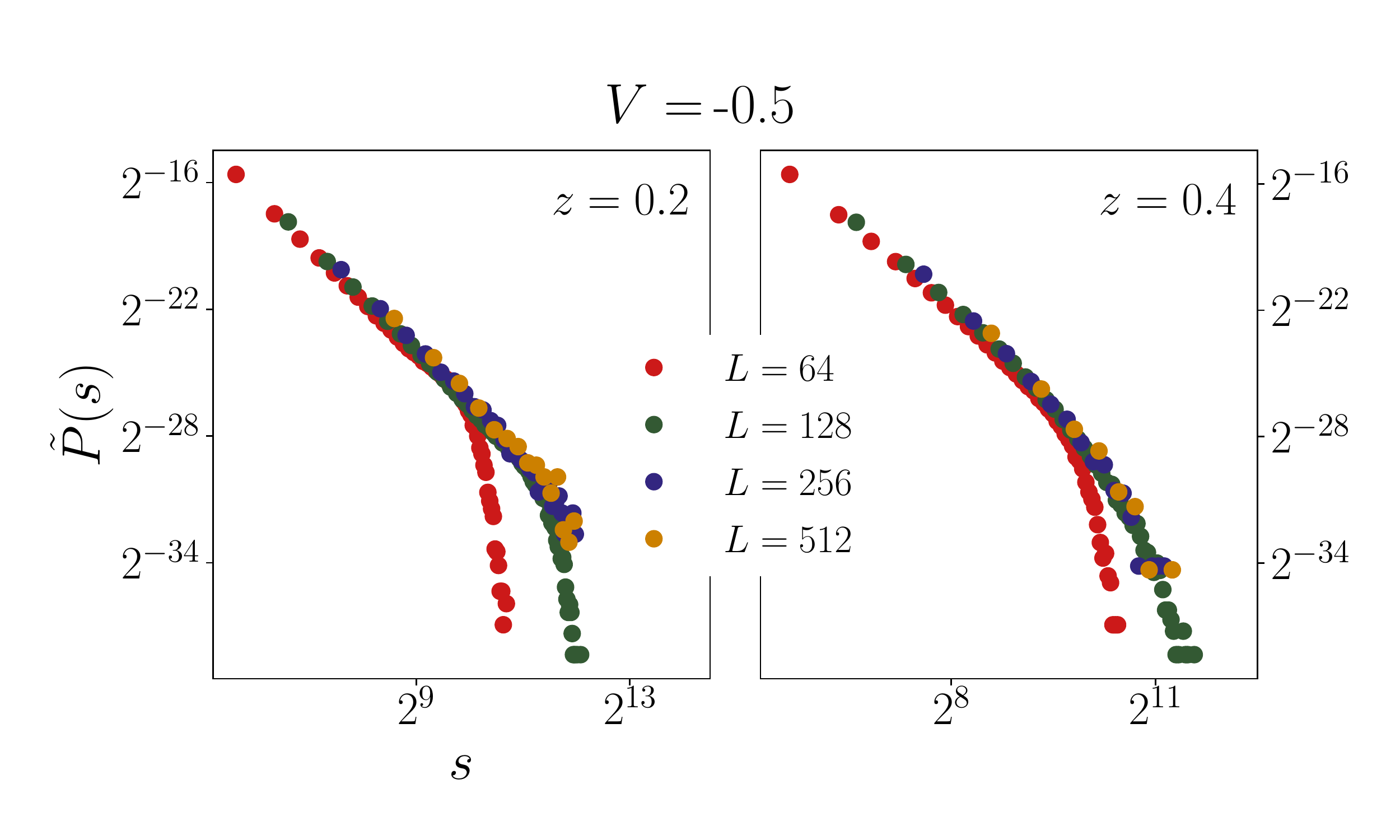}
    \caption{The probability distribution $\tilde{P}(s)$ of overlap loop lengths $s$ decays faster than a power-law for small nonzero $z$ at $V=-0.5$, consistent with our theoretical expectation that there is no stable bilayer Coulomb phase with attractive interactions. See Sec.~\ref{sec:ScalingpictureVleq0} and Sec.~\ref{sec:numerics} for a detailed discussion.}
    \label{fig:overlaploops_att}
\end{figure}
\begin{figure}[!]
    \includegraphics[width=\columnwidth]{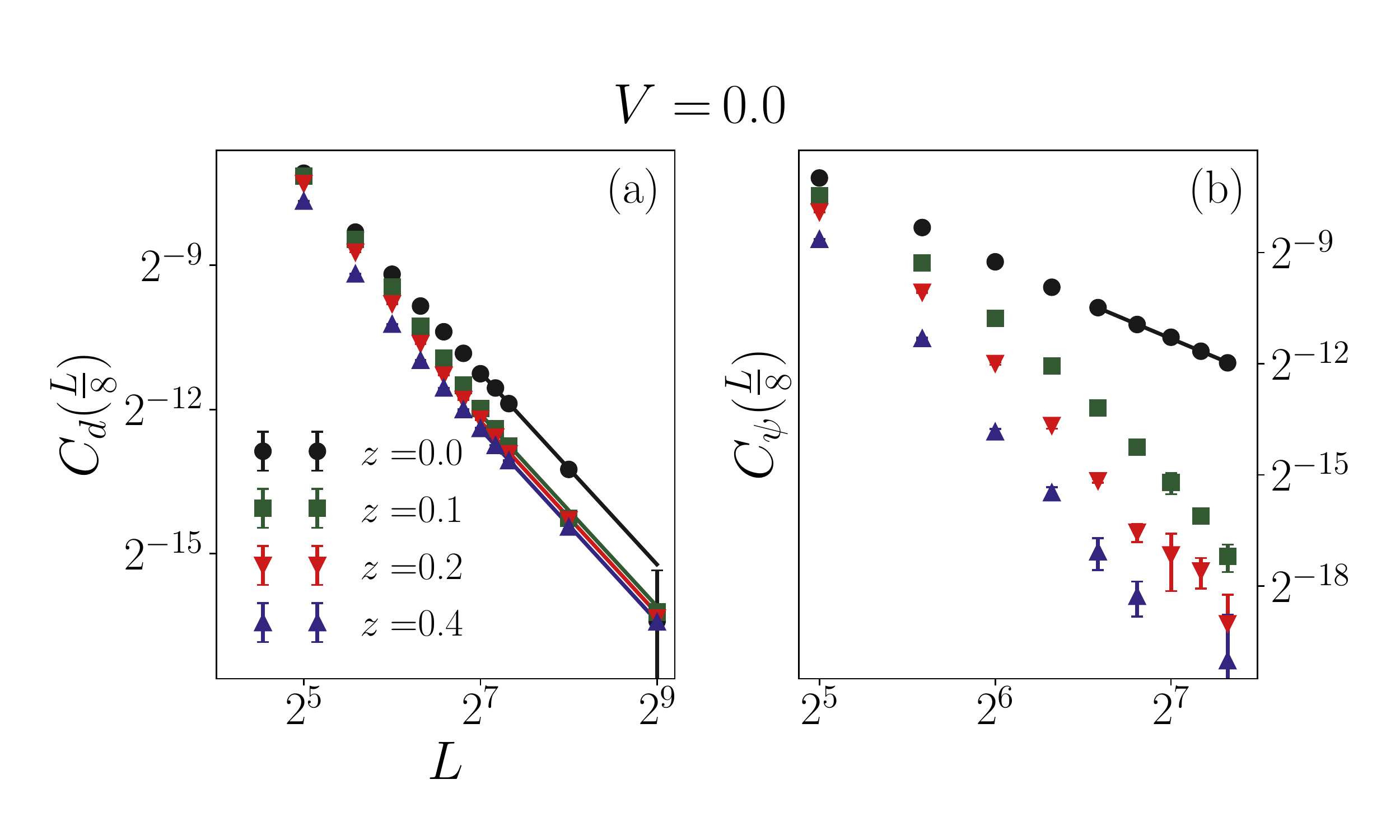}
    \caption{Left panel: The dipolar component of intralayer dimer correlations at $\mathbf{r} = (L/8,0)$ continues to show apparent power-law decay with power-law decay $\sim 1/L^2$ even at nonzero but small $z$ for $V=0$. Right panel: The corresponding columnar component for $V=0.0$ {\em at} $z=0$ has a power-law behaviour $\sim 1/L^2$, whereas the results for nonzero $z$ are better described by a crossover to $\sim 1/L^4$ behaviour at the largest $L$ available to us. Although this is qualitatively similar to the behaviour in the bilayer Coulomb phase for $V>0$, our theoretical analysis suggests that a very slow crossover from bilayer Coulomb to disordered large-$z$ behaviour is responsible for this apparent similarity. See Sec.~\ref{sec:ScalingpictureVleq0} and Sec.~\ref{sec:numerics} for a detailed discussion.}
    \label{fig:psicorr_nonint}
\end{figure}

\section{Scaling picture: $V_{cb}<V \leq 0$}
\label{sec:ScalingpictureVleq0}
Next we consider attractive interactions $V \leq 0$ that are not too large in magnitude (in a sense that is made precise here). In this case, the decoupled layers at $z=0$ are described by a $g^{*} \geq 1/2$ fixed point, which translates to fixed-point values $g_{-}^{*} = g_{+}^{*} \geq 1/4$ at $z=0$. 
Turning to the various fugacities, we see immediately that $y_{+}$ is irrelevant so long as $g_{+}^{*} < 1$. This corresponds to
$g^{*} < 2$ for the decoupled layers at $z=0$. On the other hand, $y_v$ is relevant for all $g_{+}^{*} < 1$,  {\em i.e.} for $g^{*} < 2$ for the decoupled layers at $z=0$. As $V$ becomes more and more negative $g^{*}$ increases from its $V=0$ value of
$g^{*} = 1/2$, and hits $g^{*} = 2$ for $V = V_{cb}$. From the results displayed in Fig.~31 of Ref.~\onlinecite{alet_etal_pre2006}, we estimate
$V_{cb} \approx -1.2$.

Here, we discuss the physics in this range of $V$, dealing first with nonzero attractive interactions with $|V| <  |V_{cb}|$
and next with the noninteracting $V=0$ case.

\subsection{Nonzero $|V| <  |V_{cb}|$: $z>0$ disordered phase}
\label{subsec:disorderedphase}

This is the most straightforward case from a scaling point of view: Since $g_{-}^{*} > 1/4$ for the $z=0$ decoupled layers, we see that turning on a $z$ results in a bare $y_{-}$ which is always relevant at the $z=0$ fixed point. As a result $y_{-}$
flows to strong coupling. This flow of $y_{-}$ to strong coupling also drives $g_{-}$ to larger and larger values leading to runaway flows to strong coupling. In addition, $y_{v}$ is relevant and also flows to strong coupling.
However, both $y_{+}$ and $y_{\lambda}$ are irrelevant in this regime, and expected to renormalize to zero. \begin{figure}[!]
    \includegraphics[width=\columnwidth]{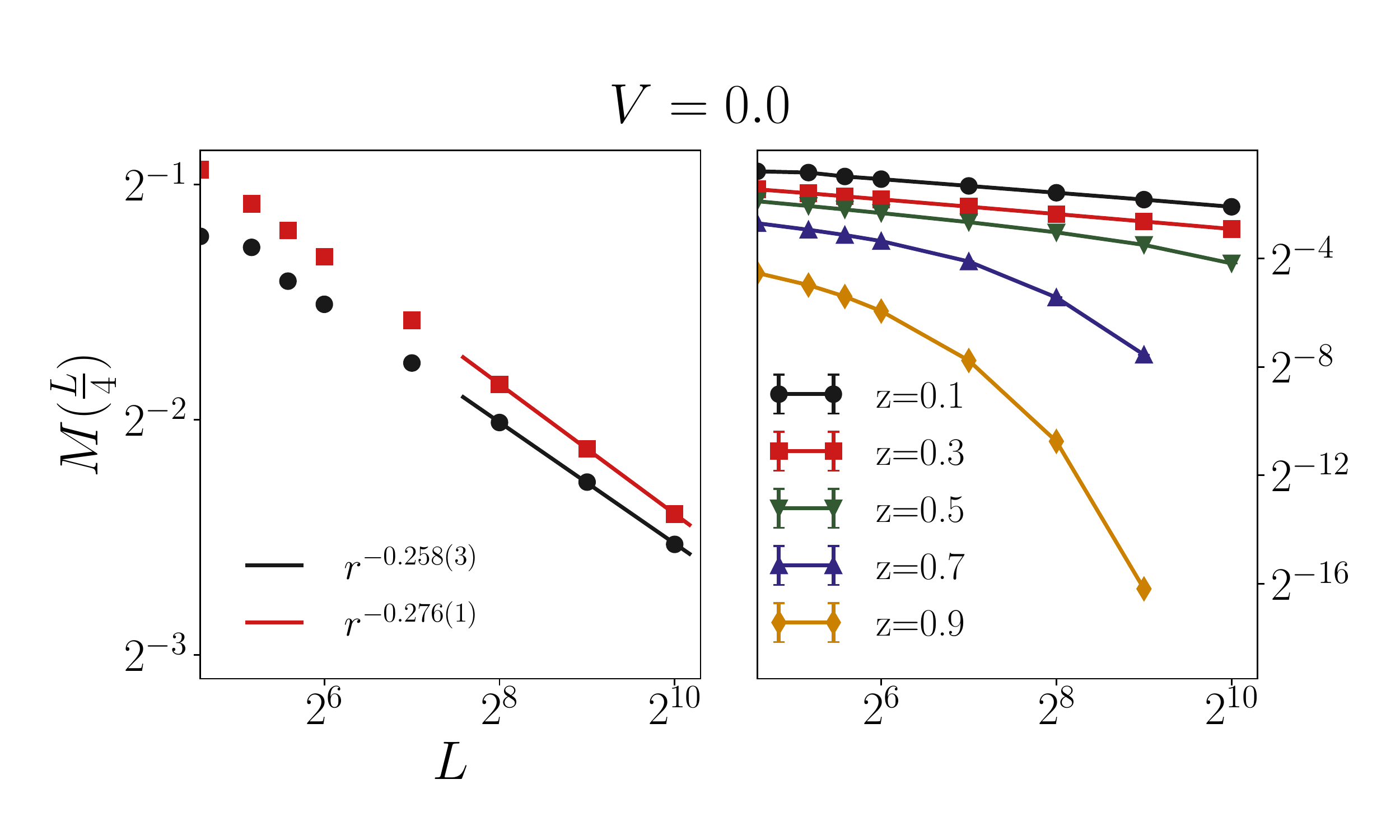}
    \caption{Left panel: The monomer-antimonomer correlation function (normalized to be unity at separation $\mathbf{r} = 0$) at separation $\mathbf{r}=(\frac{L}{4},0)$ for $V=0$ apparently shows power-law behaviour $\sim 1/L^{\eta_m(V=0,z)}$ for small nonzero $z$ in addition to $z=0$. Right panel: In contrast, the data at larger $z$ shows faster-than-power-law decay. Although this is qualitatively similar to the behaviour in the bilayer Coulomb phase for $V>0$, our theoretical analysis suggests that a very slow crossover from bilayer Coulomb to disordered large-$z$ behaviour is responsible for this apparent similarity. See Sec.~\ref{sec:ScalingpictureV>0} and  Sec.~\ref{sec:numerics} for a detailed discussion.}
    \label{fig:monomercorr_nonint}
\end{figure}

The picture at the strong coupling fixed point to which the system flow is therefore of two layers whose configurations lock together and have short-ranged correlations due to proliferation of interlayer dimers. The $z$ dependence of the length scale beyond which this description applies can be obtained by using an estimate for the bare value of $y_{-}$ and $y_v$ in conjunction with their RG eigenvalues. 

The argument is as follows: As noted earlier, $y_v$ is expected to have a bare value that is linear in $z$ since each interlayer dimer behaves as a double-vortex in $\theta_{+}$. In contrast, the entropic advantage represented by $y_{-}$ comes from the fact that two interlayer dimers on neighbouring links can be replaced by a pair of intralayer dimers on identical links in each layer. Thus, we expect the bare value of $y_{-}$ to scale as $z^2$. At RG scale $l$, these couplings thus scale as $y_{-} \sim z^2 \exp\left((2g^{*}-1)l/g^{*}\right)$ and $y_v \sim z \exp( (2-g^{*})l)$ respectively in the small $z$ limit. 

Thus, the $z=0$ power-law columnar order in each layer is expected to be disrupted for nonzero $z$ beyond a length scale $\xi_v \sim z^{-1/(2-g^{*})}$. This is the length-scale beyond which $\theta_{+}$ is disordered. On the other hand, the dimers in the two layers align with each other beyond a length scale $\xi_{-} \sim z^{-1/(1-1/2g^{*})}$. This is the length-scale beyond which $\theta_{-}$ is frozen to $0$ due to the relevance of the interlayer interaction corresponding to the fugacity $y_{-}$. For small $|V|$, $\xi_{-} >> \xi_v$ since $g^{*}$ is close to $g^{*}=1/2$. 
Thus, for very weak attractive interactions, Gaussian fluctuations of $\theta_{-}$, which are responsible for the dipolar correlations in the bilayer Coulomb phase, are not frozen out until one goes beyond the parametrically large length-scale
$\xi_{-}$. On the other hand, $\xi_{-} \ll \xi_v$ for $V$ in the vicinity of $V_{cb}$, since $g^{*}$ approaches $g^{*} = 2$
in this limit.

This strong-coupling description can be understood directly in a complementary large $z$ expansion as well, which confirms that the two regimes are continuously connected. For instance, the fact that configurations of the two layers lock together is not at all surprising at large $z$.
Indeed, it can be understood very simply in a large-$z$ strong-coupling expansion:  At $z=\infty$, there are no intralayer dimers in either layer, and all sites of both layers have interlayer dimers touching them. The leading ${\mathcal O}(1/z^2)$ corrections arise from configurations in which two dimers removed from a pair of nearest-neighbour vertical links, and the corresponding pair of intralayer links are occupied by a pair of intralayer dimers. Thus, the only terms that contribute at leading order in the $1/z$ expansion correspond to perfectly locked intralayer configurations, providing a simple picture for this limit.
\begin{figure}[!]
    \includegraphics[width=\columnwidth]{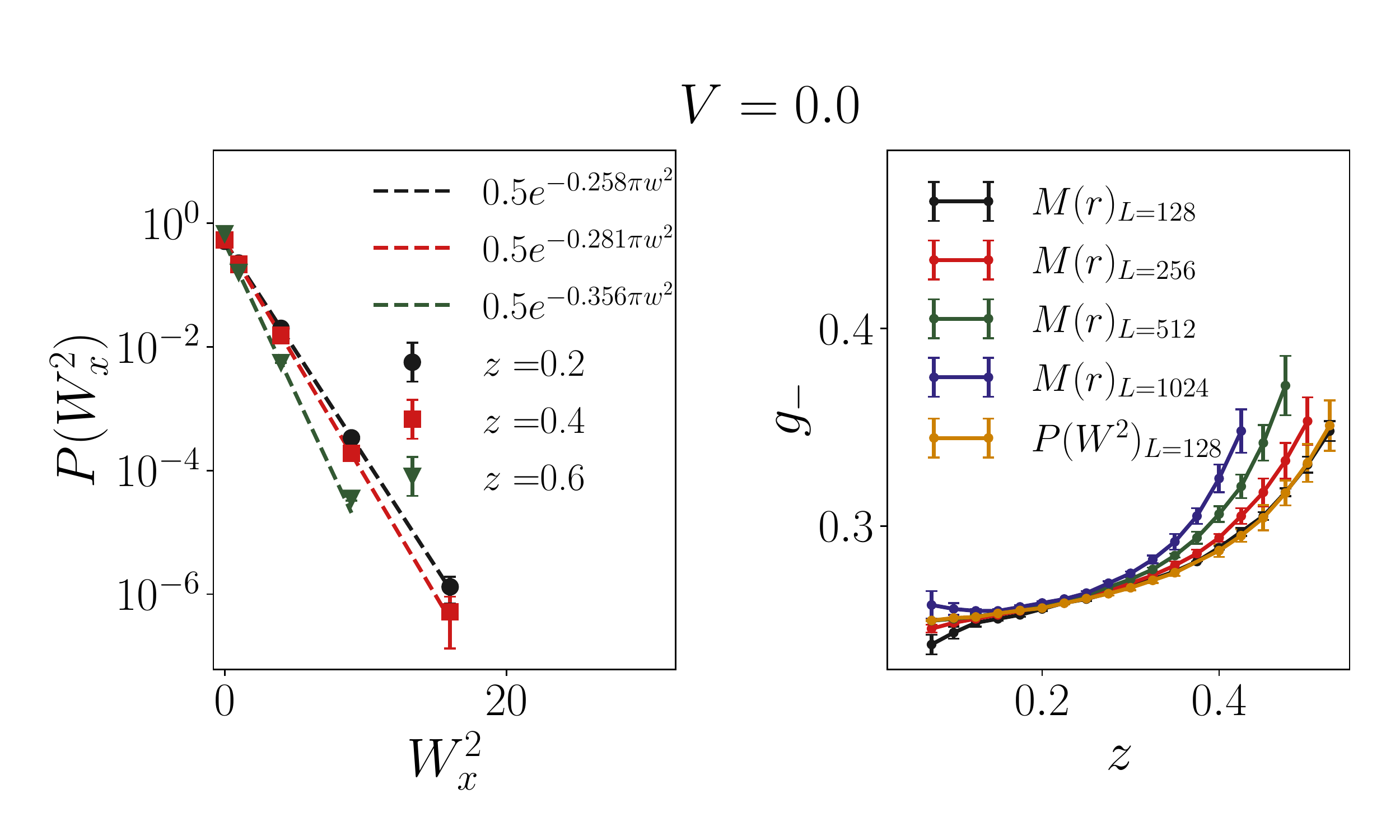}
    \caption{	 Comparison of $g_{-}^{*}$ extracted from two different analyses: (1) fitting monomer correlations to a power law, $M(r) = A r^{-g_{-}^{*}}$ and (2) fitting histograms of $W_x^2$ and $W_y^{2}$ to a common functional form $P(W^2) = C e^{- \pi g^{*}_{-} W^2}$ for $V=0.0$. The actual Gaussian fits are shown in the left panel. The monomer power law exponent has been extracted from fits to monomer-antimonomer correlations at separation $\mathbf{r}= (L/4,0)$ over a range of sizes up to $L= L_{max}$. The right panels display a comparison between the values of $g^{*}_{-}$ obtained in these two ways, for a range of choices of $L_{\max}$. The legends $M(r)_{L=L_{max}}$ in the right panels give the value of $L_{max}$ in each case. As is clear from the right panels, these estimates of $g^{*}_{-}$ are {\em not entirely consistent} with each other for nonzero $z$. This is consistent with our theoretical expectation that there is no stable bilayer Coulomb phase at nonzero $z$ for the noninteracting $V=0$ case. See Sec.~\ref{sec:ScalingpictureVleq0} and Sec.~\ref{sec:numerics} for further details.}
    \label{fig:w2gcomparison_nonint}
\end{figure}

\subsection{$V=0$: $z>0$ disordered phase}
\label{subsec:noninteracting}
Next, we consider the noninteracting bilayer. In this case, the decoupled $z=0$ system flows to the $g^{*} =1/2$ fixed
point, {\em i.e.} with $g_{-}^{*} = g_{+}^{*} = 1/4$. One might conclude that a nonzero bare $y_{-}$ of order ${\mathcal O}(z^2)$ is now marginal. However, a
nonzero $z$ also leads to a nonzero bare value of $g_{12}$ which is of order ${\mathcal O}(z^2)$. This implies that the bare value of $g_{-}$ receives a positive ${\mathcal O}(z^2)$ correction. This renders $y_{-}$ {\em relevant} at small nonzero $z$, with a positive RG eigenvalue that is ${\mathcal O}(z^2)$ in magnitude. Moreover, $y_v$ is again strongly relevant.

Thus the scaling picture in the noninteracting case is expected to be broadly the same as for the previous case with not-too-strong attractive interactions. The runaway flow of $y_v$ to strong coupling implies that $\theta_{+}$ is disordered on scales larger than a correlation length-scale which grows slowly as $\xi_v \sim z^{-2/3}$ for small $z$. However, the ${\mathcal O}(z^2)$ RG eigenvalue of $y_{-}$ implies the presence of a long crossover
in the behaviour of $\theta_{-}$, controlled by the large length-scale $\xi_{-} $. Using the flow equations and our estimate
$g_{-} = 1/4 + {\mathcal O}(z^2)$ for the bare value of $g_{-}$, we estimate this length-scale to grow very rapidly as
$\xi_{-} \sim z^{-1/\alpha z^2}$ for small $z$, where $\alpha$ is a positive constant.

For finite-size systems accessible to our numerics, this implies that it would be very hard to distinguish the behaviour of
the noninteracting system at small $z$ from the phenomenology of the stable bilayer Coulomb phase described in the previous section for bilayers with a repulsive interaction $V > 0$.

\section{$|V| > |V_{cb}|$ and $z$ small }
\label{sec:Columnarorder}
Next we consider stronger attractive interactions $|V|> |V_{cb}|$ and small $z$.  Our analysis splits naturally into two cases: $|V| \in (|V_{cb}|, |V_{c}|)$, for which the $z=0$ decoupled system flows to fixed points with $2 < g^{*} < 4$, and $|V| > |V_c|$, for which the $z=0$ system flows to fixed points with $g^{*} > 4$. The significance of the fixed point value $g^{*} =4 $ is simply the following: For $g^{*} > 4$, $y_{\lambda}$ (whose bare value is nonzero even at $z=0$) becomes relevant at the $z=0$ fixed point labeled by $g^{*}$. This signals the transition of each decoupled layer to a $z=0$ columnar ordered state for $|V| > |V_c|$, which has been studied at length in earlier work~\cite{alet_etal_pre2006,papanikolaou_luijten_fradkin_prb2007}.

\subsection{$|V| \in (|V_{cb}|, |V_{c}|)$; $z < z_{\rm AT}(V)$: Bilayer columnar order}
\label{subsec:Bilayercolumnarorder}
In this regime $2 < g^{*} < 4$, a nonzero $z$ again induces nonzero values of $y_{-}$, $y_v$ and $y_{+}$. As noted above, $y_\lambda$ remains irrelevant in this regime, and therefore not considered further in our discussion of this regime. However, $y_{-}$ remains strongly relevant and flows to strong coupling. On the other hand, $y_v$ is irrelevant in this regime, and expected to renormalize to zero, while $y_+$, which is now relevant, flows to strong coupling.

This might at first sight appear somewhat paradoxical, since $y_+$ cannot be nonzero in the absence of interlayer dimers, and a vanishing $y_{v}$ suggests the absence of interlayer dimers. However, the resolution of this apparent paradox is in fact quite clear: If $y_v$ renormalizes to $0$, it merely implies that interlayer dimers on opposite sublattices must be {\em bound on a short length scale into neutral complexes} which have no net vorticity (for instance, a pair of interlayer dimers on nearest neighbour links between the two layers). 
This is for instance the picture of the previously studied columnar-ordered states in mixture of dimers, hard-squares and holes~\cite{ramola_damle_dhar_prl2015}, or mixtures of holes and dimers, with attractive interactions between dimers\cite{alet_etal_pre2006,papanikolaou_luijten_fradkin_prb2007}. In those cases too, the hole density is nonzero, but the net vorticity at large length scales renormalizes to zero.\begin{figure}[!]
    \includegraphics[width=\columnwidth]{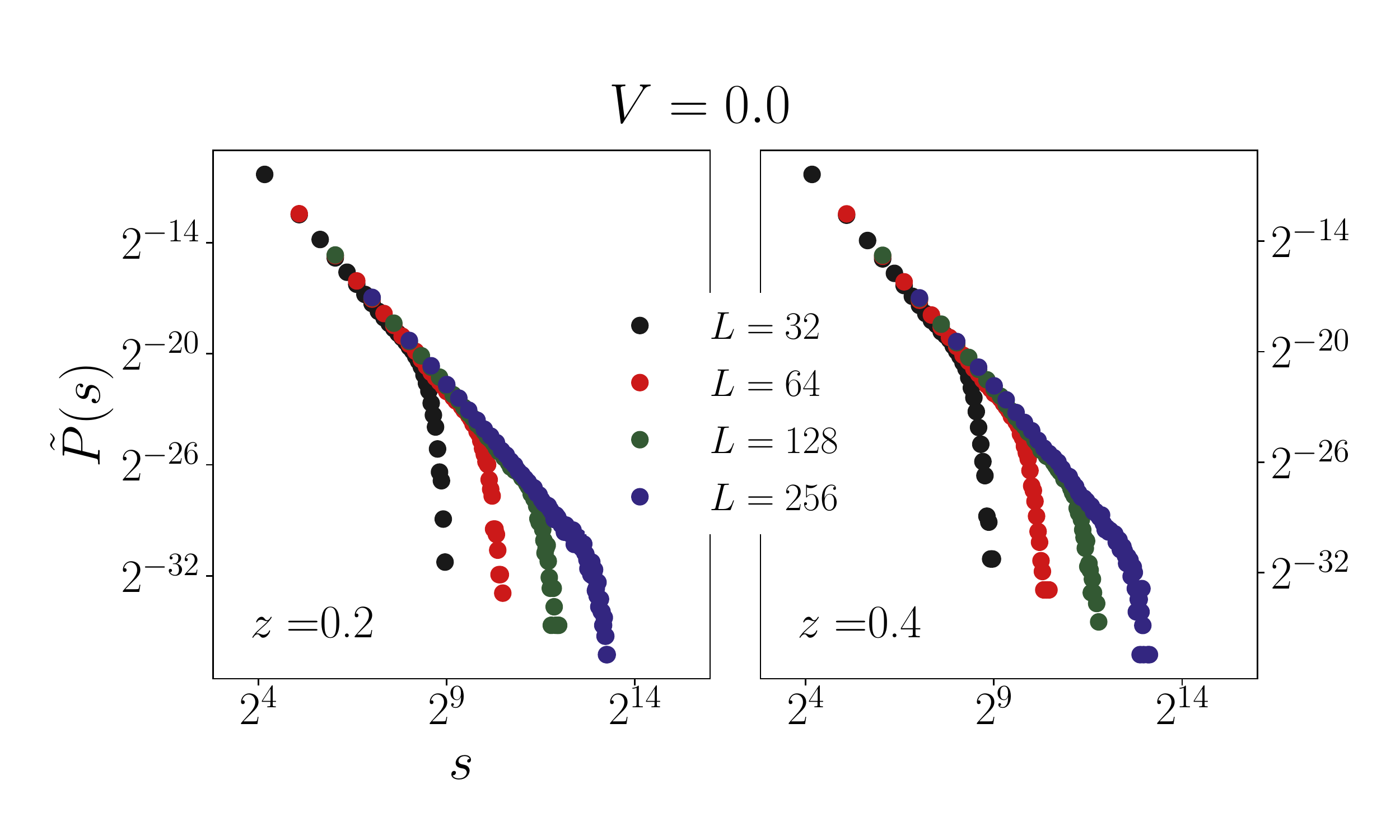}
    \caption{The probability distribution $\tilde{P}(s)$ of overlap loop lengths $s$ appears to have a power-law decay for small nonzero $z$ at $V=0$ for the range of sizes accessible to our numerical work, although our theoretical prediction is that there is no stable bilayer Coulomb phase in the non-interacting case. We ascribe this to a very slow crossover, predicted by our RG analysis for small $z$ at $V=0$, from bilayer Coulomb behaviour to disordered behaviour characteristic of the large-$z$ disordered phase. See Sec.~\ref{sec:ScalingpictureVleq0} and Sec.~\ref{sec:numerics} for a detailed discussion. }
    \label{fig:overlaploops_nonint}
\end{figure}

In this regime, the scaling picture for the $z>0$ bilayer is therefore as follows: The dimer configurations of the two layers are expected to be locked together due to the flow of $y_{-}$ to strong coupling. In effect, this implies that $h_1 - h_2 =0$, {\em i.e.} $h_1=h_2=h$ in this limit. As a consequence, the $\lambda_{+}$ term in Eq.~\ref{eq:coupledsinegordon} is minimized by $h=n/4$. Since the Coulomb-gas fugacity $y_{+}$ that corresponds to $\lambda_{+}$ also flows to strong coupling in this regime, this implies columnar order for the bilayer (with order parameters in the two layers locked to each other), since the strong-coupling theory now demands that $h_1=h_2=n/4$.

Thus, for nonzero $z$ in this range of atrractive $V$ (which corresponds to $2 <g^{*} < 4$ for individual layers in the decoupled limit), the two layers lock together to behave as a single layer that is columnar ordered in spite of the presence of a nonzero density of vertical dimers (which, in the simplest picture, come in nearest-neighbour pairs with no net vorticity). However, {\em at} $z=0$ in this regime, each decoupled layer remains in a critical state with power-law correlations of the columnar order parameter: $C_\psi(r) \sim 1/r^{1/g^{*}}$. 

Physically, the presence of bound pairs of interlayer dimers provides an entropic advantage to columnar ordering of the bilayer as a whole, and drives the system to a bilayer columnar state as soon as $z$ becomes nonzero. Naturally, in this range of $V$, we also expect that this columnar ordered state undergoes a transition to the large-$z$ disordered phase as $z$ is increased further beyond some critical value $z_{\rm AT}(V)$. Since the columnar order parameter must vanish both at $z=0$ and at $z=z_{\rm AT}(V)$ for fixed $V$ in this range of $V$, we see that this regime is characterized by an interesting nonmonotonic $z$ dependence of the columnar order parameter. 

In the next section, we argue that the long-wavelength properties of the system in the vicinity of this phase boundary $z_{\rm AT}(V)$ are described by an Ashkin-Teller critical line. 

\subsection{$|V| > |V_c|$; $z<z_{\rm AT}(V)$: Columnar order}
\label{subsec:singlelayercolumnarorder}

The final regime to consider for attractive interactions is $|V| > |V_c|$. In this regime, each decoupled layer is individually in the columnar-ordered state even at $z=0$; this ordering is driven by $y_{\lambda}$, which is relevant for $g^{*} > 4$ and flows to strong coupling. As a result, we {\em cannot} discuss the perturbative effect of a nonzero $z$ by an analysis in the vicinity of the fixed line labeled by $g^{*}$. 

In this regime, the appropriate analysis is in terms of the perturbative effect of interlayer dimers on each columnar-ordered layer. This may be understood as follows: Each vertical dimer corresponds to a monomer from the point of view of a single layer. Two such monomers can be accommodated at nearest-neighbour locations by removing a single dimer from a layer, which minimizes the disruption of the columnar order. However, since this is true in both layers, pairs of vertical dimers at nearest-neighbour locations are energetically favoured when there is a dimer each on the corresponding links of both layers. 

Thus, a nonzero fugacity $z$ for vertical dimers is expected to align the columnar ordering patterns that exist in both layers even at $z=0$. Thus, for small nonzero $z$, we again have a bilayer columnar-ordered phase in which both layers have columnar ordering patterns that line up. However, unlike in the columnar-ordered phase for $|V| < |V_c|$, the columnar order parameter of any one layer {\em does not} in this case go to zero as $z \to 0$. Instead, it goes to a nonzero constant, corresponding to the columnar order parameter of the square lattice dimer model at this value of $V$.

Although we have made a terminological distinction between the bilayer columnar ordered regime and the columnar ordered regime, we emphasize that the two regimes are continuously connected, and there is no sharp phase transition separating the two. Rather, the distinction is in terms of the $z$ dependence of the columnar order parameter at small nonzero $z$: In the bilayer columnar ordered regime, one expects a nonmonotonic dependence, since the columnar order parameter vanishes at both $z=0$ and at $z_{\rm AT}(V)$, peaking somewhere in the middle. Whereas, in the columnar ordered regime, the columnar order parameter is nonzero even at $z=0$.

\section{$z=z_{\rm AT}(V)$: Ashkin-Teller criticality }
\label{sec:Ashkin-Tellercriticality}
Next we argue that the transition line $z_{\rm AT}(V)$ that separates the small $z$ columnar-ordered phase and the disordered large-$z$ phase provides an unusual realization of an Ashkin-Teller (AT) critical line that terminates at $z=0$, $V=V_{cb}$ in the $(z,V)$ plane. This terminus corresponds to a fixed point value of $g^{*} =2$ for each decoupled layer at $z=0$. 
This is a different realization of Ashkin-Teller criticality from that found in the square lattice dimer model with attractive interactions and holes,\cite{alet_etal_pre2006,papanikolaou_luijten_fradkin_prb2007} or the corresponding critical line in a mixture of hard squares, dimers and holes.\cite{ramola_damle_dhar_prl2015}  In these cases, the AT line of transitions terminates in a Kosterlitz-Thouless transition of the fully-packed system corresponding to $g^{*}=4$. In the present case, the terminus is a decoupled system of two fully-packed layers, each described by a $g^{*}=2$ fixed point that does not correspond to a Kosterlitz-Thouless transition at full-packing, but instead lies within the power-law ordered critical phase of each fully-packed layer. 

To see how this comes about, we note that in this regime, {\em i.e.} with $g^{*} = 2 +2 \delta$ ($|\delta| \ll 1$), $y_{-}$ is strongly relevant and flows rapidly to strong coupling, while $y_{\lambda}$ is strongly irrelevant and flows rapidly to zero. Thus, beyond a relatively small crossover lengthscale $\xi_{-} \sim z^{-1/(1-1/2g^{*})} \sim 1/z$ (to leading order in $z$), $\theta_{-}$ is frozen to $0$, with the configurations in both layers locked together in terms of their coarse-grained properties. Moreover, the effective value of $y_{\lambda}$ rapidly renormalizes to zero, and its effects can therefore be neglected in our analysis of asymptotic behaviour.

Indeed, since $\theta_{-}$ is effectively frozen to $\theta_{-} = 0$ beyond the scale $\xi_{-}$, this asymptotic behaviour is controlled entirely by the fluctuations of $\theta_{+}$. These have a description that is controlled  by the competition between $y_{+}$ and $y_v$, both of which are nearly marginal when $|\delta| \ll 1$. This competition is responsible for the phase transition between the bilayer columnar ordered phase and the large-$z$ disordered phase, and our goal is to analyze this asymptotic behaviour in the vicinity of $z_{\rm AT}(V)$ {\em for small $z$ and $V$ close to $V_{cb}$}. The long-wavelength behaviour of both layers in this regime is therefore entirely determined by the theory in the $\theta_{+}$ sector, which is what we focus on in this discussion.

To this end, we write $g_{12} = 2\epsilon_{12}$ with $\epsilon_{12}$ being ${\mathcal O}(z^2)$ in the bare theory at small $z$, and focus on the flows in the $\theta_{+}$ sector of the theory. These flows in the $\theta_{+}$ sector decouple from the $\theta_{-}$ sector at large length scales due to the rapid renormalization of $y_{\lambda}$ to zero (since $y_{\lambda}$ is the only term that fugacity that couples the two sectors in our analysis). This simplifies the equations for the flows in the $\theta_{+}$ sector since we can set $y_{\lambda}$ to zero.

To analyze the flows in this $\theta_{+}$ sector, we write $g_{+} = (2+ 2\delta - 2\epsilon_{12})/2 \equiv 1- \tilde{\Delta}/2$
and $y_{v} = \epsilon_v$, where $\epsilon_v$ is ${\mathcal O}(z)$ in the bare theory for small $z$. Similarly, the bare value of $y_{+}$ scales to zero in the small $z$ limit, although it is not entirely clear how rapidly. Therefore we set $y_{+} = \epsilon_{+}$ to remind us that we are interested in a regime with a very small bare value for $y_{+}$.

Making these substitutions, setting the renormalized $y_{\lambda} = 0$, and expanding to second order in $\epsilon_{+}$, $\tilde{\Delta}$, and $\epsilon_v$, we obtain the coupled equations:
\begin{eqnarray}
\frac{d \tilde{\Delta}}{d l} & = & 16 \pi^2 \left ( \epsilon_{v}^2 - \epsilon_{+}^2 \right ) \nonumber \\
\frac{d \epsilon_v}{d l} &=& + \tilde{\Delta} \epsilon_v \nonumber \\
\frac{d \epsilon_+}{d l} &=& - \tilde{\Delta} \epsilon_+
\label{eq:ATrawform}
\end{eqnarray}
Defining 
\begin{eqnarray}
\tilde{\epsilon}_a &=& 4\pi(\epsilon_v - \epsilon_+) \nonumber \\
\tilde{\epsilon}_s &=& 4\pi (\epsilon_v+ \epsilon_+)
\end{eqnarray}
we obtain the system of equations
\begin{eqnarray}
\frac{d \tilde{\Delta}}{d l} & = & \tilde{\epsilon}_{s}\tilde{\epsilon}_{a} \nonumber \\
\frac{d \tilde{\epsilon}_s}{d l} &=&  \tilde{\Delta} \tilde{\epsilon}_a \nonumber \\
\frac{d \tilde{\epsilon}_a}{d l} &=& \tilde{\Delta} \tilde{\epsilon}_s \; .
\label{eq:ATfinalform}
\end{eqnarray}
These are readily recognized as being of exactly the form obtained by Kadanoff\cite{kadanoff_annalsofphysics1979} in his analysis of the Ashkin-Teller critical line within the renormalization group approach to multicritical behaviour in the vicinity of the Kosterlitz-Thouless point. As we have already emphasized, our analysis here finds a similar Ashkin-Teller fixed line, which, however, is not in the vicinity of the $g^{*}=4$ KT point of each individual layer. Instead, this line starts at the $g^{*} = 2$ point in the middle of the power-law columnar ordered phase of each layer.

Interesting consequences flow immediately from this proposed identification: For instance, the anomalous exponent $\eta$ for the columnar order parameter remains fixed at $\eta=1/4$ for all {\em nonzero} $z$ along this Ashkin-Teller line, although $\eta = 1/2$ precisely {\em at} $z=0$. To see that this is the case, we note that $\eta$ is expected to remain fixed along the line\cite{kadanoff_annalsofphysics1979}, and it therefore suffices to obtain the value of $\eta$ by considering nonzero but small $z$. For such $z$, $\theta_{-}$ is frozen to $\theta_{-}=0$ beyond the scale $\xi_{-}$. Therefore $\theta_1 = \theta_2 = \theta_{+}/2$, yielding $\eta =1/4$ for small but nonzero $z$ along the Ashkin-Teller line. However, it is important to note that $\eta = 1/2$ for the decoupled layers at $z=0$.

Additionally, the correlation length exponent $\nu$ for the columnar-disordered transition is expected to vary continuously along this phase boundary $z_{\rm AT}(V)$. As noted in previous computational studies of similar behaviour\cite{ramola_damle_dhar_prl2015,papanikolaou_luijten_fradkin_prb2007,alet_etal_pre2006}, this serves as a universal coordinate for the position of the system along this line. The anomalous exponent $\eta_2$ for the secondary {\em nematic} order parameter of each layer is expected to be determined entirely in terms of this universal coordinate by the Ashkin-Teller relation: $ \eta_2 = 1-1/(2\nu)$. 
The behaviour of the anomalous exponent $\eta_2$ in the $z \to 0$ limit also encodes the key difference between this realization of the Ashkin-Teller line, and previously studied Ashkin-Teller phase boundaries in single-layer systems.

To see all of this from Eq.~\ref{eq:ATfinalform}, we start by noting that these flows have three different fixed lines: i) $\tilde{\Delta} = \tilde{\epsilon}_s = 0$, ii)$\tilde{\Delta} = \tilde{\epsilon}_a = 0$, and iii) $\tilde{\epsilon}_s = 0, \tilde{\epsilon}_a = 0$. Of these, iii) represents the fixed line corresponding to the power-law ordered phase of the decoupled bilayer system at $z=0$, while i) is unphysical in our context since all bare fugacities are positive. However, ii) is physical, and represents a fixed line along which the vorticity $y_v$ of interlayer dimers is balanced by the fugacity $y_{+}$ that represents a coupling between the two layers, which is allowed by symmetry considerations for nonzero $z$. This fixed line is clearly the destination of flows starting from a critical point along the phase boundary $z_{\rm AT}(V)$.

In the vicinity of this fixed line, the flows have the structure shown in Fig.~\ref{fig:RGflows}. The relevant direction away from the fixed line corresponds to runaway flows that take the system either to the disordered fixed point describing the large-$z$ disordered phase (when $y_v$ dominates over $y_{+}$), or the ordered fixed point  that describes the columnar ordered phase (when $y_{+}$ dominates over $y_v$). The RG eigenvalue corresponding to this relevant direction is easily seen to be $\tilde{\epsilon}_s$, implying a continuously varying correlation length exponent $\nu = 1/\tilde{\epsilon}_s$. Since we expect $\tilde{\epsilon}_s \sim z$ as $z \to 0$, this implies that $\nu$ scales as
\begin{equation}
\nu \propto 1/z 
\end{equation} 
as the phase boundary $z_{\rm AT}(V)$ is crossed at successively smaller values of $z$ approaching $z=0$. This implies that $\eta_2 = 1 - 1/2\nu$ has the limit:
\begin{eqnarray}
\eta &= &1/4 \; \; {\rm for} \; \; z \neq 0 \; , \nonumber \\
\eta_2 &\to& 1 \; \; {\rm for} \; \; z \to 0 \; \; (z \neq 0) \; .
\end{eqnarray}

This encodes the key difference between our realization of the Ashkin-Teller line and other examples in the literature:\cite{alet_etal_pre2006,papanikolaou_luijten_fradkin_prb2007,ramola_damle_dhar_prl2015}
Unlike these other examples in which the behaviour of $\eta_2$ is nonsingular, here we have a singular limit: In the limit of vanishing but nonzero $z$, we have argued here that $\eta_2 \rightarrow 1$. However, the value of $\eta_2$ {\em at} $z=0$, {\em i.e.} in the problem with two decoupled layers, is given by
\begin{eqnarray}
\eta &=& 1/g^{*} = 1/2 \; \; {\rm for} \; \; z=0 \; ,\nonumber \\
\eta_2 &=& 4/g^{*} = 2 \; \; {\rm for} \; \; z=0 \; ,
\end{eqnarray}
since the $z=0$ terminus of $z_{\rm AT}(V)$ corresponds to $g^{*} = 2$.


\section{Monte Carlo Study}
\label{sec:numerics}
In the remainder of this article, we describe the results of our Monte Carlo study of the bilayer
dimer model in the $V$-$z$ plane, focusing specifically on tests that establish the broad features of the phase diagram for small $z$ and $V$, {\em i.e.} the predicted existence of a bilayer Coulomb phase for nonzero but not-too-large
$z$ and $V$, and the instability towards a large-$z$ disordered phase for not-too-large $V \leq 0$ as soon as $z$ becomes nonzero.

\subsection{MC Details and Observables}
\label{subsec:numerics_observables}

For our computational work, we use the dimer worm algorithm\cite{sandvik_moessner_prb2006,alet_etal_pre2006} to update Monte Carlo
configurations. This allows us efficient computational access to the equilibrium properties of bilayer square lattices with periodic boundary conditions and size up to
$L=1024$, {\em i.e.} with $1024 \times 1024 \times 2$ sites. 

The dimer number $n_{\mu,a}(\mathbf{r})$ is defined as the following: $n_{\mu,a}(\mathbf{r})=1$ if an intralayer dimer is present at site $\mathbf{r}$ in layer number $a$ in the direction $+\mathbf{e}_{\mu}$, otherwise $n_{\mu,a}(\mathbf{r})=0$. Here $\mu=x,y$ and $a=1,2$. Along with $n_{\mu,a}(\mathbf{r})$, we also consider the following
linear combinations: 
\begin{equation} n_{\mu,\pm}=n_{\mu,1}(\mathbf{r})\pm n_{\mu,2}(\mathbf{r}). \end{equation} 
We also track locations of interlayer dimers at site $\mathbf{r}$ via the variable $n_{z}(\mathbf{r})$ in a similar manner.

We probe equilibrium correlations via the connected intralayer dimer
correlation function 
\begin{equation}
	C_{\mu\mu}(\mathbf{r})=
\langle \left( n_{\mu,a}(\mathbf{r}) - \langle n_{\mu,a}(\mathbf{r}) \rangle \right)
\left( n_{\mu,a}(\mathbf{0}) - \langle n_{\mu,a}(\mathbf{0})) \rangle \right) \rangle
\label{eq:intralayerCdefn}
\end{equation}
This decays to zero as $\mathbf{r} \rightarrow \infty$. 
In the Coulomb phase of the usual square lattice dimer model, we expect the corresponding correlation function to have the form:
\begin{equation}
\label{eq:cxx}
C^{\rm single \; layer}_{xx}(\mathbf{r}) =  (-1)^{r_x+r_y} f_{d} (\mathbf{r}) + (-1)^{r_x} f_{\psi} (\mathbf{r}) \; ,
\end{equation}
where we expect the asymptotic behaviors:
\begin{eqnarray}
f_{d}(\mathbf{r}) & \sim & \frac{1}{r^{2}} 
\label{eq:asymptoticsofcomponents1}
\\
f_{\psi}(\mathbf{r}) & \sim & \frac{1}{r^{\eta}}
\label{eq:asymptoticsofcomponents2}
\end{eqnarray}
in the limit $r \to \infty$.

For a direct real-space test of our prediction that correlations in the bilayer Coulomb phase will be purely dipolar in their long-distance behaviour, we measure $C_{xx}$ defined in Eq.~\ref{eq:intralayerCdefn}
and perform a numerical decomposition aimed at separating the long-distance asymptotics of our data into two parts, corresponding to the decomposition of the usual Coulomb correlator displayed in Eq.~\ref{eq:cxx}. Having isolated these two pieces, we can compare the long-distance asymptotics of these individual pieces to the asymptotics expected from Eqs.~\ref{eq:cxx}, \ref{eq:asymptoticsofcomponents1}, and \ref{eq:asymptoticsofcomponents2}.
We have in fact explored two ways of separating the long-distance asymptotics of our data into a ``columnar part'' and a ``dipolar part'' to implement this test.
As we now detail, together these two analyses provide fairly conclusive evidence in favour of the unusual pattern of correlations predicted by our analysis of the bilayer Coulomb phase.

First, we make the linear combinations $C_d(\mathbf{r}_L)$ and $C_\psi(\mathbf{r}_L)$, where $\mathbf{r}_L$ lies on the $x$-axis and $|\mathbf{r}_L|$ scales linearly with the system size (in the results we display, $\mathbf{r}_L=(L/8,0)$): 
\begin{align}
 C_{d}(\mathbf{r}_L)=(-1)^{r_x}[C_{xx}(\mathbf{r}_L) - C_{xx}(\mathbf{r}_L+\mathbf{e_y})]
\label{eq:dipolar_linear_combo}
 \\
 C_{\psi}(\mathbf{r}_L)=(-1)^{r_x}[C_{xx}(\mathbf{r}_L) + C_{xx}(\mathbf{r}_L+\mathbf{e_y})]  \label{eq:columnar_linear_combo_1}
\end{align}
To see what to expect for the long-distance behaviour of these linear combinations, we assume that $C_{xx}$ has a decomposition of the form Eq.~\ref{eq:cxx} and expand the smooth functions $f_d$ and $f_\psi$ in a Taylor series to obtain the leading result:
 \begin{align}
  & C_d(\mathbf{r}_L) = 2 f_d(\mathbf{r}_L) + \ldots \label{eq:subleading1}
\\
  & C_\psi(\mathbf{r}_L) =  2 f_\psi(\mathbf{r}_L) +\ldots
  \label{eq:subleading2}
 \end{align}
 where the subleading contributions denoted by ellipses arise from {\em second (y)-derivatives} of the
 smooth functions $f_d$ and $f_{\psi}$ since we have chosen $\mathbf{r}_L$ to lie on the $x$-axis . More precisely, we see that the subleading contributions in Eqs.~\ref{eq:subleading1} and \ref{eq:subleading2} will fall off as $1/L^{p_d}$ and $1/L^{p_\psi}$ where $p_d = p_{\psi} = {\rm min} (4, \eta+2)$ if the asymptotic behaviour of {\em both} pieces $f_d$ and $f_\psi$ is of the respective power-law form displayed in Eqs.~\ref{eq:asymptoticsofcomponents1}~\ref{eq:asymptoticsofcomponents2}. On the other hand, if columnar correlations are not critical (which is what we expect in the bilayer Coulomb phase from RG considerations
for small $z$ in the repulsive regime),  {\em i.e.} if $f_\psi$ is short-ranged and falls off exponentially, then the long-distance behavior of $C_\psi$ will 
be dominated by the sub-leading term that scales as $1/r^{p_\psi}$ with $p_\psi = 4$.



An alternative approach to isolating the columnar part can also be used, and serves as a check on the approach outlined above. This alternate approach uses a slightly different linear combination:
 \begin{align}
 C'_\psi(\mathbf{r}_L) = &\:
  (-1)^{r_x} \Bigg[ \frac{9}{8} [C_{xx}(\mathbf{r}_L)
 + C_{xx}(\mathbf{r}_L+\mathbf{e_y})] \nonumber \\
&\: -
 \frac{1}{8}
 [C_{xx}(\mathbf{r}_L)
 + C_{xx}(\mathbf{r}_L+3\mathbf{e_y})] \Bigg] \; .
\label{eq:columnar_linear_combo_2}
\end{align}
Here, the coefficients are arranged to cancel off the subleading term arising from the second (y)-derivatives of $f_d$ and $f_\psi$. As a result, if $f_\psi$ is rapidly decaying, we expect $C'_\psi$ to scale as the {\em fourth} (y)-derivative of the dipolar piece $f_d$, and therefore fall off as $1/L^6$:
 This linear combination has the asymptotic behavior
\begin{align}
  & C'_\psi(\mathbf{r}_L) \sim 2 f_\psi(\mathbf{r}_L) + {\mathcal O}(|\mathbf{r}_L|^{-6})
\end{align}

Thus, if we find that $C_\psi(\mathbf{r}_L) $ falls off as $1/L^4$ and $C'_\psi(\mathbf{r}_L) $ falls off as $1/L^6$ at large $L$, and $C_d(\mathbf{r}_L)$ scales as $1/L^2$ at large $L$ for small $z \neq 0$, while $C_\psi$ and $C_{\psi}^{'}$ both scale as $1/L^{1/g^{*}}$ at $z=0$ for a range of $V>0$, we may take this as essentially conclusive evidence in favour of the predicted bilayer Coulomb phase with purely dipolar dimer correlations.


In reciprocal space, we measure the structure factors of the dimers
defined as the expectation value 
\begin{equation}
S_{\mu \nu,aa}(\mathbf{k}) \equiv\langle \hat{n}_{\mu,a}(-\mathbf{k}) 
\hat{n}_{\nu,a}(\mathbf{k}) \rangle
\end{equation}
where
$\hat{n}_{\mu,a}(\mathbf{k}) \equiv \frac{1}{L^2} \sum_\mathbf{r} n_{\mu,a}(\mathbf{r})
e^{i \mathbf{k} \cdot \mathbf{r}}$, $\mu \in \{x,y\}$ and $a \in \{+,-\}$.
In the bilayer Coulomb phase, we expect to see a pinch-point singularity in the vicinity of the dipolar vector ${\mathbf Q} = (\pi,\pi)$ for $S_{xx,--}$, whereas $S_{xx,++}$ is expected to be smooth and singularity-free in this vicinity.

We also measure the test monomer-antimonomer correlation function $M(\mathbf{r})$,
where $\mathbf{r}$ is separation between the lattice locations of a monomer and an antimonomer introduced into an otherwise fully-packed bilayer, with the monomer and the antimonomer (a site at which two dimers touch) located on the same (opposite) sublattice if they are in the same (opposite) layer (here, we are using a convention whereby two sites connected by an interlayer link both have the same sublattice index). This can be measured without any reweighting in worm algorithm simulations~\cite{Rakala_Damle_PRE2017,Rakala_Damle_Dhar_arXiv2018}, and therefore provides a convenient way of measuring vortex-antivortex correlations of the effective field theory. We use this procedure since the more well-known method, \cite{alet_etal_pre2006} which keeps track of monomer-monomer correlators during worm algorithm simulations, involves a reweighting (see for instance Sec. IVA of Ref. \onlinecite{alet_etal_pre2006}) which we wish to avoid.
Since both approaches measure the vortex-antivortex correlations in the effective field theory for $\theta_{-}$, we expect the long-distance behaviour obtained in both approaches to be the same. As noted in Sec.~\ref{sec:ScalingpictureV>0}, this 
vortex-antivortex correlator is expected to fall off as $1/r^{g^{*}_{-}}$ in the bilayer Coulomb phase, providing us a way of measuring the fixed-point stiffness constant $g^{*}_{-}$ directly.

We also measure the statistics of the winding numbers $W_x$ and $W_y$ of the height field $h_{-}(\mathbf r) = h_1(\mathbf r) - h_2(\mathbf r)$  corresponding to dimer configurations obtained in our Monte Carlo simulation. We define the mean square winding $\langle W^2 \rangle = \langle W^2_x + W^2_y \rangle/2$, where the windings $W_x$ and $W_y$ are given by the corresponding fluxes of the divergence-free field $B_{\mu,-} = B_{\mu, 1} - B_{\mu, 2}$ in the $\hat{x}$ and $\hat{y}$ directions.  As we have seen in our discussion of the bilayer Coulomb phase in Sec.~\ref{subsec:BilayerCoulombphase}, a nonzero value for this mean-square winding 
in the thermodynamic limit is indicative of a Coulomb phase. As noted there, we expect this Coulomb phase to give way to a large-$z$ disordered phase when $g_{-}^{*}(z)$ increases beyond the critical value $g_{\rm inv. KT}^{*} \equiv 1/4$, at which $y_{-}$ becomes relevant and drives the system to the disordered large-$z$ phase.
As noted earlier, $\langle W^2 \rangle$ can be computed within the fixed point description of the bilayer Coulomb phase to yield a prediction
\begin{eqnarray}
\langle W^2 \rangle &=& {\mathcal J}(g^{*}_-) \; ,
\label{eq:W2fromJ}
\end{eqnarray}
with ${\mathcal J}(g^{*}_{-})$ given by Eq.~\ref{eq:Jdefn}.
As we have already emphasized in Sec.~\ref{subsec:BilayerCoulombphase}, $\langle W^2 \rangle$ thus provides a second convenient way to measure the fixed point stiffness constant $g_{-}^{*}$. In particular, the transition to the large-$z$ disordered phase is signalled by $\langle W^2 \rangle$ decreasing to the critical value of
$\langle W^2 \rangle_{\rm inv. KT} = {\mathcal J}(1/4)$.

In addition, we study another geometric quantity: As already reviewed in Sec.~\ref{sec:ScalingpictureV>0}, if one superimposes the dimer configuration of one layer on to the corresponding configuration of the second layer (leaving out interlayer dimers), this defines a configuration of non-intersecting fully-packed loops. In our Monte Carlo simulations, we keep track of the statistics of these loops. In fact, since the overlap loops can also be classified according to their winding number, we separately study the statistics of overlap loops of a given winding number. Since our estimate of the distribution of loop lengths $s$ in the zero winding sector is statistically the most reliable (since most loops do not wind around the sample), we focus in our numerical work on this sector.
In other words, we test whether our measured histograms of lengths of non-winding loops for samples of various size $L$ exhibit data-collapse when scaled as predicted by the finite-size scaling ansatz discussed in Sec.~\ref{sec:ScalingpictureV>0}:
\begin{eqnarray}
\tilde{P}(s,L) &=& \frac{{\mathcal C}}{L^{D_f\tau}} \Phi\left(\frac{s}{L^{D_f}} \right )
\label{eq:scalingformforoverlaps}
\end{eqnarray}
where $\tau = 7/3$, $D_f = 3/2$, and $\Phi(x) \sim x^{-\tau}$ for $x \ll 1$. 

Parenthetically, we note that part of our motivation for this analysis comes from the fact that these scaling ideas do not appear to have been subjected to any previous numerical tests in the Coulomb phase of a two-dimensional dimer or spin model.
Since this appears to be the case, we have tested this for the $z=0$ decoupled layers for a variety of values of $V$. The results are shown in Fig.~\ref{fig:scalingformforz_zero_overlaps}. As is clear from this figure, we find that this scaling form provides an excellent account of the data for the noninteracting case, as well as in the presence of attractive or repulsive interactions that place the system within the Coulomb phase of each layer. Below, in our discussion of our numerical results for $z>0$, we will return to the statistics of these overlap loops and discuss their behaviour again.

\subsection{Results}
\label{subsec:numerics_results}

We begin our discussion of the numerical results by first noting that the density of interlayer dimers $n_z$ remains oblivious to the complexities of the phase diagram. Indeed,
as the interlayer dimer fugacity is increased, the density $n_z$ of interlayer dimers increases smoothly from zero. From the data displayed 
in Fig. \ref{fig:interlayer_dimer_corr}, we see that this evolution
of $n_z$ is quite featureless for the three values of interactions shown, increasing monotonically with the 
fugacity $z$ as expected.




In sharp contrast to this, our results for $\langle W^2 \rangle$ provide a bird's eye view of the phase diagram for repulsive, attractive and non-interacting bilayers: In Fig. \ref{fig:w2vsL}, we display the $L$ and $z$ dependence of the mean square winding $\langle W^2 \rangle = (\langle W^2_x \rangle+\langle W^2_y) \rangle/2$,
where the windings $W_x$ and $W_y$ are given by the corresponding flux of the divergence free field $B_{\mu,-} = B_{\mu, 1} - B_{\mu, 2}$.
For the repulsive $V > 0$ case (left panel), we clearly
see that $\langle W^2 \rangle $ extrapolates to a nonzero thermodynamic limit for small $z$. However, as $z$ is increased beyond a threshold value, $\langle W^2 \rangle $ vanishes in the thermodynamic limit. The separatrix that signals the transition is seen to match quite closely with the expected value of
${\mathcal J}(1/4) = 0.636\dots$.

On the other hand for the attractive
case, we see very clearly that any nonzero $z$ leads to a vanishing $\langle W^2\rangle $ in the thermodynamic limit. 
Finally, in the noninteracting case, the data seems to
indicate the presence of a slow crossover to disordered behaviour, signalled by a $\langle W^2 \rangle$ that is always below ${\mathcal J}(1/4) = 0.636\dots$, but does not readily extrapolate to zero at accessible sizes. This is consistent with our theoretical prediction in Sec.~\ref{subsec:noninteracting}
 of an extremely slow crossover to disordered behaviour, expected for the bilayer system at arbitrarily small nonzero $z$.

\subsubsection{Repulsive $V > 0$}

We now present numerical results that provide compelling evidence for a bilayer Coulomb phase at small nonzero $z$ and not-too-large $V>0$. 
The structure factor $S_{\mu \nu, --}$ of $n_{\mu,-}$
in this regime is shown in Fig. \ref{fig:struc_fac_repulsive}. 
The left panel of Fig.~\ref{fig:struc_fac_repulsive} a) shows the characteristic
bow-tie-like structure arising from the dipolar pinch-point singularity of the structure factor in Coulomb systems\cite{Henley_review} in the vicinity of $\mathbf{Q}$, i.e. $\mathbf{k}=\mathbf{Q} + \mathbf{q}$
for small $\mathbf{q}$.
This pinch-point singularity is explored further
in the right panel of Fig.  \ref{fig:struc_fac_repulsive} a) along the indicated path in the Brillouin zone. Note that the value of the structure factor {\em at}
${\mathbf Q}$ is identical to $\langle W^2 \rangle$, as is evident from the definitions of both quantities. In the vicinity of ${\mathbf Q}$, the $\mathbf{q}$-dependence of this structure factor is seen in  Fig.  \ref{fig:struc_fac_repulsive} b)to be fit well by
a lattice-discretized finite-size version (see Appendix~\ref{app:struc_fac}) of the asymptotic prediction for the pinch-point singularity obtained in Sec.~\ref{sec:ScalingpictureV>0} for small nonzero $q$:
\begin{equation}
\frac{1}{2 \pi g^*_-} \left(\delta_{\mu\nu} - \frac{q_\mu q_\nu}{q^2}\right)
\end{equation} 
obtained
from the fixed-point effective action (Eq.~\ref{eq:fixedpointdefinition1}) 
\begin{equation}
S_{-} = \pi g^*_- \int |\nabla h_-|^2 
\end{equation}
Also shown in Fig. \ref{fig:struc_fac_repulsive} c) is the corresponding data for structure factor of $n_{\mu,+}$. As is clear from the right panel, this is fit well by the form derived in Appendix~\ref{app:struc_fac} from a simple phenomenology for the short-range correlations of $n_{\mu +}$ at nonzero $z$ in this regime.

As already noted in Sec.~\ref{sec:ScalingpictureV>0}, this implies unusual singular structure in the vicinity of $\mathbf{Q}$ in the interlayer and intralayer structure factors $S_{\mu \nu,11}(\mathbf{k}) = \langle \hat{n}_{\mu,1}(-\mathbf{k}) \hat{n}_{\nu,1}(\mathbf{k}) \rangle$ and $S_{\mu \nu,12}(\mathbf{k}) = \langle \hat{n}_{\mu,1}(-\mathbf{k}) \hat{n}_{\nu,2}(\mathbf{k}) \rangle $ in the vicinity of the pinch-point at $\mathbf{Q}$. As noted there,
since $g^{*}_{-} \to g^{*}/2$ as $z \to 0$,
the strength of the pinchpoint singularity of $S_{\mu \nu,11}(\mathbf{k})$ in the limit of small but nonzero $z$  tends to a value that is exactly half of the corresponding $z=0$ result for decoupled layers. On the other hand, the pinch-point singularity in $S_{\mu \nu,12}(\mathbf{k})$ has the same magnitude in this limit as the corresponding singularity of $S_{\mu \nu,11}(\mathbf{k})$, but is opposite in sign. Data for this is shown in Fig. \ref{fig:pnchpntintra_rep}, and we see that these predictions are borne out by the data.

We have also implemented the strategy outlined in the previous section to test for the purely dipolar nature of intralayer dimer correlations in the bilayer Coulomb phase, and found that our data conforms to these predictions. As is clear from
Fig. \ref{fig:psicorr_rep}a), b), the dipolar and columnar linear combinations
$C_d(\mathbf{r}_L)$ and $C_{\psi}^{'}(\mathbf{r}_L)$ (see Eqs.~\ref{eq:dipolar_linear_combo}, \ref{eq:columnar_linear_combo_1} \ref{eq:columnar_linear_combo_2}) indeed follow the expected power-law forms $1/L^2$ and
$1/L^{6}$ respectively at nonzero $z$, while $C^{'}_{\psi}(\mathbf{r}_L)$ at $z=0$ has a power-law decay with exponent $\eta = 1/g^{*}$. In Fig. \ref{fig:psicorr_rep} c), d), we also see that $C_{\psi}(\mathbf{r}_L)$ falls off as expected, with a slower decay $1/L^4$, whenever $C^{'}_{\psi}(\mathbf{r}_L)$ falls off as $1/L^6$. All of this provides compelling evidence for the unusual nature of dimer correlations in the bilayer Coulomb phase.

Turning to the monomer correlation function, we see in Fig. \ref{fig:monomercorr_rep} that the monomer-antimonomer correlations for small $z$ have a clear power-law behavior with a floating exponent, consistent with our predictions for the bilayer Coulomb phase. In contrast, they fall off much more rapidly at large $z$, as expected in the large-$z$ disordered phase. 
A curious feature of the power-law exponent for these monomer correlations in the bilayer Coulomb phase is the fact that this exponent $\eta_m$ is predicted to have a singular $z \rightarrow 0$ limit. To see this, note that $\eta_m = g^{*}_{-}(V,z)$ for $z>0$ in the bilayer Coulomb phase, while $\eta_m = g^{*}(V)$ at $z=0$. Since $g_{12} \sim {\mathcal O}(z^2)$ in the $z \to 0$ limit, we expect $g^{*}_{-} \to g^{*}/2$ as $z \to 0$, implying that $\lim_{z \to 0} \eta_m(V,z) = \eta_m(V)/2$. As is clear from the comparison shown in Fig. \ref{fig:monomerLby4fug0} of the best-fit values of $\eta_m$ for $z=0$ and $z=0.1$ over a range of $V>0$, our data is entirely consistent with this expectation.

The value of $g^{*}_{-}$ extracted from such fits to the monomer correlation function can be directly compared with fits of the distribution of winding numbers to a Gaussian form, as in the summand in Eq.~\ref{eq:windingpartitionsum}. This is shown in Fig. \ref{fig:w2gcomparison_rep}. The values of $g_-^{*}$ from the monomer correlations
and the winding data are seen to agree with each other rather well
for a range of not-too-large $z$ for nonzero $V>0$. Thus, all of our computational results in this regime have a quantitatively consistent and natural explanation in terms of the fixed point action (Eq.~\ref{eq:fixedpointdefinition2}) that governs the long-wavelength behaviour of the bilayer Coulomb phase. This conclusively establishes the central claim made earlier, regarding the presence of a bilayer Coulomb phase in this part of the $(z,V)$ plane.

In Fig. \ref{fig:repulsivezoom}, we display the $L$ dependence of $\langle W^2 \rangle$ near the transition out of bilayer Coulomb phase.
We see that the lowest nonzero value to which $\langle W^2 \rangle$ extrapolates in the thermodynamic limit is rather close to ${\mathcal J}(1/4)$, which is the expected
value of $\langle W^2 \rangle$ at the inverted Kosterlitz-Thouless transition separating the bilayer Coulomb phase from the large-$z$ disordered phase.
This also provides compelling evidence in favour of our scaling theory for this
transition.

Finally, we turn to the distribution of the lengths $s$ of overlap loops defined earlier.
As noted in the previous section, this distribution is expected to have the same tail at large loop sizes as the distribution of contour lines of a scalar Gaussian free field that represents the height fluctuations of a random surface. From the work of
Henley and Kondev~\cite{Henley_Kondev}, this is expected to have a power-law form, with power-law exponent $\tau =7/3$. As we have already noted, this prediction, along with the value of $D_f=3/2$ for the corresponding fractal exponent, can be tested in a convenient way by asking if our data for the histogram of the lengths of these ovlerlap loops collapses onto the scaling ansatz displayed in Eq.~\ref{eq:scalingformforoverlaps}.
In Fig.~\ref{fig:scalingformforoverlaps}, we see that this form indeed provides a very good description of our data for non-winding loops. The properties of winding loops also deserve a more detailed study, which we defer to follow-up work.

\subsubsection{Attractive $V < 0$}

Our numerical results for not-too-large attractive interactions $V<0$ provide a clear contrast to these earlier results on the repulsive side. Since the quantities being studied and our methods of analysis remain the same, we now summarize these results in brief: First, from a study of the dipolar and the columnar components ($C_d(\mathbf{r}_L)$ and $C_\psi(\mathbf{r}_L)$) of the dimer correlations, we see immediately that any nonzero $z$ leads immediately to a dipolar component that decays faster than $1/L^2$, and has a downward curvature on the log-log plot, consistent with our prediction that any nonzero $z$ leads immediately to the destruction of the $z=0$ Coulomb phase. Likewise, any nonzero $z$ also leads to a similar faster-than-power-law decay for the columnar part. This is displayed in  Fig. \ref{fig:psicorr_att}. Note that the length-scale beyond which this destruction of Coulomb phase power-laws is visible in Fig. \ref{fig:psicorr_att} can be correlated with the sample-size beyond which winding fluctuations are visibly suppressed in Fig. \ref{fig:w2vsL}. 

The destruction of Coulomb
correlations at nonzero $z$ is also reflected in monomer correlations and
overlap loop size histograms shown in 
Fig. \ref{fig:monomercorr_att} and \ref{fig:overlaploops_att}, which show clear faster-than-power-law decays.


\subsubsection{Non-interacting $V = 0$}


We finally come to the case of zero interactions which is
hardest to interpret numerically. We believe this is related to the presence of a very slow crossover at small nonzero $z$, from intermediate-scale physics that looks Coulomb-like, to asymptotically-large length-scale physics characteristic of the disordered large-$z$ phase. As noted in Sec.~\ref{sec:ScalingpictureVleq0}, the crossover length-scale $\xi_{-}$ corresponding to this is parametrically large at
small $z$, with $z$ dependence given as: $\log(\xi_{-}) \sim (\log(1/z))/z^2$ in the limit $z \ll 1$. Indeed, we note parenthetically that our detailed renormalization group analysis presented in previous sections, and our detailed study of the repulsive and attractive cases, were both motivated by the conundrum presented by our original data on non-interacting bilayers, to which this section is devoted.

Some indications of the difficulties involved are easily gleaned from our results for winding fluctuations, shown in Fig. \ref{fig:w2vsL} earlier. From the non-interacting panel of this figure, we see that although $\langle W^2 \rangle$ does not get appreciably suppressed for small nonzero $z$ over the range of sizes $L$ available to us, its {\em value} is always lower than ${\mathcal J}(1/4) \equiv 0.636\dots$. On the other hand, our RG analysis implies that the smallest possible value for $\langle W^2 \rangle$ is ${\mathcal J}(1/4)$, since this value, corresponding to $g^{*}_{-} = 1/4$ is characteristic of the inverted Kosterlitz-Thouless transition point separating the bilayer Coulomb phase from the large-$z$ disordered phase. This is seen more clearly in the non-interacting panel of Fig. \ref{fig:repulsivezoom} as well. Thus, although the measured winding fluctuations over the range of sizes accessible to numerics ``look Coulomb-like'', our RG analysis suggests that no consistent Coulomb description of the full data set would be possible.

This effect is also visible in the
dipolar correlations shown in Fig. \ref{fig:psicorr_nonint}.
The disappearance of power-law behaviors is eventually seen
for those fugacities ($z > 0.6$) 
for which the winding fluctuations have
been suppressed enough at the finite sizes studied.
Thus, in the columnar correlations, we see initial trends quite
similar to the repulsive case for small fugacities. Again, this is consistent with our RG prediction of a long crossover.

The difficulty in data interpretation 
at finite sizes
is also reflected in the monomer correlations
as shown in Fig. \ref{fig:monomercorr_nonint}. 
For the smaller values of $z$, the data
can be fit to a power-law form over the range of sizes studied. 
However, when we extract $g_-^{*}$ from this fit, 
and compare this estimate of $g^{*}_{-}$ to the value of $g_-^{*}$ extracted from winding data as shown in
the right panel of Fig. \ref{fig:w2gcomparison_nonint}, 
we find that
the agreement is strongly system-size dependent, 
with the discrepancy increasing if we use data from larger system sizes. This should be contrasted with the system-size independent and consistent values of $g^{*}_{-}$ obtained via a similar procedure in the repulsive case (Fig. \ref{fig:w2gcomparison_rep}).
Finally, the measured histograms of the overlap loop sizes in our finite-size systems are shown in Fig. \ref{fig:overlaploops_nonint}. We see that these too can be fit to power-law forms even at nonzero $z$ as large as $z=0.4$.

This confusing-at-first-glance state of affairs underscores the importance of the systematic RG analysis presented in previous sections, as well as our results for bilayers with a nonzero value for the interaction $V$. With the perspective provided by these additional inputs, we see that all these results in the non-interacting case can be explained in terms of a long crossover from bilayer Coulomb behaviour at intermediate length scales to behaviour characteristic of a large-$z$ disordered phase in the asymptotic long-distance limit, which, however, is not accessible to us. 



\section{Discussion}
\label{sec:Discussion}
Our work has led us to identify a bilayer Coulomb phase of dimers, with purely dipolar correlations between dimers. The dimer correlation functions in this phase are distinguished from those of the usual Coulomb phase of two-dimensional bipartite dimer models by the absence of a second power-law piece, with a floating exponent that depends on details such as the nature and strength of interactions between the dimers. This expands our understanding of the possibilities for correlated liquid states of strongly interacting systems in two dimensions. Several natural and interesting questions arise immediately from our work. Some of these provide promising avenues for follow-up work, and we close our discussion by highlighting these below. Additionally, it is instructive to contrast our results with those of Wilkins and Powell~\cite{wilkins_powell2020}; as we see below, this helps clarify exactly what feature of our our system leads specifically to the existence of this new phase with purely dipolar dimer correlations.

\subsection{Aside: Interacting square bilayer {\em without} interlayer dimers}
\label{Aside}
In very recent and interesting work that appeared as our manuscript was in preparation, Wilkins and Powell~\cite{wilkins_powell2020} consider (among other things) a bilayer square lattice with intralayer interaction $J$ (entirely equivalent to our interaction $V$), and interlayer interaction $K <0$ which assigns a lower energy to dimers occuring simultaneously on corresponding links of the two layers, thereby favouring identical dimer configurations in the two layers. Interlayer dimers are notably absent in the system they study.

It is instructive to examine their system from the coarse-grained effective field theory perspective developed here. Within this approach, nonzero values of the interaction $K$ are again expected to give rise to nonzero values for the couplings $\lambda_{-}$ and $\lambda_{+}$; indeed, the former directly captures the energetic preferences resulting from a nonzero $K$, and the latter must be included since it is allowed by the symmetries of the coupled system at nonzero $K$ (exactly as in our case). As in our case, the coupling $\lambda$ is however expected to be nonzero even when $K$ is zero. 

The crucial difference between our system and the one studied by Wilkins and Powell~\cite{wilkins_powell2020} is that the vortex fugacity $y_v$ must be set to zero for the bilayer studied by them, in order to represent the fact that interlayer dimers are disallowed in their study.
This crucial difference completely changes the long-wavelength physics. Since $y_v$ is strictly zero, the RG flows are those of the vortex-free theory. 

For small $K < 0$ and not-too-large repulsive $J \equiv V > 0$ in their case, $\lambda$, $\lambda_+$ and $\lambda_-$ are all irrelevant along the fixed-line that describes the power-law columnar ordered phase of the decoupled layers at $K=0$. In this regime, we thus expect that their coupled bilayer with a small $K < 0$ will be in a Coulomb phase whose long-wavelength physics is described by {\em two} independently fluctuating scalar fields $h_{+}$ and $h_{-}$ with stiffnesses $g_{+}^*$ and $g_{-}^*$. A quick calculation then predicts that dimer correlations in each layer will have both a dipolar piece, and a second piece that represents power-law columnar order with a floating exponent that depends on both $g^{*}_+$ and $g^{*}_-$. 

This is in sharp contrast to the physics of the bilayer Coulomb phase displayed by our bilayer system in the corresponding regime. In our case, although $\lambda$, $\lambda_+$ and $\lambda_-$ are all irrelevant for not-too-large repulsive $V$ and small $z$, $y_v$ is strongly relevant and flows off to strong coupling. This implies that correlations of $h_{+}$ are short-ranged and decay exponentially to zero. As a result, the dimer correlation function is purely dipolar in nature. As mentioned already in the Introduction and detailed in Sec.~\ref{subsec:BilayerCoulombphase}, this is because the two-point correlation function at the columnar ordering wavevector ${\mathbf K}$ is a {\em product} of a power-law factor arising from correlations of $\exp(2\pi i h_{-})$ and an exponentially-decaying factor arising from the short-ranged correlations of $\exp(2 \pi i h_{+})$. Whereas the two-point correlation function at the dipolar pinch point wavevector ${\mathbf Q}$ is a {\em sum} of a short-ranged correlated piece arising from correlations of $\nabla h_{+}$ and a dipolar power-law term arising from the correlations of $\nabla h_{-}$.

For small $K<0$ and not-too-strong attractive interactions $J \equiv V < 0$ in the bilayer studied by Wilkins and Powell~\cite{wilkins_powell2020}, $\lambda_{+}$ and $\lambda$ remain irrelevant, but $\lambda_-$ is now relevant. This leads to their `synchronized' phase in which the dimer configurations in the two layers lock together. In this synchronized phase, our coarse-grained approach implies that dimer correlations again have two pieces, a dipolar piece and a power-law columnar ordered piece, with the floating exponent of the latter piece being controlled entirely by the fixed-point value $g^{*}_{+}$ of the stiffness of the fluctuating Gaussian field $h_{+}$ (since $h_{-}$ is frozen to $h^{-}=0$ at such `synchronized' fixed points). 

In contrast, in the corresponding regime of small $z$ and not-too-strong attractive interactions $V < 0$ in our case, $y_v$ is also relevant in addition to $\lambda_{-}$ being relevant. As a result, {\em both} flow to strong-coupling, leading to a disordered phase that is {\em continuously connected to the large-$z$ regime of our bilayer system}.

Thus, in the bilayer studied by Wilkins and Powell~\cite{wilkins_powell2020}, their Coulomb and synchronized phases are distinguished by the central charge~\cite{Cardybook} of the corresponding long-wavelength field theory: The long-wavelength physics of their Coulomb phase is expected to be described by two independently fluctuating Gaussian fields $h_{+}$ and $h_{-}$, each with their own critical correlations, while the corresponding physics in their synchronized phase will be described by a single fluctuating Gaussian field $h_{+}$ with critical correlations. In our case, the distinction is quite different: It is the distinction between a bilayer Coulomb phase with {\em purely dipolar correlations} on the one hand, and a disordered phase continuously connected to the large-$z$ regime on the other hand.

\subsection{Outlook}
\label{Outlook}
Our work suggests several potentially fruitful avenues for future work. We close by describing some of these.
First, our RG analysis suggests that a similar bilayer honeycomb lattice system may host interesting physics; this provides motivation for follow-up computational work aimed at elucidating the phase diagram of such a bilayer system. Second,  the present work suggests it would be interesting to study bilayer variants of a system of hard-squares and rods studied earlier~\cite{ramola_damle_dhar_prl2015}.  Another natural line of thought involves the identification of quantum dimer models whose ground state wavefunctions map on to such classical bilayer systems with purely dipolar dimer correlations. Another natural question has to do with the physics of trilayers as well as systems made up of four layers. We hope our detailed analysis of this simplest bilayer realization of the purely dipolar Coulomb liquid phase of two-dimensional dimer models motivates follow-up studies aimed at resolving some of these questions.

\begin{acknowledgments}
We acknowledge stimulating discussions with S. Bhattacharjee, S. Biswas, R. Kaul, S. Kundu, G. Murthy, R. Sensarma, G. Sreejith, and V. Tripathi.
The numerical results were obtained using the computational resources of
XSEDE (DMR-150037) and the Arts and Sciences Computational Cluster of the Univ. of Kentucky, as well as the computational facilities of the Department of Physics, Indian Institute of Technology (IIT) Bombay. N.D. was supported by NSF grant DMR-1611161
and a Keith B. Macadam Graduate Excellence Fellowship in Physics and Astronomy
(2018) at the Univ. of Kentucky during a major part of this work, and by a National Postdoctoral Fellowship of SERB, DST India (NPDF/2020/001658) at the Tata Institute of Fundamental Research (TIFR) during the final stages of this work.
The work of S.P. was supported in the intial conception stages by postdoctoral fellowships at the Univ. of Kentucky (NSF grant DMR-1056536) and at the TIFR, and later by IRCC, IIT Bombay (17IRCCSG011) and SERB, DST India (SRG/2019/001419).  KD is supported at the TIFR by DAE, India, and in part by a J. C. Bose Fellowship (JCB/2020/000047) of SERB, DST India, and by the Infosys Foundation under the aegis of the Infosys-Chandrasekharan Random Geometry Center. SP and KD gratefully acknowledge the Mumbai-Pune qCMT Workshop-2018 at IISER Pune and the YIMQCMT Workshop-2018 at S.N. Bose Center Kolkata for facilitating part of this work. All authors gratefully acknowlege the 2nd Asia-Pacific Workshop on Quantum Magnetism-2018 (ICTS/apfm2018/11) at ICTS-TIFR Bengaluru for facilitating another part of this work.

\end{acknowledgments}

\bibliography{bibliography}
\bibliographystyle{apsrev4-1}
\appendix

\newpage

\appendix
\onecolumngrid
\newpage

\appendix
\onecolumngrid

\section{Structure Factor Formulae}
\label{app:struc_fac}

In order to obtain predictions that can be directly compared with our Monte Carlo results, we re-discretize the fixed point action back onto a square lattice to write
\begin{align}
	S & = \pi g_-   \sum_{\mathbf{r}}
	|\mathbf{\Delta} h_-(\mathbf{r}))|^2 
\end{align}
where $\Delta_\mu h_-(\mathbf{r})$ represents the lattice approximation to $\partial_\mu h_{-}$ in terms of the difference of $h_{-}$ between $\mathbf{r}$ and its neighbour in the $\mu$ direction, and we use periodic boundary conditions on $h$ after separating out the winding part as discussed in Sec.~\ref{sec:ScalingpictureV>0}.

Transforming to reciprocal space via a discrete Fourier transform, this Gaussian theory gives us the lattice-level structure factor for $\mathbf{q} \neq 0$:
\begin{align}
    \label{eq:bminuscorrfit}
    \langle \hat{n}_{x,-}(-\mathbf{Q}-\mathbf{q}) \hat{n}_{x,-}(\mathbf{Q}+\mathbf{q}) \rangle
    &= \frac{1}{2 \pi g^{*}_-}
    \frac{\sin^2 \frac{q_y}{2}}{\sin^2 \frac{q_x}{2}+\sin^2 \frac{q_y}{2}} \\
    \langle \hat{n}_{y,-}(-\mathbf{Q}-\mathbf{q}) \hat{n}_{y,-}(\mathbf{Q}+\mathbf{q}) \rangle
    &= \frac{1}{2 \pi g^{*}_-}
    \frac{\sin^2 \frac{q_x}{2}}{\sin^2 \frac{q_x}{2}+\sin^2 \frac{q_y}{2}} \\
    \langle \hat{n}_{x,-}(-\mathbf{Q}-\mathbf{q}) \hat{n}_{y,-}(\mathbf{Q}+\mathbf{q}) \rangle
    &= \frac{1}{2 \pi g^{*}_-} 
    \: \frac{f(q_y)^* f(q_x)}{4\sin^2 \frac{q_x}{2}+4\sin^2 \frac{q_y}{2}} \nonumber \\ 
	&= 
    \langle \hat{n}_{y,-}(-\mathbf{Q}-\mathbf{q}) \hat{n}_{x,-}(\mathbf{Q}+\mathbf{q}) \rangle^* \; ,
\end{align}
where $f(x) = 1-\exp(ix)$
{\em At} $\mathbf{q} = 0$, {\em i.e.} at the pinch-point wavevector $\mathbf{Q}$, the first two of these reduce to the mean-square winding ${\mathcal J}(g_{-}^{*}) $, while the third is zero.
These are the functional forms used for fitting in the structure factor data displayed in the main text (Fig. \ref{fig:struc_fac_repulsive}).

For the $h_{+}$ sector, we do not have an asymptotically exact fixed-point description within a renormalization group framework. However, since $y_v$ flows to strong coupling, we may model the short-ranged correlations of $h_{+}$ by a simple phenomenological action that correctly encodes the fact that our description of this strong-coupling regime must be in terms of an action that contains the effects of a nonzero density of mobile double vortices. Denoting this density of mobile vortices by $\rho(\mathbf{r})$, we thus write:
\begin{align}
F =  \sum_{\mathbf{r}} \left[
\pi g_+
|\polkd_+(\mathbf{r}))|^2 
- 
\log y_0 \:
\rho(\mathbf{r}))^2 \right]
\end{align}
where again $\Delta_\mu h_+(\mathbf{r}) = B_{\mu,+}(\mathbf{r}) =
B_{\mu,1}(\mathbf{r}) + B_{\mu,2}(\mathbf{r})$ and we have the constraint
\begin{align}
  \Delta_\mu B_{\mu,+}(\mathbf{r})) = 2 \rho(\mathbf{r})
\label{eq:lattice_divergence_3}
\end{align}
that encodes the fact that each interlayer dimer is seen as a double-vortex in $B_{\mu,+}(\mathbf{r})) $.
After transforming Eq. \ref{eq:lattice_divergence_3} to reciprocal space, we arrive at
\begin{align}
    \hat{B}_{x,+} (\mathbf{q}) f(q_x) + \hat{B}_{y,+} (\mathbf{q}) f(q_y) = 
     2 \hat{\rho}(\mathbf{q})
\end{align}
Thus, the action can now be written as
\begin{align}
    \sum_\mathbf{q} 
    \left( \begin{matrix}
    \hat{B}_{x,+}(-\mathbf{q}) & \hat{B}_{y,+}(-\mathbf{q})
    \end{matrix} \right)
    \cdot A(\mathbf{q}) \cdot
    \left( \begin{matrix}
    \hat{B}_{x,+}(\mathbf{q}) \\ \hat{B}_{y,+}(\mathbf{q})
    \end{matrix} \right)
\end{align}
where $A(\mathbf{q})=\left( \begin{matrix}
    \pi g_+ + \frac{\log (1/y_0)}{4} |f(q_x)|^2 & \frac{\log(1/ y_0)}{4} f(q_y)^*f(q_x) \\
    \frac{\log(1/ y_0)}{4} f(q_x)^*f(q_y) & \pi g_+ + \frac{\log(1/ y_0)}{4} |f(q_y)|^2
    \end{matrix} \right)$,
and consequently the $n_{+}$ correlators have the following expressions:
\begin{align}
	\label{eq:bpluscorrfit}    
    \langle \hat{n}_{x,+}(-\mathbf{Q}-\mathbf{q}) \hat{n}_{x,+}(\mathbf{Q}+\mathbf{q})\rangle 
    & = \frac{1}{2 \pi g_+} \frac{1 + \frac{\log (1/y_0)}{4 \pi g_+} |f(q_y)|^2}
    {1 + \frac{\log(1/ y_0)}{4 \pi g_+} \left(|f(q_x)|^2 + |f(q_y)|^2\right)} \nonumber \\
    \langle \hat{n}_{y,+}(-\mathbf{Q}-\mathbf{q}) \hat{n}_{y,+}(\mathbf{Q}+\mathbf{q})\rangle 
    & = \frac{1}{2 \pi g_+}  \frac{1 + \frac{\log(1/ y_0)}{4 \pi g_+} |f(q_x)|^2}
    {1 + \frac{\log(1/ y_0)}{4 \pi g_+} \left(|f(q_x)|^2 + |f(q_y)|^2 \right)} \nonumber \\
    \langle \hat{n}_{x,+}(-\mathbf{Q}-\mathbf{q}) \hat{n}_{y,+}(\mathbf{Q}+\mathbf{q})\rangle 
    & = \frac{1}{2 \pi g_+}  \frac{ \frac{\log(1/ y_0)}{4 \pi g_+} f(q_y)^* f(q_x)}
    {1 + \frac{\log(1/ y_0)}{4 \pi g_+} \left(|f(q_x)|^2 + |f(q_y)|^2 \right)} 
    \nonumber \\
    \langle \hat{\rho}(-\mathbf{q}) \hat{\rho}(\mathbf{q})\rangle 
    & = \frac{1}{2 \pi g_+}  \frac{\frac{1}{4}\left(|f(q_x)|^2 + |f(q_x)|^2\right)}
    {1 + \frac{\log(1/y_0)}{4 \pi g_+} \left(|f(q_x)|^2 + |f(q_y)|^2\right)}
\end{align}

In the main text, we have used these expressions to fit the $n_{+}$ correlators and extract the effective parameters $g_+$ and $\log(1/ y_0)$ by using these expressions;  these fits also work well and correctly capture the short-ranged correlations of $n_{+}$ (Fig.~\ref{fig:struc_fac_repulsive}) in
the bilayer Coulomb phase. 

In the limit $q \to 0$, these reduce to the more transparent expressions quoted in Sec.~\ref{sec:ScalingpictureV>0}  for the altered pattern of singular behaviour which is characteristic of the pinch-point phenomenology of the bilayer Coulomb phase. To see this, we note that in this limit, we have for nonzero but small $q$:
\begin{align}
	\langle \hat{n}_{\mu,-}(-\mathbf{Q}-\mathbf{q}) \hat{n}_{\nu,-}(\mathbf{Q}+\mathbf{q}) \rangle & =
\frac{1}{2 \pi g^{*}_-} 
\left( \delta_{\mu\nu} - \frac{q_\mu q_\nu}{q^2}\right) \\
\langle \hat{n}_{\mu,+}(-\mathbf{Q}-\mathbf{q}) \hat{n}_{\nu,+}(\mathbf{Q}+\mathbf{q}) \rangle & =
\frac{1}{2 \pi g_+}  \: \delta_{\mu\nu} \\
\langle \krho(-\mathbf{q})
\krho(\mathbf{q}) \rangle & =
\frac{q^2}{8 \pi g_+}
\label{eq:transverse_structure_factor_k12}
\end{align}
which implies the following altered singularity structure in the layer-resolved structure factor for nonzero but small $q$:
\begin{align}
\text{intralayer:  } \langle \hat{n}_{\mu,a} & (-\mathbf{Q}-\mathbf{q}) \hat{n}_{\mu,a}(\mathbf{Q}+\mathbf{q}) \rangle \nonumber \\
& = \frac{1}{4} 
\left[ \frac{1}{2 \pi g_+}    
\delta_{\mu\nu} + \frac{1}{2 \pi g^{*}_-} \left( \delta_{\mu\nu} - \frac{q_\mu q_\nu}{q^2} \right)
 \right]  
\label{eq:structure_factor1_k12a} \\
\text{interlayer:  } \langle \hat{n}_{\mu,a} & (-\mathbf{Q}-\mathbf{q}) \hat{n}_{\nu,b}(\mathbf{Q}+\mathbf{q}) \rangle \nonumber \\
& = \frac{1}{4} 
\left[ \frac{1}{2 \pi g_+}    
\delta_{\mu\nu} - \frac{1}{2 \pi g^{*}_-} \left( \delta_{\mu\nu} - \frac{q_\mu q_\nu}{q^2} \right)
 \right]  
\label{eq:structure_factor1_k12b}
\end{align}
As already noted in Sec.~\ref{sec:ScalingpictureV>0}, these expressions do not carry over smoothly to $z=0$, since the $z \to 0$ limit does not commute with the $q \to 0$ limit.

\end{document}